\documentclass[%
 superscriptaddress,
 amsmath,amssymb,
 aps,
 pra,
 longbibliography,
 twocolumn
]{revtex4-2}
\pdfoutput=1
\usepackage[utf8]{inputenc}
\usepackage{graphicx}
\usepackage{dcolumn}
\usepackage{bm}
\usepackage{physics}
\usepackage[svgnames]{xcolor} 
\usepackage[colorlinks=true,citecolor=blue,linkcolor=blue]{hyperref}
\hypersetup{
    allcolors  = {blue},
}
\usepackage{listings}
\usepackage{tikz}

\usepackage{color}
\usepackage{comment}
\usepackage[normalem]{ulem}
\usepackage{soul}

\definecolor{ColorNoteDenis}{RGB}{100, 0, 200}
\definecolor{ColorModDenis}{RGB}{150, 0, 255}
\definecolor{ColorRemove}{RGB}{255, 0, 0}
\definecolor{ColorModDenis2}{RGB}{150,0,255}
\definecolor{ColorNoteDenis2}{RGB}{150, 0, 255}
\definecolor{ColorModDenis3}{RGB}{150,0,255}

\begin{document}
\title{Modular and Looping Quantum Photonic Processing Unit for Quantum Computing}

\author{Denis Stanev}
\affiliation{Dipartimento di Fisica, Sapienza Universit\`{a} di Roma, Piazzale Aldo Moro 5, I-00185 Roma, Italy}
\affiliation{Gran Sasso Science Institute, Viale Francesco Crispi 7, I-67100 L’Aquila, Italy}
\author{Nicol\`o Spagnolo}
\email{nicolo.spagnolo@uniroma1.it}
\affiliation{Dipartimento di Fisica, Sapienza Universit\`{a} di Roma, Piazzale Aldo Moro 5, I-00185 Roma, Italy}
\author{Fabio Sciarrino}
\affiliation{Dipartimento di Fisica, Sapienza Universit\`{a} di Roma, Piazzale Aldo Moro 5, I-00185 Roma, Italy}

\begin{abstract}
Photonic processors are emerging as promising platforms in the race towards the realization of practical universal quantum computers. Several architectures with universal capabilities have been proposed for their implementation via a photonic platform, each with their own set of benefits and drawbacks. In this work we propose an architecture for a universal photonic quantum computer. The proposed architecture is modular, allowing for a high degree of flexibility and scalability, and relies on a loop-based structure, significantly reducing the number of physical components required to implement it experimentally. We prove that our architecture is universal, deterministic, and compatible with the quantum circuital model, as well as reprogrammable and capable of adaptive measurements. Furthermore, we showcase its functioning via numerical simulation on specific examples, by using machine learning techniques on a set of fundamental linear and nonlinear optical components to obtain some of the building blocks used in the architecture. The obtained results can be used to highlight the benefits and the drawbacks of the proposed architecture, as well as the route for future developments towards its physical implementation.
\end{abstract}

\maketitle

\section{Introduction}

In recent years an increasing interest has been dedicated to quantum computing, due to the prospect of a significant boost in performance in certain tasks when compared to classical computation. This has led to the development of multiple architectures for potential quantum computers, utilizing different platforms. In particular, the final aim is the development of a large-scale universal quantum computer, where universality stands for the capability of implementing in principle any computation \cite{divincenzo,chuang2010}. This requires a large amount of resources, in order also to encompass the required step of implementing error-correcting codes \cite{errorcorrection_review}.

As an intermediate step before the development of a larger scale universal quantum computer, the noisy intermediate scale quantum era has been focused in the last years on reaching the quantum advantage regime \cite{harrow17}, namely the scenario where a quantum device solves a problem faster than any classical counterpart, with a non-universal dedicated hardware. A few experiments have been recently reported tackling this scenario, employing superconducting and photonic platforms. For the superconducting platform, the \textit{Sycamore} quantum processor by Google had originally reached unprecedented size to solve the random circuit sampling problem \cite{QuantumSupGoogle}, and more recently they showed that the 105 qubit Willow processor reached an error rate lower than the threshold needed to implement quantum error correction via specific surface codes \cite{GoogleWillow}. IBM recently demonstrated the measurement of accurate expectation values for circuit volumes at a scale beyond brute-force classical computation using their noisy 127-qubit \textit{IBM Eagle} quantum processor \cite{QuantumSupIBM}, as well as showing adaptive monitored quantum circuits with conditional feedback using local mid-circuit measurements on up to 100 qubits \cite{IBMAdaptive}.

Such quantum advantage regime has been also tackled with photonic platforms \cite{OpticalQuantumComputingReview}. In particular, two optical quantum devices have reached significant size levels, that is, the \textit{Jiuzhang} \cite{QuantumSupChina, Jiuzhang3, Jiuzhang4}, and the \textit{Borealis} \cite{XanaduQAdv} systems. Both of these optical quantum devices are Gaussian Boson Samplers (GBS) \cite{BosonSampling1, BosonSampling2, GBSHardness1, GBSHardness2, GBSHardness3, yu2023universal, Graphs_GBS_china}, a type of non-universal quantum processor. More specifically, GBS is a limited model of quantum computation, useful to prove an early quantum advantage, but whose architectures intrinsically does not allow the implementation of universal quantum computing. 

\begin{figure}[ht!]
    \centering
    \includegraphics[width=1\linewidth]{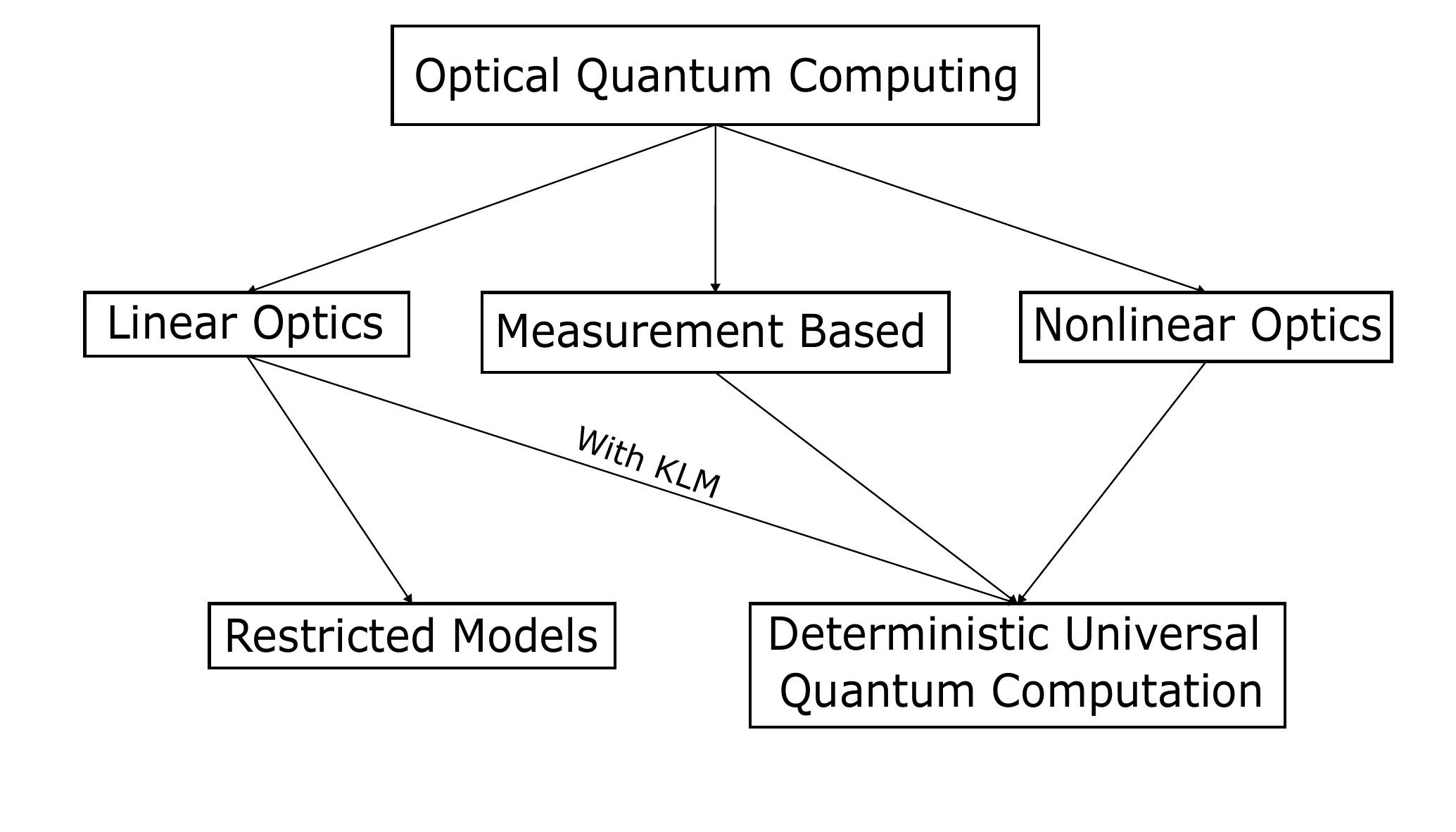}
    \caption{\textbf{Approaches for optical quantum computing.} Conceptual map representing the different architectures tailored for optical quantum computation. The architecture proposed in this manuscript uses nonlinear optical components to cause the photon-photon interaction.}
    \label{fig:ArchitecturesCompare}
\end{figure}

Among the different platforms for the implementation of universal quantum computing, photonic systems present some peculiarities, such as the capability of encoding the information in different degrees of freedom, leading to the possibility of exploiting time-bin encoding \cite{LoopLinear, BartlettLoop, TimeBinCPHASE, TimeBinCarosini}, dual-rail encoding \cite{BartlettUniversal, DualRail1}, polarization encoding \cite{flamini18}, orbital angular momentum encoding \cite{allen92} or frequency encoding \cite{lukens17}. Furthermore, other approaches beyond the qubit structure have also been developed \cite{QuantuCompCV, GKP_Orig, GKP_Error1, GKP_Error2, CVPfister, LoopLinearMultiTime}. Aside from the different possible encodings, photonic systems need to address the challenge of implementing photon-photon interactions, required for the realization of two-qubit gates. 
Some of these architectures consider only the use of linear optical components, leading for instance to restricted computational models such as Boson Sampling \cite{BosonSampling1}. When moving towards universal quantum computation, the implementation of quantum gates can be performed with linear optics in a probabilistic fashion \cite{LinearOpticalUniversalQuantumGate}. Near-deterministic universal quantum computation can be achieved with linear optics through the use of schemes such as the KLM protocol \cite{KLMScheme}, which mimic the necessary non-linearities by the measurement stage. This method is accompanied by an additional resource overhead, though the scaling is still polynomial in the number of resources needed. An alternative approach, named measurement based quantum computation \cite{MeasurementFirst, MeasurementFirst2, Measurement1, MeasurementReview, MeasurementPhotonic, PsiquantumFusion, MeasurementSakaguchi, ORCAGHZ}, allows for deterministic quantum computing and is based on preparing initially a highly entangled resource, that is, a cluster state with a large number of qubits. The computation is then performed by measuring part of the qubits of this large entangled structure. This method moves the challenge from the implementation of non-linear gates to the preparation of the large entangled resource, though some recent approaches such as fusion-based quantum computation \cite{PsiquantumFusion} have provided a significant advance in this direction. Recently, there has also been a renewed interest in quantum optical architectures that go beyond the linear optics restriction and that make use of optical nonlinearities to achieve the necessary photon-photon interaction \cite{NonlinearBSSpagnolo, QONN_Orig, ClonerQONN, QONN_Realistic, NonlinearCNOTPavia}. Using optical nonlinearities allows for the creation of deterministic universal optical quantum computers that do not rely on probabilistic gates or on measurement based gates. However, the drawback is that the physical realization of such optical nonlinearities provides additional challenges, though there have been recent developments in this direction \cite{PhotonicNonlinearity, PhotonSortingNonlinearity}. Finally, architectures specifically tailored for different encodings have been proposed \cite{LoopLinear, BartlettLoop, TimeBinCPHASE, lukens17}. All of these approaches have shown strengths and drawbacks, where the associated challenges are currently being addressed. 

The development of quantum computing architectures has shown an increasing attention to modularity, scalability, and fault tolerance, as those are considered to be fundamental requirements to achieve a practical quantum advantage. As such, several studies have been performed to develop architectures with these aspects as core elements \cite{XanaduGKP, QuandelaSpin, ORCAGHZ, PsiquantumFusion, XanaduAurora}. There have also been developments with regard to implementations of some of these architectures, with significant steps forward in the integration of multiple optical elements on a single integrated optical chip \cite{PsiQuantumFab, QuandelaAscella, DiMicco2026BosonSampling128}, paving the way for the ability to mass-produce these types of optical architectures.

Here, we introduce an architecture for an optical quantum computer where the necessary interactions can be implemented with a limited number of nonlinear components, that presents its peculiarities with respect to other approaches, while maintaining the theoretical capability of a universal quantum computer. We will discuss such an architecture for an optical quantum computer, the Modular and Looping Quantum Photonic Processing Unit (ML-QPPU), whose structure allows a fully optical system to run programs written for the quantum circuital model. It also comes with the benefit of having full connectivity between all qubits while needing relatively few components, relying on a loop-based structure, and being designed to be modular and expandable.

In Sec. \ref{sec:Architecture} we will give an overview of the structure of the architecture and of its operation, along with giving a short proof of the universality of the approach. Then, in Sec. \ref{sec:QONN} we will discuss possible realization of some of the needed components, mainly the optical CNOT gate, and give an example of operation of the architecture. Finally, in Sec. 
\ref{sec:OptionalModulesAndDiscussion} we showcase some optional modifications and variations to the architecture, and discuss how the losses and number of components needed scale with the size of the computation, as well as highlighting the main strengths and challenges of this architecture.

\section{Architecture with fiber loops}
\label{sec:Architecture}

In this Section we will provide a detailed description of the proposed architecture. More specifically, in Sec. \ref{sec:Layout_blocks} we will introduce the layout of our architecture, and describe the corresponding fundamental building blocks. Then, in Sec. \ref{sec:Universality} we will provide proof of its capacity to work as a fully optical universal quantum computer.

\subsection{Architecture layout and building blocks}
\label{sec:Layout_blocks}

\begin{figure*}[ht!]
    \centering
    \includegraphics[width=0.99\textwidth]{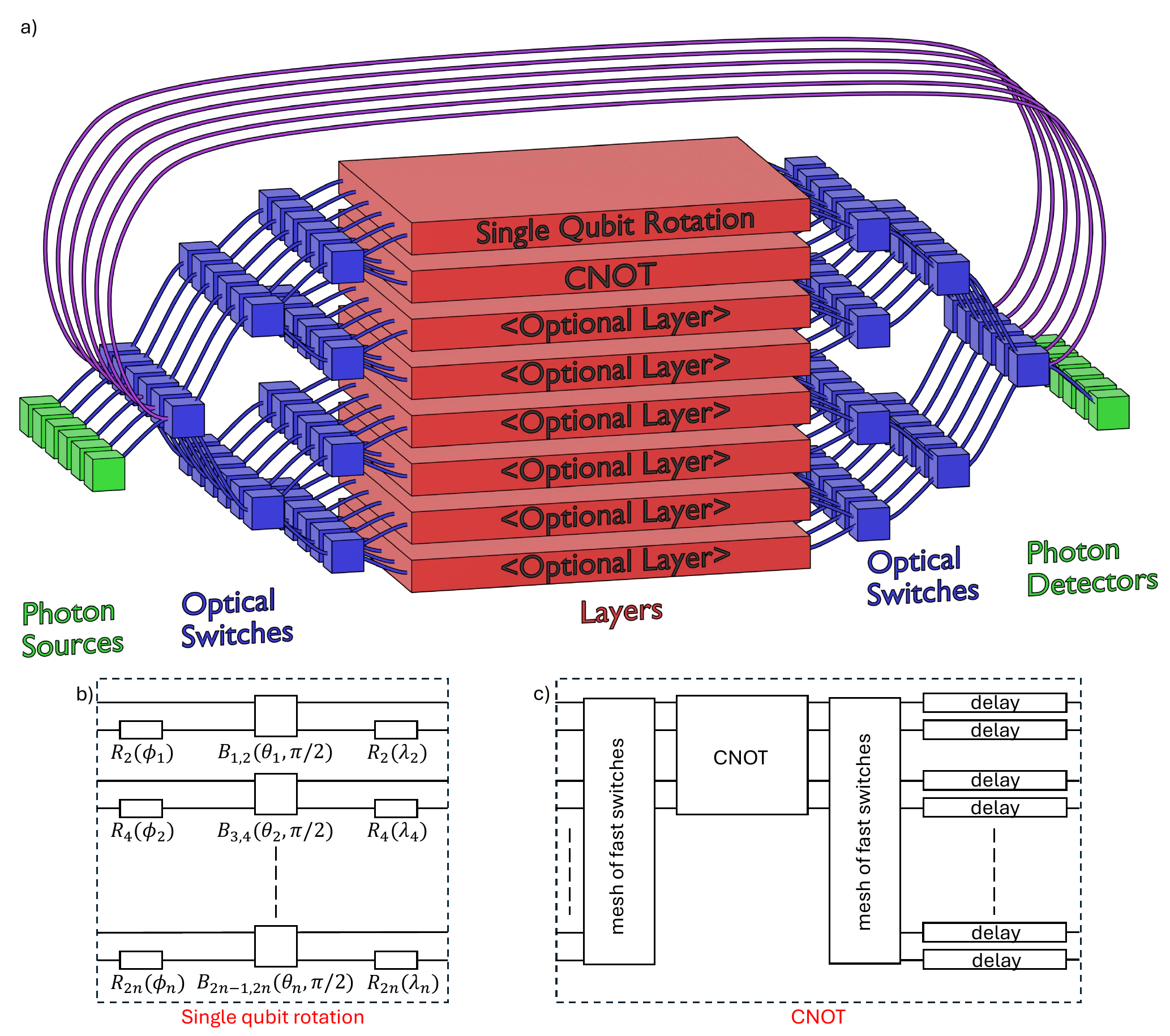}
    \caption{\textbf{Proposed architecture with 3 qubits (6 modes) and 8 layers.} a) The input resource is provided by a set of single-photon sources (green boxes, left), while the output modes after the complete computation are measured via single-photon detectors (green boxes, right). The layers section (red boxes) embeds the fundamental blocks for the computation, and the order in which photons pass through the various layers depends on the instructions given to the device based on the required computation. Two sets of programmable optical switches (blue boxes) are used to route the photons to and from the required layers. While in principle each mode can be sent to a different layer, our architecture operates using dual-rail encoding, and thus both modes corresponding to the same qubit need to be routed at each step to the same layer by the optical switch mesh. For each mode, the output of the device is connected to the input (purple wires), thus allowing the photons to pass multiple times through the system. b) Internal structure of the single-qubit rotation layer. c) Structure of the CNOT layer, comprising two switch meshes to route the desired modes at the input of the CNOT operation. The additional set of delays is required to ensure that the overall time length of the CNOT layer is $N \tau + (N-1) \epsilon$.}
    \label{fig:QuantCompArch_Side}
\end{figure*}

The proposed design of a modular architecture for quantum computing is shown in Fig. \ref{fig:QuantCompArch_Side}. Qubits are encoded in single-photon states via a dual-rail encoding, thus the $\vert 0 \rangle$ and $\vert 1 \rangle$ logical states for each qubit corresponds to the presence of the single photon in one of two distinct spatial modes.
The requirement for the input resource generation is then a set of single-photon sources, generating indistinguishable single-photon states. This step can be alternatively performed via a single emitter, generating a train of single photons in a single spatial mode and temporally separated bins, followed by an appropriate demultiplexing module \cite{Pont22,rodari2024}. Then, two sets of fast optical switches (one after the input, and one before the output) are included to route qubits into specific layers to perform each round of the computation. Indeed, the computation is performed internally in a set of layers with specific functionalities, that we detailed below. The proposed architecture is designed so that photons can pass multiple times through the device, for instance via a network of single-mode fibers acting as wires in the logical scheme, allowing the same physical components to be reused multiple times during a single computation, and thus lessening the hardware requirements. Finally, a set of single-photon detectors is placed at the output to measure the output state of the photons in a chosen basis after the computation has been performed.

The layer structure intrinsically allows to have a certain degree of modularity, thus allowing the possibility to build the system with different configurations. However, it is possible to define a minimum set of layers needed for universal computation, which are respectively: 
(i) 1 rotation layer, with tunable phase shifters on each mode and a tunable beamsplitter on each pair of modes of the same qubit, thus allowing to perform an arbitrary single-qubit rotation on each qubit; (ii) 1 CNOT layer, to perform the CNOT operation between two arbitrary qubits, with a possible implementation  using few optical nonlinearities discussed in Sec. \ref{sec:CNOT}. More specifically, at the input and output of this layer there is an arrangement similar to the two set of optical switches placed before and after the layers section. This permits any 2 qubits, corresponding to 4 spatial modes, to be routed to the input ports of the CNOT, and requires four meshes of optical switches in input and four in output, one for each of the modes of the CNOT. Such an architecture enables performing the CNOT operation on any pairwise combination of qubits by using only 1 physical device acting as the CNOT gate.

All of the layers need to have an overall propagation time of the form $N \tau + (N-1) \epsilon$, with $N$ an integer, ideally as low as possible. Here, $\tau$ is the time required by the photons to propagate through the rotation layer, while $\epsilon$ is the time required for a photon at the output of the layers section to be routed back at its input. According to this architecture, the propagation time of the photons through the CNOT layer is required to be $N \tau + (N-1) \epsilon$. This arrangement is required to ensure that, regardless of which layer the photons pass through, they are still synchronized at the output of the layers section.  As an important note, the proposed layout enables full direct connectivity between all modes, without the need to use SWAP operations such as those used in other architectures for superconducting quantum computers \cite{swap1,swap2}. The absence of SWAP operations thus allows to mitigate the amount of losses present in the system, as long as the losses introduced by the optical switches are kept to be small. 

Starting from the minimal layout discussed above, additional optional layers can be added to reduce losses or speed up specific tasks, similar to an instruction set on a CPU. We will discuss these optional layers in Sec. \ref{sec:OptionalModules}.

\subsection{Proof of universality}
\label{sec:Universality}
We now discuss the universality of the architecture described in Fig. \ref{fig:QuantCompArch_Side}. To prove that this layout enables the functionalities of a universal quantum computer, we need to argue that the system enables (i) arbitrary single qubit rotations, (ii) CNOTs between each pairwise combination of qubits and (iii) possibility of concatenating these operation with an arbitrary sequence. In such a way, the system is shown to reproduce a quantum hardware that implements a complete set of gates for universal quantum computation according to standard theory \cite{UniversalQuantumGates}.

We first start by discussing the layer regarding single-qubit rotations. Since we use the dual-rail encoding, such a layer can be constructed by relying on well-established results on the implementation of arbitrary single qubit unitary $U^s_{2i-1,2i}$, implemented for a qubit encoded in mode pair $(2i-1,2i)$. In such an encoding, this can be performed by using a beam-splitter with arbitrary transmission coefficients on the two modes of a qubit, and two phase shifters placed on one of the two modes, positioned before and after the beam-splitter operation. Considering that the beam-splitter and the phase shift operations are defined as:
\begin{eqnarray}
B_{2i-1,2i}(\theta, \varphi) &=& 
\begin{pmatrix} 
\cos(\theta) & -e^{-\imath \varphi} \sin(\theta) \\
e^{\imath \varphi} \sin(\theta) & \cos(\theta)
\end{pmatrix}, \\
R_{2i}(\varphi) &=& \begin{pmatrix} 
1 & 0 \\
0 & e^{\imath\varphi}
\end{pmatrix},
\end{eqnarray}
the overall sequence for an arbitrary single-qubit operation can be written as:
\begin{equation}
\begin{split}    
&U^s_{2i-1,2i}(\theta, \phi, \lambda) = R_{2i}(\lambda) B_{2i-1,2i}\left(\theta, \frac{\pi}{2}\right) R_{2i}(\phi) = \\
&= \begin{pmatrix} 
1 & 0 \\
0 & e^{\imath\lambda} \\
\end{pmatrix} 
\begin{pmatrix} 
\cos(\theta) & -e^{-\imath \frac{\pi}{2}} \sin(\theta) \\
e^{\imath \frac{\pi}{2}} \sin(\theta) & \cos(\theta)\\
\end{pmatrix}
\begin{pmatrix} 
1 & 0 \\
0 & e^{\imath\phi} \\
\end{pmatrix} = \\
&= 
\begin{bmatrix} 
\cos(\theta) & -e^{-\imath \frac{\pi}{2} + \imath \phi} \sin(\theta) \\
e^{\imath \frac{\pi}{2} + \imath\lambda} \sin(\theta) & e^{\imath\phi + \imath\lambda} \cos(\theta)\\
\end{bmatrix} = \\
&=\begin{bmatrix} 
\cos(\theta) & -e^{\imath\phi'}\sin(\theta) \\
e^{\imath\lambda'}\sin(\theta) & e^{\imath(\phi'+\lambda')}\cos(\theta). \\
\end{bmatrix}
\end{split}
\end{equation} 
Here $R_{2i}(\varphi)$ indicates that the phase shifter is applied to mode $2i$, while in the last step we defined $\phi' = \phi - \frac{\pi}{2}$ and $\lambda' = \lambda + \frac{\pi}{2}$. Indeed, since the rotation angles $\phi$ and $\lambda$ are arbitrary, we can instead take $\phi'$ and $\lambda'$ as parameters and freely choose their values by appropriately setting $\phi$ and $\lambda$. For an experimental implementation, the beam-splitter operation $B_{2i-1,2i}\left(\theta, \pi/2\right)$ can be alternatively implemented (up to phase factors than can be compensated via $\phi$ and $\lambda$) as a sequence of two symmetric beam-splitters, interspersed with a phase factor $2 \theta$. In such a way, all the parameters ($\theta$, $\phi$, $\lambda$) can be actively modified in the apparatus by tuning the corresponding phase shifts inserted in the modes.

The CNOT layer, as discussed above, requires the implementation of a CNOT gate. One possible solution for the optical implementation of such a gate is found in training a Quantum Optical Neural Network (QONN) \cite{QONN_Orig, ClonerQONN, QONN_Realistic} to perform this operation. This will be discussed in higher detail, both regarding the architecture and the training process, in Sec. \ref{sec:CNOT}. As previously stated, our architecture needs only one physical CNOT and additional switches that are used to route the modes corresponding to any two qubits in the input ports of the gate.

Finally, concatenation of these operations is obtained via the fast switches and the rotation layer, where the latter element can be used by setting the parameters so that no rotation is applied on the desired qubit, and thus causes only a delay in time for the associated photon. This is required to keep temporal synchronization between the photons. This is a relevant feature for the proposed scheme, given that the implementation of the CNOT gate via the QONN is based on interference between the photons witihin the circuit. In summary, the ML-QPPU can perform any arbitrary single qubit rotation and the CNOT gate, thus implementing a set of universal quantum gates \cite{UniversalQuantumGates}. Since, due to the loop-based design, the architecture can be programmed to repeat these gates in arbitrary sequence and order depending on the desired computation, this design can reproduce any quantum circuit.

\section{Optical scheme for the CNOT gate via Quantum Optical Neural Networks}
\label{sec:QONN}

We now discuss the implementation of a deterministic optical CNOT gate (Sec. \ref{sec:CNOT}), and show an example of operation of the architecture above by discussing a specific task (Sec. \ref{sec:cloning}).

\begin{figure}[ht!]
    \centering
    \includegraphics[width=1\linewidth]{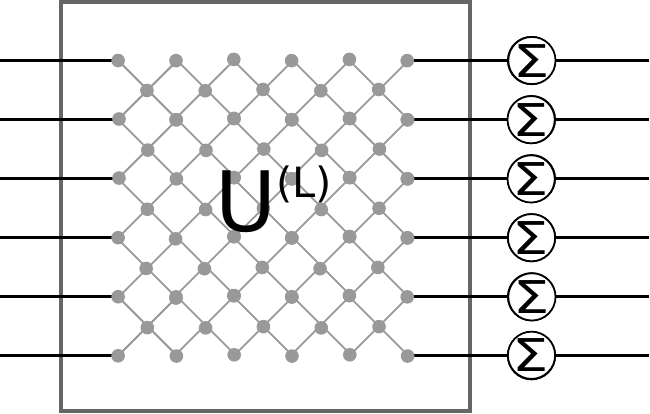}
    \caption{\textbf{Structure of a single layer of the quantum optical neural network}. The full QONN is composed by multiple sets of these layers placed in sequence. Each layer has a linear part, which performs the linear evolution $U^{(L)}_{k}$, and a nonlinear part, which is composed of single-site nonlinearities with a Kerr-type interaction $\Sigma(\chi)$. The general linear transformation $U^{(L)}_{k}$ can be implemented optically using a sequence of beamsplitters $B_{j,l}(\theta_{i}^{k},\phi_{i}^{k})$ and phase shifters $R_{j}(\varphi_{i}^{k})$, according to the structure defined by Clements et al. \cite{Clements2017Optimal}.}
    \label{fig:QONNStruct}
\end{figure}

Before going into details of this analysis, we first discuss the Quantum Optical Neural Network paradigm, introduced and analyzed in Refs. \cite{QONN_Orig, ClonerQONN, QONN_Realistic}. QONN is a class of optical network designed specifically to represent the quantum optical version of classical neural networks. Like its classical counterpart, it is composed of multiple layers, each composed of a linear unitary transformation $U^{(L)}_{k}$ and a set of nonlinear single-site transformations, in our case single mode Kerr nonlinearities $\Sigma(\chi)$. Input states into the network are photons representing qubits through dual-rail encoding, which associates each pair of optical modes to a qubit. We can see a representation of a layer of the QONN structure in Fig. \ref{fig:QONNStruct}. In Refs. \cite{QONN_Orig, ClonerQONN, QONN_Realistic} it was shown that such an architecture can be trained to implement different classes of transformation, thus being a versatile tool for its adoption as building block in a more complex quantum optical architecture.

\subsection{Training and verification for the CNOT gate}
\label{sec:CNOT}
The possibility of obtaining a CNOT by training a QONN has already been shown in a previous work by Steinbrecher et al. \cite{QONN_Orig}. Here, we provide a refined version of the training process, which allows to compensate additional phase shifts which are present in the output modes of the circuit according to the original procedure. Indeed, compensating for such a phase shifts is relevant if one needs to embed the CNOT gate within a more complex computation composed of a sequence of gates.

More specifically, the QONN is trained to implement the following gate:
\begin{equation}
\mathrm{CNOT} = \begin{pmatrix} 
1 & 0 & 0 & 0 \\
0 & 1 & 0 & 0 \\
0 & 0 & 0 & 1 \\
0 & 0 & 1 & 0
\end{pmatrix}.
\end{equation}
Training of the QONN is performed via a training set composed by a subset of all possible input states, and by progressively tuning the parameters of the system to obtain the desired operation. More specifically, the parameters to be tuned are those related to the beam-splitter transformations and to the phase shifters of the linear part, while the Kerr non-linearities are kept fixed to $\pi/2$ (more details can be found in \cite{ClonerQONN}). A relevant aspect regards an appropriate choice for the set of training states. If one limits this choice only to the states of the computational basis $[ \ket{00}, \ket{01}, \ket{10}, \ket{11}]$, the training process cannot distinguish relative phases between the states, thus obtaining an output circuit of the form:
\begin{equation}
\label{eq:CNOTprime}
\mathrm{CNOT}^{\prime} = \begin{pmatrix} 
1 & 0 & 0 & 0 \\
0 & e^{i\alpha} & 0 & 0 \\
0 & 0 & 0 & e^{i\beta} \\
0 & 0 & e^{i\gamma} & 0
\end{pmatrix},
\end{equation}
where $(\alpha, \beta, \gamma)$ are phase factors which can vary between different trainings.

To obtain the correct phase shifts, one can employ an extended set of states for the training, sensitive to the relative phases between the modes. This can be performed by adding 6 states to the computational basis set, two for each phase of the set $(\alpha, \beta, \gamma)$. In principle, only a single additional state for each phase is strictly necessary. However, having a slightly bigger dataset can help with the training process. The setups used to prepare and measure the additional 6 states are reported in Fig. \ref{fig:QONNCNOT}, while more details on the form of the states can be found in App. \ref{sec:CNOTCalc}.

\begin{figure}[ht!]
\centering
\includegraphics[width=0.49\textwidth]{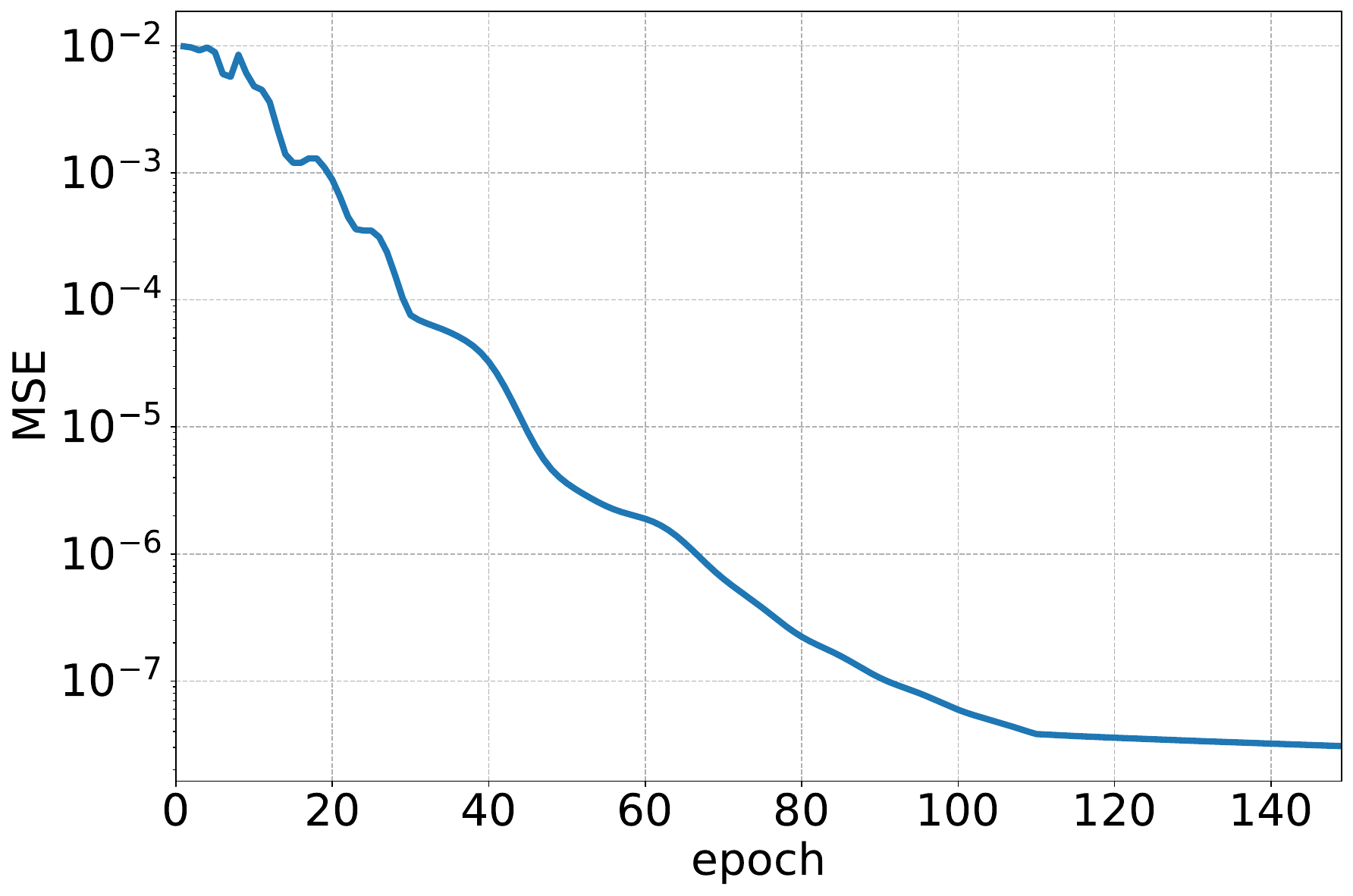}
\caption{\textbf{Training of the CNOT gate.} Evolution of the mean square error during the epochs of the numerical simulation of a 2-layer QONN to act as a CNOT gate.}
\label{fig:training_cnot}
\end{figure}

We have then simulated numerically the training process to obtain a CNOT transformation, by using a QONN architecture composed of 2 layers. Since the output state has at most one photon in each mode, the last layer of nonlinearity does not take any role in producing the desired output state, and thus the resources to implement a CNOT transformation can be reduced to only one nonlinear layer, and two linear layers. In Fig. \ref{fig:training_cnot} we show an example of evolution of the loss function during the epochs of the training process. As loss function, the mean square error $\mathrm{MSE} = \frac{1}{n} \sum_{i=1}^{n} (p_{i} - \hat{p}_{i})^2$ is employed, being $p_{i}$ the target probabilities of the different output configurations and $\hat{p}_{i}$ the ones obtained from the QONN at a given epoch. As discussed in \cite{ClonerQONN}, the training stage can be further improved by adding a refinement using the mean absolute error $\mathrm{MAE} = \frac{1}{n} \vert p_{i} - \hat{p}_{i} \vert$. The performances are quantified by the error rate $p_{\mathrm{err}} = 1 - p_{\mathrm{target}}$, being $p_{\mathrm{target}}$ the probability of obtaining the correct target rate. The trained CNOT via the performed numerical simulations is found to have an average error rate of $\overline{p}_{\mathrm{err}} \sim 3.9 \times 10^{-7}$, with the maximum error rate amongst the states on which it was trained being $p^{\mathrm{max}}_{\mathrm{err}} \sim 7.7 \times 10^{-7}$.

This architecture has also a potential advantage in terms of long term stability of the CNOT operation. First, additional steps of the training process can be performed directly on the physical device, to compensate for any manufacturing defects. This could also be done periodically, acting as a device calibration. In such a way, it is possible to compensate for any small variation of the parameters in the experimental platforms caused by external factors. Furthermore, regarding the effect of physical imperfections on the functioning of the CNOT trained with a QONN, other works have already shown that the QONN still works even in the presence of imperfections. These include small errors in the linear part of the evolution $U^{(L)}_{k}$, or when weaker Kerr nonlinearities ($\chi < \pi/2$) are used, albeit with the cost of adding more layers to the QONN to achieve a comparable performance in those scenarios \cite{QONN_Realistic,ClonerQONN}.

\subsection{Example of operation: optimal universal quantum cloning}
\label{sec:cloning}
Here we show how to program a device built according to the proposed architecture, to implement a universal $1 \rightarrow 2$ optimal quantum cloning circuit. This provides a specific example of operation for the proposed architecture, as well as a further validation of the trained CNOT via QONN. The numerical simulation of the device is performed by flattening the corresponding logical circuit. However, this approach produces an equivalent result to the theoretical operation of the device.

\begin{figure}[ht!]
    \centering
    \includegraphics[width=1\linewidth]{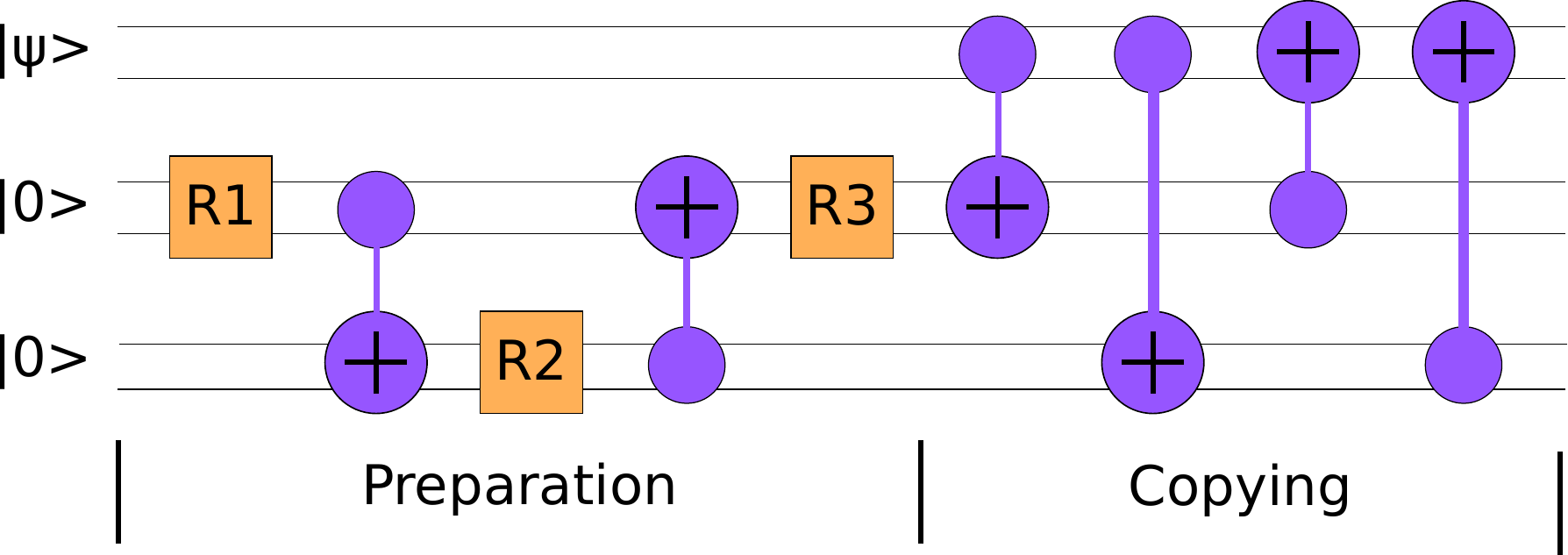}
    \caption{\textbf{Quantum circuit for the optimal universal quantum cloning machine}. Each qubit is encoded in 2 optical modes. Here, $\mathrm{R1}=(\theta_1, \phi_1, \lambda_1)$, $\mathrm{R2}=(\theta_2, \phi_2, \lambda_2)$ and $\mathrm{R3}=(\theta_3, \phi_3, \lambda_3)$ are the set of parameters representing the appropriate values of the rotation for the single qubit rotations. The copies of the input qubit are encoded at the outputs in qubits 1 and 2.}
    \label{fig:ClonerCircuit}
\end{figure}

The layout for the universal optimal quantum cloning machine in the circuit model \cite{QuantumClonerStructure} is shown in Fig. \ref{fig:ClonerCircuit}. With our architecture, we can program the device to implement in our fiber-loop architecture the sequence of operations corresponding to the logical circuit. More specifically, an instruction to perform a gate (be it a single qubit rotation, a CNOT, or anything else) is performed by sending the corresponding photons/modes to the layer that performs said gate, through the use of the optical switch mesh before the layer. Then, the photons are retrieved from the end of the layer through the corresponding optical switch mesh after the layers, so that they can be used again for the subsequent operation. The complete set of instructions for this operation is discussed in App. \ref{sec:OperationDetails}.

\begin{figure}[ht!]
    \centering
    \includegraphics[width=0.49\textwidth]{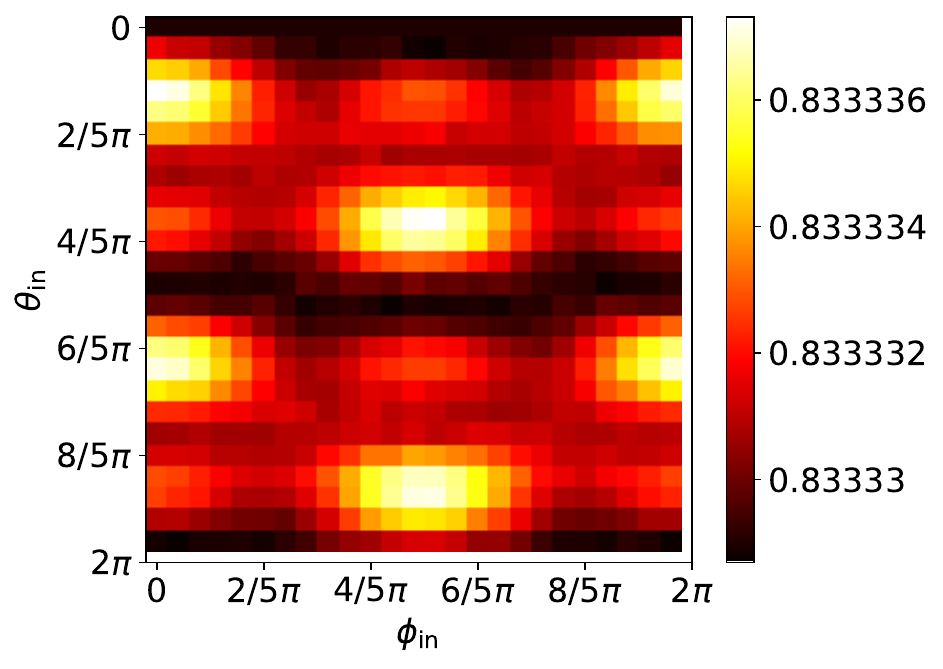}
    \caption{\textbf{Fidelity heatmap for the universal optimal quantum cloner}. Fidelities obtained for the numerically simulated circuit based on the proposed architecture with the trained QONN-based CNOT gate, evaluated with a set of equally spaced $25 \times 25$ input states over the Bloch sphere parameter $(\theta_{\mathrm{in}}, \phi_{\mathrm{in}})$.}
    \label{fig:Heatmap_Cloner}
\end{figure}

Implementing this circuit on a simulation of our device we obtain a deterministic optical quantum cloner. Verification of its operation is found by testing it on 1000 random input states, obtaining an average fidelity of the cloned states equal to $\overline{F} \sim 0.8333315$, where the optimal fidelity from the theory is $F_{\mathrm{opt}} = 5/6$, and the minimum fidelity found among the 1000 different states was $F_{\mathrm{min}} \sim 0.833328$. Here, the fidelity is estimated by projecting the output states in the basis corresponding to the target output state, without applying post-selection. In Fig. \ref{fig:Heatmap_Cloner} we also show a fidelity heatmap for a grid of $25 \times 25$ different values of the parameters $(\theta_{\mathrm{in}}, \phi_{\mathrm{in}})$ of a generic input qubit $\vert \psi \rangle = \cos(\theta_{\mathrm{in}}/2) \ket{0} + e^{\imath \phi_{\mathrm{in}}} \sin(\theta_{\mathrm{in}}/2) \ket{1}$. These results give an additional validation to the CNOT that we trained through the use of the QONN, given that such a building block is embedded in a more complex circuit where the transformation phases are relevant.

\section{Optional modules, scalability and discussion}
\label{sec:OptionalModulesAndDiscussion}

The proposed architecture is intrinsically modular, thanks to the layer-based structure. This means that, besides the fundamental elements described above enabling universal quantum computation, additional modules can be inserted for specific applications. Adding such functionalities may be useful in certain scenarios\textcolor{ColorModDenis}{,} for instance to boost the computational time, as well as reducing the number of losses in case less steps are required. In this Section, we discuss possible integrations to the basic architecture, and then provide an analysis on the scalability of this approach in terms both of number of optical components and losses. We then conclude by discussing the main advantages and disadvantages of the architecture.

\subsection{Optional modules}
\label{sec:OptionalModules}

Here, we provide some examples of additional modules that could be added to the architecture. 

\textbf{Additional CNOT layers.} A first possible class of optional layers would correspond to adding additional CNOT layers. This option would allow for multiple CNOT operations in parallel, thus speeding up execution of instructions for those algorithms where multiple CNOT gates between distinct pairs of qubits are present in a round of computation.

\textbf{Specialized tasks.} A second option is to include in the architecture specialized layers that perform a given task. An example is provided by a layer that performs deterministic quantum cloning, that analogously to the CNOT gate could be implemented by an appropriately trained QONN \cite{ClonerQONN}. In general, specialized layers could be added to the device as a way to speed up the solution of recurring sub-routines in a given algorithm, thus allowing to solve such an operation in shorter time and with reduced losses. In case these specialized layers are implemented via a QONN structure, an additional step of calibration can be done directly on the device to compensate for fabrication imperfections and fluctuations in the performance.

\textbf{Delay lines layer.} A third optional layer is made of delay lines, to induce delay between the photons. This layer can be used to minimize the losses and the errors in the cases where some of the qubits need to wait for other qubits to finish their step of computation before they can be used, in order to maintain the temporal synchronization between the photons. As with every other layer, this layer requires a time duration of $N \tau + (N-1) \epsilon$. In the simplest configuration, its duration can be $\tau$, so as to replace the rotation layer in cases where the desired qubits need to wait for others, and can lead to lower errors and losses when compared to passing through a rotation layer that applies no rotation. It is also possible to have delay lines layers with longer delays, still following the constraint of having the duration of the delay be in the form $N \tau + (N-1) \epsilon$. This can be useful in cases where some qubits may not need to be used in the computation for a longer period of time, and can serve to further minimize the losses by avoiding going through the optical switch meshes multiple times when no additional operation needs to be performed on said qubits. This layer does not expand the computational capabilities of the device, since the rotation layer can perform the same operation by setting the parameters so that it does not apply any rotation, but it can help in minimizing losses and errors.

\textbf{Interfacing multiple devices.} Another possible optional layer in the architecture would be an input-output one. This layer, while not changing the computational capability of the architecture, would allow to send and receive qubits, and thus a device according to the proposed architecture could be interfaced with more specialized devices. Furthermore, an input-output layer would also enable interfacing multiple devices following this architecture. This would allow to interface multiple small units through the use of a routing system. This could provide advantages in terms of losses, and allow for circuits with a higher number of qubits to be run on several connected smaller devices, at the cost of decreased interconnectivity. In general, the benefit of such a multi-device approach would depend on the algorithm that one needs to execute. The corresponding implementation can be obtained by physically connecting the input and output layers of multiple devices with a central router. Furthermore, we note that the photon sources and photon detectors can also be integrated into a single layer using a similar method. 

\textbf{Perspective: quantum memory.} Another long term perspective layer which can be added would correspond to adding short-term photon storage capabilities, via a quantum memory \cite{QuantumMemory1}. This module would be accompanied with a mesh of optical switches (similar to the one at the input and output of the CNOT layer), used to route the photons coming from the chosen modes into free memory slots. Adding layers of memory/storage could enable running circuits that require a larger number of modes than those physically available on the device. This could be performed by using only some of the photons within each loop while storing the rest of them in the photon storage layer. The photons used at each loop are then swapped depending on the algorithm instructions. In case the storage layer is implemented with an amount of losses smaller than having the photons propagate through fiber switches and loops, such module could also be used to minimize losses by storing photons that are not used for many cycles in the storage layer.

\textbf{Reconfigurable Quantum Optical Neural Network.} In the presence of reconfiguration capabilities, a layer composed by a QONN layer can be in principle used multiple times in sequence by progressively reconfiguring the operation of the linear part. By passing multiple times through such a layer programmed with different parameters, one could obtain a QONN with many layers, while using a limited number of components. Depending on the response time of the reconfigurable parts of the QONN layer, more than one physical layer might be needed, with photons passing through the various layers in order to wait for the response time of the reconfigurable elements. In general, with such an approach, one could reproduce a QONN with an arbitrary number of layers by having only a fixed number of physical layers in the device.

\subsection{Scalability: number of optical components}

We can now  discuss how the number of required resources and components scales with increasing size of the computation. On one side, scaling of the number of optical modes is a direct consequence of the dual-rail encoding used within this architecture. Regarding the number of components, this is one of the main strengths of the proposed architecture. The resources needed can be divided in three groups, (a) those elements whose number is being independent from the number of photons/qubits $n$ and the number of layers $L$, (b) those elements whose number scales linearly with $n$ and (c) those elements whose number scales linearly with $n$ and $L$. 

Regarding group (a), this design and the use of QONN for the CNOT implementation requires only 4 single-mode nonlinearities, independently from the number of qubits $n$ and iterations $N$. A similar argument can be used for the linear layers of the QONN, where again the number of elements do not depend on the number of qubits $n$ and on the number of iterations $N$. More specifically, considering the need of implementing two arbitrary 4-mode interferometers, this leads to the requirement of an overall number of 12 tunable beamsplitters for the CNOT. On top of that, there are ways to use only a single physical nonlinearity to obtain multiple logical nonlinearities, like switching from a space encoding to a time encoding before passing through the nonlinearity, passing all of the modes through the single nonlinearity at different times, and then reconverting to spatial encoding after the nonlinearity \cite{PhotonicNonlinearity}. Thus only a single physical nonlinearity would be needed for the implementation of the 4 logical nonlinearities needed for the CNOT. This would come at the cost of a longer CNOT with corresponding lower transmission efficiency, and thus a trade-off needs to be established between the reduced number of physical components and the larger amount of losses. Finally, a possible route for large scale circuits could also involve the possibility of having multiple CNOTs in the device, so as to be able to execute multiple CNOT operations in parallel. This would come at the cost of an increased number of logical single mode nonlinearities, that is, 4 for each additional CNOT.

Next, we can note that the number of linear elements needed for the single-qubit operations belong to group (b), since their number scales linearly with the number of qubits. Indeed, for each new qubit added a single beamsplitter and 2 phase shifters need to be added for the single-qubit rotation.

Finally, the remaining set of elements is formed by the fast optical switches, which belong to group (c) since the required number scales linearly with $n$ and $L$. To be more precise, for each mode $m$, the meshes needed at the layer selection stage require a number of switches proportional to $L$. Since each mode needs 2 of these meshes, namely one at the input and one at the layer outputs, the overall number of switches needed to route the photons to the different layers is approximately $2mL = 4nL$. To this we must also add the number of switches needed for the meshes in the CNOT, which scales linearly with the number of qubits. Given that there are 8 such meshes, 4 in input and 4 in output, the number of switches needed for the CNOT scales like $8n$. Finally, 2 additional switches for each mode are needed to route the photon to the desired mesh, one in input and one in output, leading to an additional number of $2m = 4n$ switches. Thus, the total number of switches scales as $O(4nL + 12n)$, assuming the use of a single CNOT layer. We can thus see that the number of switches scales linearly with the number of qubits $n$ and layers $L$, but it does not depend from the number of iterations $N$.

\subsection{Scalability: Losses}
One of the main aspects that needs to be addressed for optical quantum computing applications is the amount and impact of losses within the architecture. Here, we address how the presence of losses scales in ML-QPPU systems, with respect to the number of resources used for a specific computation. More specifically, we consider the scaling with the number of photons/qubits $n$ and with the number of iterations within the loop structure $N$. The latter parameter depends on the specific algorithm. We discuss below the loss mechanism per photon propagating through the structure.

On a first note, one needs to consider the efficiency $\eta_{\mathrm{source}}$ of the photon generation system, and the overall detection efficiency $\eta_{\mathrm{det}}$. In both cases, this parameter represents a constant loss for each photon involved in the computation, and thus scales as $O(1)$ per photon, while corresponding to an overall efficiency for $n$ qubit computation as $\eta_{\mathrm{source}}^{n}$ and $\eta_{\mathrm{det}}^{n}$ respectively. Conversely, these mechanisms are independent from the number of iterations $N$. 

The second set of loss mechanism depends in general from the number of iterations $N$ required by the performed computation. These include the losses due to the fiber loops (transmission efficiency $\eta_{\mathrm{loop}}$), which scales as $O(N)$. Analogously, other mechanisms that depend on the number of iterations are those occurring within the single layers, such as the ones due to (i) the number of times ($N_{\mathrm{sq}}$) a single-qubit gate is applied (transmission efficiency $\eta_{\mathrm{sq}}$) and (ii) the number of applications of the CNOT gates ($N_{\mathrm{CNOT}}$) in the corresponding layer (transmission efficiency $\eta_{\mathrm{CNOT}}$). Regarding the latter efficiency, in case the CNOT layer is implemented via the QONN, this corresponds to the overall transmission efficiency deriving from the sequences of linear and nonlinear optical elements composing that architecture. The effective number of all terms ($N_{\mathrm{sq}}, N_{\mathrm{CNOT}}$) depends on the specific algorithm. However, on average all terms can be considered to be a fraction of the number of iterations $N$, and thus all terms scale as $O(N)$.

Finally, a separate discussion needs to be performed for the losses occurring due to the optical switches (transmission efficiency $\eta_{\mathrm{switch}}$), which are present in the ML-QPPU architecture in different points, namely at the input/output of the layer structure and within the CNOT layer. In the first case, the number of times a photon passes through a switch is $2 N \log_2 L$. Conversely, by optimizing the positioning of the optical switches in the CNOT layer, each photon undergoing $N_{\mathrm{CNOT}}$ applications of the CNOT gate has to propagate through $2 N_{\mathrm{CNOT}} + 2 N_{\mathrm{CNOT}} \log_2 n$ switches. Overall, this  corresponds to having an overhead per photon of $\eta_{\mathrm{switch}}$ scaling as $O(2 N \log_2 L + 2 N_{\mathrm{CNOT}} + 2 N_{\mathrm{CNOT}} \log_2 n)$.

We have thus seen that the losses per photon scale linearly with the number of iterations of the ML-QPPU, and logarithmically with the number of layers and of photons (see summary in Table \ref{tab:loss}). With this, we can argue that the component whose losses provides a larger impact in the overall efficiency are those occurring at the optical switches, since a $n$-qubit computation with $N$ iterations is performed with a transmission efficiency of $\sim \eta_{\mathrm{switch}}^{n  N (2 \log_2 L + 2 + 2 \log_2 n)}$. Hence, fast low-loss components of this kind are an essential requirement for this architecture.

\begin{table}[ht!]
\begin{tabular}{|c|c|}
\hline
loss mechanism efficiency & scaling per photon \\
\hline
$\eta_{\mathrm{source}}$ & $O(1)$ \\
$\eta_{\mathrm{det}}$ & $O(1)$ \\
$\eta_{\mathrm{loop}}$ & $O(N)$ \\
$\eta_{\mathrm{sq}}$ & $O(N)$\\
$\eta_{\mathrm{CNOT}}$  & $O(N)$\\
$\eta_{\mathrm{switch}}$ & $O(2 N \log_{2} L + 2 N+ 2 N \log_{2} n)$\\
\hline
\end{tabular}
\caption{Summary of the main losses occuring in the architecture, and the corresponding scaling per photon. As an example, the $O(N)$ scaling per photon for $\eta_{\mathrm{sq}}$ corresponds to a transmission efficiency per photon of $\sim \eta_{\mathrm{sq}}^{N}$, and thus an overall efficiency for a $n$ qubit computation of $\sim \eta_{\mathrm{sq}}^{nN}$.}
\label{tab:loss}
\end{table}

When targeting to perform fault-tolerant quantum computation, one clearly needs to properly take into account the presence of optical losses analogously to other photonic architectures. Since the ML-QPPU is compatible with the circuital model, this means that we can also leverage error correction techniques developed for the circuital model \cite{errorcorrection_review}, such as CSS codes \cite{CSSCode1, CSSCode2}, Bacon-Shor codes \cite{BaconCode}, toric codes \cite{ToricCode}, and surface codes \cite{SurfaceCodeBase}. For instance, it has been shown that surface codes can correct up to $50\%$ qubit loss in the absence of other errors \cite{SurfaceCodeCorrection50P1, SurfaceCodeCorrection50P2}. These error correction processes require additional overhead in the architecture. Typically, error-correcting codes are based on encoding logical qubits for the computation, rather than acting directly on physical qubits. This means that implementation of a single logical qubit requires using multiple physical qubits. Thus, the use of such approaches to improve the robustness to loss of the architecture can be performed at the cost of an increase to the number of physical qubits required for a specific computation. However, it should be noted that the high level of connectivity of our architecture is likely to be of benefit in some scenario. For instance, when implementing surface codes, the high connectivity can allow operations between distant qubits without the need for swap gates, thus decreasing the overall losses incurred in those situations \cite{SurfaceCodeConnectivity}.

\subsection{Discussion}
\label{sec:Discussion}

This architecture has its own set of strengths and challenges, that we aim to summarize in this section. On a first note, as previously discussed, one of the main strengths of this architecture is that it requires relatively few physical components, and that their number scales at most linearly with the number of qubits and layers, and does not scale with the length of the circuit, due to its looping nature. Thus, this can allow the creation of circuits of arbitrary length with just a limited number of physical components.

Due to the structure of this architecture, it is compatible with the representation of quantum information as qubits in the circuital model. As such, this architecture has access to the full set of algorithms designed for the quantum circuital model. The architecture also presents full connectivity between all qubits and, as we have proven in Section \ref{sec:Universality}, it is capable of universal quantum computation. It is also programmable, although that comes at the cost of potentially higher losses for certain circuits when compared to fixed designs, due to the need to pass through the loop and the mesh of optical switches at each iteration. This architecture also has the ability to perform adaptive operations, and even alter the length and structure of circuits, based on measurements performed during the operation of the device. This can be useful to run circuits which require adaptive measurements \cite{QML_Adaptive}. 

Furthermore, the modularity of the architecture allows to adapt its design to embed novel components, and gives the possibility of adding specialized layers for common operations, thus leading to potentially faster operation and lower losses in specific scenarios. Furthermore, as we have already noted in Section \ref{sec:CNOT}, the layers obtained using a QONN, including the CNOT layer, could compensate for manufacturing defects and oscillations in performance caused by external factors by periodically performing a calibration process directly on the physical device. This feature can be useful to provide a first layer of error reduction mechanism. 

Conversely, one of the main challenges for this architecture is the need for the Kerr nonlinearities, required for the QONN component implementing the CNOT gate. While recent developments in photon-photon nonlinearities have shown significant progress in the field, such as a recent paper by Nielsen et al. \cite{PhotonicNonlinearity} in which a programmable nonlinear quantum photonic circuit is reported, further work is still required in this direction. Other significant experimental challenges require the need of precise and possibly tunable delays, given the sensibility of the architectures to mismatch in the optical length which is enhanced by its looping structure. Additionally, a further experimental requirement is the need of maintaining the phase stability of the photons, especially in longer circuits, which require careful design as well as possible active phase stabilization hardware.

Another challenge in scaling the device, which it has in common with most quantum optical devices which use single photons, has to do with generating a large number of indistinguishable single photons \cite{PhotonIndistinguishability1, PhotonIndistinguishability2, PhotonIndistinguishabilityDifferentSources}. Starting from the significant advances performed in the last period, further technological improvements are expected in the next period to reach almost deterministic single-photon generation.

Finally, we leave as an open question what effect will correlated errors have on the proper functioning of the device \cite{CorrelatedNoise}, and especially if they could interfere with potential future error-correction attempts on the device. Indeed, given that the photons pass through the same elements multiple times during the computation, the errors at each step present a certain degree of correlation, and thus standard error correction protocols devised for uncorrelated noise may not be effective, requiring the development of novel protocols.

\section{Conclusions}

In this work we have introduced an architecture for a universal quantum computer on an optical platform, the ML-QPPU. We have described its structure and given some examples of operation, and have also showcased some potential variations and additions to the device. We have proven that it is capable of performing universal quantum computation, that it is reprogrammable, and that it is capable of adaptive measurements. We have also discussed the use of machine learning, more specifically QONNs, to obtain some of the required components by using only a limited number of optical nonlinearities. Furthermore, we have illustrated how the losses and the number of components scale with the size of the computation. This architecture relies on a loop-based structure, thus allowing for a low number of overall components needed for its physical implementation. Furthermore, being a modular design, it allows for high flexibility, given that future improvements to the design can be readily introduced following technological advances. 

As one of the main features, this architecture could benefit from connecting quantum optical computation with the quantum circuital model, thus allowing the quantum optical platform to benefit from all of the optimizations and algorithms that have been developed for the circuital model. This would also allow hardware independent quantum languages \cite{HardwareIndipendent} to be used on the quantum optical platform, thus allowing for a more unified and intercompatible quantum computing landscape.

Future work towards its actual implementation is required. On one side, besides the recent advances further improvements are necessary for the implementation of the required nonlinearities for the QONN module. Other challenges can be found in tackling the problem of phase stability for longer circuits. On the theoretical side, there is room for further analysis to evaluate the impact of experimental imperfections and correlated errors on the device. 

\begin{acknowledgments}
This work is supported by ICSC – Centro Nazionale di Ricerca in High Performance Computing, Big Data and Quantum Computing, funded by European Union – NextGenerationEU and by the European Union’s Horizon Europe research and innovation program under EPIQUE Project (Grant Agreement No. 10113528).
\end{acknowledgments}

\appendix

\section{Details on the training of the CNOT}
\label{sec:CNOTCalc}
The training procedure of the CNOT gate implemented via a QONN architecture is performed by using a set of states which includes at least a subset whose output is sensitive to the relative phases. More specifically, there are three phases $(\alpha, \beta, \gamma)$ to be trained in the $\mathrm{CNOT}^{\prime}$ transformation of Eq. \eqref{eq:CNOTprime}. The training set is thus composed by a set of 10 states, among which the 4 states of the computational basis $(\vert 00\rangle, \vert 01 \rangle, \vert 10 \rangle, \vert 11 \rangle)$, and 6 additional states.

\begin{figure}[ht!]
    \centering
    \includegraphics[width=0.49\textwidth]{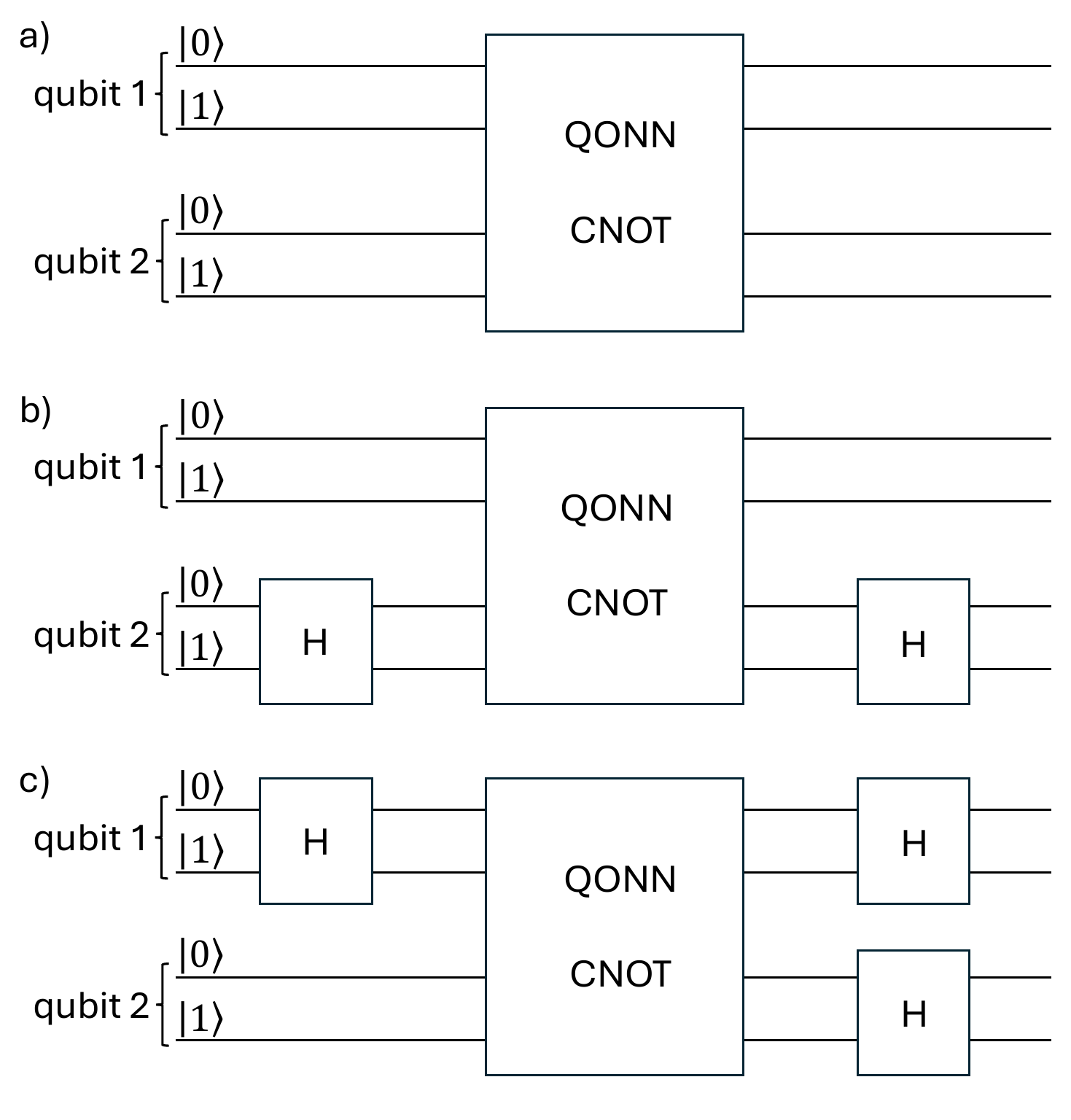}
    \caption{\textbf{Circuit layouts to train the QONN to act as a CNOT}. Here, $H$ represent the Hadamard gate, and is realized with a beam-splitter and two phase shifters. a) The input states are those belonging to the computational basis $ \ket{00}, \ket{01}, \ket{10}$ and $\ket{11}$, while the target output states the expected output from the CNOT gate, namely $ \ket{00}, \ket{01}, \ket{11}$ and $\ket{10}$ respectively. b) The input states are those belonging to the computational basis $ \ket{00}, \ket{01}, \ket{10}$ and $\ket{11}$, and the target output states are $ \ket{00}, \ket{01}$, $\ket{10}$ and $\ket{11}$ respectively. The first pair of states are sensitive to phase $e^{\imath\alpha}$ in $\mathrm{CNOT}^{\prime}$, while the second pair of states are sensitive to the difference between the phases $e^{\imath\beta}$ and $e^{\imath\gamma}$. c) The input states are $\ket{00}$ and $\ket{10}$, while the target output states are $\frac{\ket{00}+\ket{11}}{\sqrt{2}}$ and $\frac{\ket{01}+\ket{10}}{\sqrt{2}}$, and are sensitive to the phase $e^{\imath\gamma}$.}
    \label{fig:QONNCNOT}
\end{figure}

The part of the training performed with the states of the computational basis is performed according to the configuration of Fig. \ref{fig:QONNCNOT}a. More specifically, for each input state the target is chosen according to the element of the CNOT gates, namely $\vert 00 \rangle \rightarrow \vert 00 \rangle$, $\vert 01 \rangle \rightarrow \vert 01 \rangle$, $\vert 10 \rangle \rightarrow \vert 11 \rangle$, $\vert 11 \rangle \rightarrow \vert 10 \rangle$. However, such a procedure is not sensitive to the phases $(\alpha, \beta, \gamma)$, namely any value of these parameters leads to the same probabilities when the output is measured in the computational basis.

To train the circuit to apply the correct phase factor $e^{\imath \alpha}$ (having as target $\alpha = 0$), we employed the configuration shown in Fig. \ref{fig:QONNCNOT}b. More specifically, by injecting the state $\ket{00}$ the output state is found to be:
\begin{equation}
\begin{split}  
\vert \psi'_{00} \rangle &= (I \otimes H) \mathrm{CNOT}^{\prime} (I \otimes H) \ket{00} = \\
&= \frac{1}{2} [(1 + e^{i\alpha})\ket{00} + (1 - e^{i\alpha})\ket{01}],
\end{split}
\end{equation}
which is equal to $\ket{00}$ if and only if $e^{\imath\alpha} = 1$.  Analogous result is found by sending the state $\ket{01}$, and setting the target output state to be $\ket{01}$. Thus, inserting those two states in the training set allows to set the phase $\alpha$ to 0.

The same configuration can be used to have states in the training set sensitive to the difference $e^{\imath \beta} - e^{\imath\gamma}$. This is obtained by injecting state $\ket{10}$ in the circuit of Fig. \ref{fig:QONNCNOT}b, which leads to an output state:
\begin{equation}
\begin{split}  
\vert \psi'_{10} \rangle &= (I \otimes H) \mathrm{CNOT}^{\prime} (I \otimes H) \ket{10} = \\
&= \frac{1}{2}  [(e^{\imath\beta} + e^{\imath\gamma})\ket{10} + (e^{\imath\beta} - e^{\imath\gamma})\ket{11}]
\end{split}
\end{equation}
which is equal to $\ket{10}$ if and only if $e^{\imath \beta} = e^{\imath\gamma}$. Analogous result is obtained by sending the state $\ket{11}$, and setting the target output state to $\ket{11}$.

The remaining set of two states is chosen to be sensitive to phase factor $e^{\imath \gamma}$. This is performed by considering the configuration shown in Fig. \ref{fig:QONNCNOT}c. By injecting state $\ket{00}$, we obtain at the output:
\begin{equation}
\begin{split}  
\vert \psi'' \rangle &= (H \otimes I) \mathrm{CNOT}^{\prime} (H \otimes H) \ket{00} = \\
&= \frac{1}{2\sqrt{2}}[(1 + e^{\imath \gamma})\ket{00} + (1 - e^{\imath\gamma})\ket{01} + \\
&+ (1 - e^{\imath \gamma})\ket{10} + (1 + e^{\imath\gamma})\ket{11}]
\end{split}
\end{equation}
which is equal to $\frac{\ket{00}+\ket{11}}{\sqrt{2}}$ if and only if $e^{\imath\gamma} = 1$. The training stage for this state is then performed by having as target outputs $\ket{00}$ and $\ket{11}$, with probability $p=1/2$ each. Analogous result is obtained by injecting state $\ket{10}$, which is found in the output state $\frac{\ket{01}+\ket{10}}{\sqrt{2}}$ if and only if $e^{\imath\gamma} = 1$.

Overall, the training performed with these 6 additional states allows to set the three phase values to be $\alpha = \beta = \gamma = 0$, and thus effectively training the QONN to implement the CNOT gate with chosen phase values.

\section{Example of operation: details and basic instruction set}
\label{sec:OperationDetails}

In this Appendix we describe the basic instruction set for the proposed architecture, and then explain in more detail how the universal optimal quantum cloning circuit (see Sec. \ref{sec:cloning}) would run on a physical realization based on this approach.

\textbf{Instruction set.} We start by first defining the symbols $x_n$ in the instructions below, that represent the parameters that it requires. The parameter in square bracket after each instruction indicates the time it takes to run that instruction (if not provided, the execution time is negligible or not relevant).
\begin{itemize}
    \item INPUT($x_1$): Injects a single photon from the photon sources (or the input layer, if that layer is present) into the modes associated with qubit $x_1$, making sure that the photon is temporally synchronized with the other photons in the device.
    \item MEASURE($x_1$): Qubit $x_1$ is routed to the photon detectors to be measured.
    \item ROTATE($x_1$, R) [$1 \tau$]: This instruction applies a single qubit rotation to qubit $x_1$. The set of parameters $R = (x_2, x_3, x_4)$ represents the parameters of the rotation. This is physically handled by routing that qubit to the rotation layer, and by setting the phase shifters and beam splitter relative to that qubit on that layer to the values required to perform the transformation $U^{s}_{2x_1-1, 2x_1}(x_2, x_3, x_4)$.
    \item CNOT($x_1$, $x_2$) [$N \tau + (N-1) \epsilon$]: Applies a CNOT gate to qubits $x_1$ (control qubit) and $x_2$ (target qubit). It is physically implemented by sending qubits $x_1$ and $x_2$ to the CNOT layer, and arranging the meshes of optical switches at the input and output of the CNOT layer to direct the modes corresponding to these qubits to the correct inputs of the CNOT circuit.
    \item DELAY($x_1$) [$1 \tau$]: Adds a $1 \tau$ time delay on qubit $x_1$. In the minimal device configuration, this is achieved by sending the desired qubit to the single qubit rotation layer and setting the parameters to apply no rotation, thus being equivalent to the instruction ROTATE($x_1$, R=(0,0,0)). If an optional delay lines layer is present, sends the qubit to the delay lines layer instead.
\end{itemize}
Please note that the CNOT layer is written as having a duration of $N \tau + (N-1) \epsilon$, taking into account that its duration might depend on the implementation used. Nonetheless, for a given machine, the value of $N$ will be fixed.

Finally, one can insert in the instruction set those corresponding to eventual optional modules. For instance, if one needs to connect two devices with such an architecture, this additional instruction is required:
\begin{itemize}
    \item OUTPUT($x_1$): Routes the photon from modes associated with qubit $x_1$ to the output layer, to be sent to another device.
\end{itemize}

\textbf{Programming a universal optimal quantum cloning machine}. Using the instructions that we have defined here above, we can write the code for the quantum cloner shown in Sec. \ref{sec:cloning} as described above:
\begin{lstlisting}[mathescape=true]
Universal optimal quantum cloning machine

Instructions #1:
INPUT(1)
INPUT(2)
INPUT(3)

Instructions #2:
DELAY(1)
ROTATE(2, R1)
DELAY(3)

Instructions #3:
DELAY(1)$\times N$
CNOT(2, 3)

Instructions #4:
DELAY(1)
DELAY(2)
ROTATE(3, R2)

Instructions #5:
DELAY(1)$\times N$
CNOT(3, 2)

Instructions #6:
DELAY(1)
ROTATE(2, R3)
DELAY(3)

Instructions #7:
CNOT(1, 2)
DELAY(3)$\times N$

Instructions #8:
CNOT(1, 3)
DELAY(2)$\times N$

Instructions #9:
CNOT(2, 1)
DELAY(3)$\times N$

Instructions #10:
CNOT(3, 1)
DELAY(2)$\times N$

Instructions #11:
MEASURE(1)
MEASURE(2)
MEASURE(3)
\end{lstlisting}
Here the sets $\mathrm{R1}=(\theta_1, \phi_1, \lambda_1)$, $\mathrm{R2}=(\theta_2, \phi_2, \lambda_2)$ and $\mathrm{R3}=(\theta_3, \phi_3, \lambda_3)$ are the 3 parameters representing the appropriate values of the rotation for the $U^s$ gates. 
The final MEASURE(3) operation is needed, even though we do not use the information regarding the measurement of qubit 3, to remove all of the photons from the QPPU, to ensure that they do not interfere with subsequent computations.

In the set above, we note that the DELAY instructions are necessary to maintain temporal synchronization of the photons. To simplify the device programming, and to make sure that there are no photons left inside the QPPU at the end of a program, such an aspect of adding required delay can be handled by a compiler module that automatically takes it into account. This compiler would be also needed to make sure that there are no conflicts between operations, and to optimize their sequence.

\newpage


\begin{thebibliography}{80}%
\makeatletter
\providecommand \@ifxundefined [1]{%
 \@ifx{#1\undefined}
}%
\providecommand \@ifnum [1]{%
 \ifnum #1\expandafter \@firstoftwo
 \else \expandafter \@secondoftwo
 \fi
}%
\providecommand \@ifx [1]{%
 \ifx #1\expandafter \@firstoftwo
 \else \expandafter \@secondoftwo
 \fi
}%
\providecommand \natexlab [1]{#1}%
\providecommand \enquote  [1]{``#1''}%
\providecommand \bibnamefont  [1]{#1}%
\providecommand \bibfnamefont [1]{#1}%
\providecommand \citenamefont [1]{#1}%
\providecommand \href@noop [0]{\@secondoftwo}%
\providecommand \href [0]{\begingroup \@sanitize@url \@href}%
\providecommand \@href[1]{\@@startlink{#1}\@@href}%
\providecommand \@@href[1]{\endgroup#1\@@endlink}%
\providecommand \@sanitize@url [0]{\catcode `\\12\catcode `\$12\catcode
  `\&12\catcode `\#12\catcode `\^12\catcode `\_12\catcode `\%12\relax}%
\providecommand \@@startlink[1]{}%
\providecommand \@@endlink[0]{}%
\providecommand \url  [0]{\begingroup\@sanitize@url \@url }%
\providecommand \@url [1]{\endgroup\@href {#1}{\urlprefix }}%
\providecommand \urlprefix  [0]{URL }%
\providecommand \Eprint [0]{\href }%
\providecommand \doibase [0]{https://doi.org/}%
\providecommand \selectlanguage [0]{\@gobble}%
\providecommand \bibinfo  [0]{\@secondoftwo}%
\providecommand \bibfield  [0]{\@secondoftwo}%
\providecommand \translation [1]{[#1]}%
\providecommand \BibitemOpen [0]{}%
\providecommand \bibitemStop [0]{}%
\providecommand \bibitemNoStop [0]{.\EOS\space}%
\providecommand \EOS [0]{\spacefactor3000\relax}%
\providecommand \BibitemShut  [1]{\csname bibitem#1\endcsname}%
\let\auto@bib@innerbib\@empty
\bibitem [{\citenamefont {DiVincenzo}(2000)}]{divincenzo}%
  \BibitemOpen
  \bibfield  {author} {\bibinfo {author} {\bibfnamefont {D.~P.}\ \bibnamefont
  {DiVincenzo}},\ }\href
  {https://doi.org/10.1002/1521-3978(200009)48:9/11<771::AID-PROP771>3.0.CO;2-E}
  {\bibfield  {journal} {\bibinfo  {journal} {Fortschr. Phys.}\ }\textbf
  {\bibinfo {volume} {48}},\ \bibinfo {pages} {771} (\bibinfo {year}
  {2000})}\BibitemShut {NoStop}%
\bibitem [{\citenamefont {Nielsen}\ and\ \citenamefont
  {Chuang}(2010)}]{chuang2010}%
  \BibitemOpen
  \bibfield  {author} {\bibinfo {author} {\bibfnamefont {M.~A.}\ \bibnamefont
  {Nielsen}}\ and\ \bibinfo {author} {\bibfnamefont {I.~L.}\ \bibnamefont
  {Chuang}},\ }\href@noop {} {\emph {\bibinfo {title} {Quantum Computation and
  Quantum Information}}}\ (\bibinfo  {publisher} {Cambridge University Press},\
  \bibinfo {year} {2010})\BibitemShut {NoStop}%
\bibitem [{\citenamefont {Roffe}(2019)}]{errorcorrection_review}%
  \BibitemOpen
  \bibfield  {author} {\bibinfo {author} {\bibfnamefont {J.}~\bibnamefont
  {Roffe}},\ }\href {https://doi.org/10.1080/00107514.2019.1667078} {\bibfield
  {journal} {\bibinfo  {journal} {Contemp. Phys.}\ }\textbf {\bibinfo {volume}
  {60}},\ \bibinfo {pages} {226} (\bibinfo {year} {2019})}\BibitemShut
  {NoStop}%
\bibitem [{\citenamefont {Harrow}\ and\ \citenamefont
  {Montanaro}(2017)}]{harrow17}%
  \BibitemOpen
  \bibfield  {author} {\bibinfo {author} {\bibfnamefont {A.~W.}\ \bibnamefont
  {Harrow}}\ and\ \bibinfo {author} {\bibfnamefont {A.}~\bibnamefont
  {Montanaro}},\ }\href {https://doi.org/10.1038/nature23458} {\bibfield
  {journal} {\bibinfo  {journal} {Nature}\ }\textbf {\bibinfo {volume} {549}},\
  \bibinfo {pages} {203} (\bibinfo {year} {2017})}\BibitemShut {NoStop}%
\bibitem [{\citenamefont {Arute}\ \emph {et~al.}(2019)\citenamefont {Arute},
  \citenamefont {Arya}, \citenamefont {Babbush}, \citenamefont {Bacon},
  \citenamefont {Bardin}, \citenamefont {Barends}, \citenamefont {Biswas},
  \citenamefont {Boixo}, \citenamefont {Brandao}, \citenamefont {Buell},
  \citenamefont {Burkett}, \citenamefont {Chen}, \citenamefont {Chen},
  \citenamefont {Chiaro}, \citenamefont {Collins}, \citenamefont {Courtney},
  \citenamefont {Dunsworth}, \citenamefont {Farhi}, \citenamefont {Foxen},
  \citenamefont {Fowler}, \citenamefont {Gidney}, \citenamefont {Giustina},
  \citenamefont {Graff}, \citenamefont {Guerin}, \citenamefont {Habegger},
  \citenamefont {Harrigan}, \citenamefont {Hartmann}, \citenamefont {Ho},
  \citenamefont {Hoffman}, \citenamefont {Huang}, \citenamefont {Humble},
  \citenamefont {Isakov}, \citenamefont {Jeffrey}, \citenamefont {Jiang},
  \citenamefont {Kafri}, \citenamefont {Kechedzhi}, \citenamefont {Kelly},
  \citenamefont {Klimov}, \citenamefont {Knysh}, \citenamefont {Korotov},
  \citenamefont {Kostritsa}, \citenamefont {Landhuis}, \citenamefont
  {Lindmark}, \citenamefont {Lucero}, \citenamefont {Lyakh}, \citenamefont
  {Mandr\`a}, \citenamefont {McClean}, \citenamefont {McEwen}, \citenamefont
  {Megrant}, \citenamefont {Mi}, \citenamefont {Michielsen}, \citenamefont
  {Mohseni}, \citenamefont {Motus}, \citenamefont {Naaman}, \citenamefont
  {Neeley}, \citenamefont {Neill}, \citenamefont {{Yuezhen Niu}}, \citenamefont
  {Ostby}, \citenamefont {Petukhov}, \citenamefont {Platt}, \citenamefont
  {Quintana}, \citenamefont {Rieffel}, \citenamefont {Rousham}, \citenamefont
  {Rubin}, \citenamefont {Sank}, \citenamefont {Satzinger}, \citenamefont
  {Smelyanskiy}, \citenamefont {Sung}, \citenamefont {Trevithick},
  \citenamefont {Vainsencher}, \citenamefont {Villalonga}, \citenamefont
  {White}, \citenamefont {Yao}, \citenamefont {Yeh}, \citenamefont {Zalcman},
  \citenamefont {Neven},\ and\ \citenamefont {Martinis}}]{QuantumSupGoogle}%
  \BibitemOpen
  \bibfield  {author} {\bibinfo {author} {\bibfnamefont {F.}~\bibnamefont
  {Arute}}, \bibinfo {author} {\bibfnamefont {K.}~\bibnamefont {Arya}},
  \bibinfo {author} {\bibfnamefont {R.}~\bibnamefont {Babbush}}, \bibinfo
  {author} {\bibfnamefont {D.}~\bibnamefont {Bacon}}, \bibinfo {author}
  {\bibfnamefont {J.~C.}\ \bibnamefont {Bardin}}, \bibinfo {author}
  {\bibfnamefont {R.}~\bibnamefont {Barends}}, \bibinfo {author} {\bibfnamefont
  {R.}~\bibnamefont {Biswas}}, \bibinfo {author} {\bibfnamefont
  {S.}~\bibnamefont {Boixo}}, \bibinfo {author} {\bibfnamefont {F.~G. S.~L.}\
  \bibnamefont {Brandao}}, \bibinfo {author} {\bibfnamefont {D.~A.}\
  \bibnamefont {Buell}}, \bibinfo {author} {\bibfnamefont {B.}~\bibnamefont
  {Burkett}}, \bibinfo {author} {\bibfnamefont {Y.}~\bibnamefont {Chen}},
  \bibinfo {author} {\bibfnamefont {Z.}~\bibnamefont {Chen}}, \bibinfo {author}
  {\bibfnamefont {B.}~\bibnamefont {Chiaro}}, \bibinfo {author} {\bibfnamefont
  {R.}~\bibnamefont {Collins}}, \bibinfo {author} {\bibfnamefont
  {W.}~\bibnamefont {Courtney}}, \bibinfo {author} {\bibfnamefont
  {A.}~\bibnamefont {Dunsworth}}, \bibinfo {author} {\bibfnamefont
  {E.}~\bibnamefont {Farhi}}, \bibinfo {author} {\bibfnamefont
  {B.}~\bibnamefont {Foxen}}, \bibinfo {author} {\bibfnamefont
  {A.}~\bibnamefont {Fowler}}, \bibinfo {author} {\bibfnamefont
  {C.}~\bibnamefont {Gidney}}, \bibinfo {author} {\bibfnamefont
  {M.}~\bibnamefont {Giustina}}, \bibinfo {author} {\bibfnamefont
  {R.}~\bibnamefont {Graff}}, \bibinfo {author} {\bibfnamefont
  {K.}~\bibnamefont {Guerin}}, \bibinfo {author} {\bibfnamefont
  {S.}~\bibnamefont {Habegger}}, \bibinfo {author} {\bibfnamefont {M.~P.}\
  \bibnamefont {Harrigan}}, \bibinfo {author} {\bibfnamefont {M.~J.}\
  \bibnamefont {Hartmann}}, \bibinfo {author} {\bibfnamefont {A.}~\bibnamefont
  {Ho}}, \bibinfo {author} {\bibfnamefont {M.}~\bibnamefont {Hoffman}},
  \bibinfo {author} {\bibfnamefont {T.}~\bibnamefont {Huang}}, \bibinfo
  {author} {\bibfnamefont {T.~S.}\ \bibnamefont {Humble}}, \bibinfo {author}
  {\bibfnamefont {S.~V.}\ \bibnamefont {Isakov}}, \bibinfo {author}
  {\bibfnamefont {E.}~\bibnamefont {Jeffrey}}, \bibinfo {author} {\bibfnamefont
  {Z.}~\bibnamefont {Jiang}}, \bibinfo {author} {\bibfnamefont
  {D.}~\bibnamefont {Kafri}}, \bibinfo {author} {\bibfnamefont
  {K.}~\bibnamefont {Kechedzhi}}, \bibinfo {author} {\bibfnamefont
  {J.}~\bibnamefont {Kelly}}, \bibinfo {author} {\bibfnamefont {P.~V.}\
  \bibnamefont {Klimov}}, \bibinfo {author} {\bibfnamefont {S.}~\bibnamefont
  {Knysh}}, \bibinfo {author} {\bibfnamefont {A.}~\bibnamefont {Korotov}},
  \bibinfo {author} {\bibfnamefont {F.}~\bibnamefont {Kostritsa}}, \bibinfo
  {author} {\bibfnamefont {D.}~\bibnamefont {Landhuis}}, \bibinfo {author}
  {\bibfnamefont {M.}~\bibnamefont {Lindmark}}, \bibinfo {author}
  {\bibfnamefont {E.}~\bibnamefont {Lucero}}, \bibinfo {author} {\bibfnamefont
  {D.}~\bibnamefont {Lyakh}}, \bibinfo {author} {\bibfnamefont
  {S.}~\bibnamefont {Mandr\`a}}, \bibinfo {author} {\bibfnamefont {J.~R.}\
  \bibnamefont {McClean}}, \bibinfo {author} {\bibfnamefont {M.}~\bibnamefont
  {McEwen}}, \bibinfo {author} {\bibfnamefont {A.}~\bibnamefont {Megrant}},
  \bibinfo {author} {\bibfnamefont {X.}~\bibnamefont {Mi}}, \bibinfo {author}
  {\bibfnamefont {K.}~\bibnamefont {Michielsen}}, \bibinfo {author}
  {\bibfnamefont {M.}~\bibnamefont {Mohseni}}, \bibinfo {author} {\bibfnamefont
  {J.}~\bibnamefont {Motus}}, \bibinfo {author} {\bibfnamefont
  {O.}~\bibnamefont {Naaman}}, \bibinfo {author} {\bibfnamefont
  {M.}~\bibnamefont {Neeley}}, \bibinfo {author} {\bibfnamefont
  {C.}~\bibnamefont {Neill}}, \bibinfo {author} {\bibfnamefont
  {M.}~\bibnamefont {{Yuezhen Niu}}}, \bibinfo {author} {\bibfnamefont
  {E.}~\bibnamefont {Ostby}}, \bibinfo {author} {\bibfnamefont
  {A.}~\bibnamefont {Petukhov}}, \bibinfo {author} {\bibfnamefont {J.~C.}\
  \bibnamefont {Platt}}, \bibinfo {author} {\bibfnamefont {C.}~\bibnamefont
  {Quintana}}, \bibinfo {author} {\bibfnamefont {E.~G.}\ \bibnamefont
  {Rieffel}}, \bibinfo {author} {\bibfnamefont {P.}~\bibnamefont {Rousham}},
  \bibinfo {author} {\bibfnamefont {N.~C.}\ \bibnamefont {Rubin}}, \bibinfo
  {author} {\bibfnamefont {D.}~\bibnamefont {Sank}}, \bibinfo {author}
  {\bibfnamefont {K.~J.}\ \bibnamefont {Satzinger}}, \bibinfo {author}
  {\bibfnamefont {V.}~\bibnamefont {Smelyanskiy}}, \bibinfo {author}
  {\bibfnamefont {K.~J.}\ \bibnamefont {Sung}}, \bibinfo {author}
  {\bibfnamefont {M.~D.}\ \bibnamefont {Trevithick}}, \bibinfo {author}
  {\bibfnamefont {A.}~\bibnamefont {Vainsencher}}, \bibinfo {author}
  {\bibfnamefont {B.}~\bibnamefont {Villalonga}}, \bibinfo {author}
  {\bibfnamefont {T.}~\bibnamefont {White}}, \bibinfo {author} {\bibfnamefont
  {Z.~J.}\ \bibnamefont {Yao}}, \bibinfo {author} {\bibfnamefont
  {P.}~\bibnamefont {Yeh}}, \bibinfo {author} {\bibfnamefont {A.}~\bibnamefont
  {Zalcman}}, \bibinfo {author} {\bibfnamefont {H.}~\bibnamefont {Neven}},\
  and\ \bibinfo {author} {\bibfnamefont {J.~M.}\ \bibnamefont {Martinis}},\
  }\href {https://doi.org/10.1038/s41586-019-1666-5} {\bibfield  {journal}
  {\bibinfo  {journal} {Nature}\ }\textbf {\bibinfo {volume} {574}},\ \bibinfo
  {pages} {505–510} (\bibinfo {year} {2019})}\BibitemShut {NoStop}%
\bibitem [{\citenamefont {Acharya}\ \emph {et~al.}(2024)\citenamefont
  {Acharya}, \citenamefont {Abanin}, \citenamefont {Aghababaie-Beni},
  \citenamefont {Aleiner}, \citenamefont {Andersen}, \citenamefont {Ansmann},
  \citenamefont {Arute}, \citenamefont {Arya}, \citenamefont {Asfaw},
  \citenamefont {Astrakhantsev}, \citenamefont {Atalaya}, \citenamefont
  {Babbush}, \citenamefont {Bacon}, \citenamefont {Ballard}, \citenamefont
  {Bardin}, \citenamefont {Bausch}, \citenamefont {Bengtsson}, \citenamefont
  {Bilmes}, \citenamefont {Blackwell}, \citenamefont {Boixo}, \citenamefont
  {Bortoli}, \citenamefont {Bourassa}, \citenamefont {Bovaird}, \citenamefont
  {Brill}, \citenamefont {Broughton}, \citenamefont {Browne}, \citenamefont
  {Buchea}, \citenamefont {Buckley}, \citenamefont {Buell}, \citenamefont
  {Burger}, \citenamefont {Burkett}, \citenamefont {Bushnell}, \citenamefont
  {Cabrera}, \citenamefont {Campero}, \citenamefont {Chang}, \citenamefont
  {Chen}, \citenamefont {Chen}, \citenamefont {Chiaro}, \citenamefont {Chik},
  \citenamefont {Chou}, \citenamefont {Claes}, \citenamefont {Cleland},
  \citenamefont {Cogan}, \citenamefont {Collins}, \citenamefont {Conner},
  \citenamefont {Courtney}, \citenamefont {Crook}, \citenamefont {Curtin},
  \citenamefont {Das}, \citenamefont {Davies}, \citenamefont {De~Lorenzo},
  \citenamefont {Debroy}, \citenamefont {Demura}, \citenamefont {Devoret},
  \citenamefont {Di~Paolo}, \citenamefont {Donohoe}, \citenamefont {Drozdov},
  \citenamefont {Dunsworth}, \citenamefont {Earle}, \citenamefont {Edlich},
  \citenamefont {Eickbusch}, \citenamefont {Elbag}, \citenamefont {Elzouka},
  \citenamefont {Erickson}, \citenamefont {Faoro}, \citenamefont {Farhi},
  \citenamefont {Ferreira}, \citenamefont {Burgos}, \citenamefont {Forati},
  \citenamefont {Fowler}, \citenamefont {Foxen}, \citenamefont {Ganjam},
  \citenamefont {Garcia}, \citenamefont {Gasca}, \citenamefont {Genois},
  \citenamefont {Giang}, \citenamefont {Gidney}, \citenamefont {Gilboa},
  \citenamefont {Gosula}, \citenamefont {Dau}, \citenamefont {Graumann},
  \citenamefont {Greene}, \citenamefont {Gross}, \citenamefont {Habegger},
  \citenamefont {Hall}, \citenamefont {Hamilton}, \citenamefont {Hansen},
  \citenamefont {Harrigan}, \citenamefont {Harrington}, \citenamefont {Heras},
  \citenamefont {Heslin}, \citenamefont {Heu}, \citenamefont {Higgott},
  \citenamefont {Hill}, \citenamefont {Hilton}, \citenamefont {Holland},
  \citenamefont {Hong}, \citenamefont {Huang}, \citenamefont {Huff},
  \citenamefont {Huggins}, \citenamefont {Ioffe}, \citenamefont {Isakov},
  \citenamefont {Iveland}, \citenamefont {Jeffrey}, \citenamefont {Jiang},
  \citenamefont {Jones}, \citenamefont {Jordan}, \citenamefont {Joshi},
  \citenamefont {Juhas}, \citenamefont {Kafri}, \citenamefont {Kang},
  \citenamefont {Karamlou}, \citenamefont {Kechedzhi}, \citenamefont {Kelly},
  \citenamefont {Khaire}, \citenamefont {Khattar}, \citenamefont {Khezri},
  \citenamefont {Kim}, \citenamefont {Klimov}, \citenamefont {Klots},
  \citenamefont {Kobrin}, \citenamefont {Kohli}, \citenamefont {Korotkov},
  \citenamefont {Kostritsa}, \citenamefont {Kothari}, \citenamefont
  {Kozlovskii}, \citenamefont {Kreikebaum}, \citenamefont {Kurilovich},
  \citenamefont {Lacroix}, \citenamefont {Landhuis}, \citenamefont {Lange-Dei},
  \citenamefont {Langley}, \citenamefont {Laptev}, \citenamefont {Lau},
  \citenamefont {Le~Guevel}, \citenamefont {Ledford}, \citenamefont {Lee},
  \citenamefont {Lee}, \citenamefont {Lensky}, \citenamefont {Leon},
  \citenamefont {Lester}, \citenamefont {Li}, \citenamefont {Li}, \citenamefont
  {Lill}, \citenamefont {Liu}, \citenamefont {Livingston}, \citenamefont
  {Locharla}, \citenamefont {Lucero}, \citenamefont {Lundahl}, \citenamefont
  {Lunt}, \citenamefont {Madhuk}, \citenamefont {Malone}, \citenamefont
  {Maloney}, \citenamefont {Mandr\`{a}}, \citenamefont {Manyika}, \citenamefont
  {Martin}, \citenamefont {Martin}, \citenamefont {Martin}, \citenamefont
  {Maxfield}, \citenamefont {McClean}, \citenamefont {McEwen}, \citenamefont
  {Meeks}, \citenamefont {Megrant}, \citenamefont {Mi}, \citenamefont {Miao},
  \citenamefont {Mieszala}, \citenamefont {Molavi}, \citenamefont {Molina},
  \citenamefont {Montazeri}, \citenamefont {Morvan}, \citenamefont {Movassagh},
  \citenamefont {Mruczkiewicz}, \citenamefont {Naaman}, \citenamefont {Neeley},
  \citenamefont {Neill}, \citenamefont {Nersisyan}, \citenamefont {Neven},
  \citenamefont {Newman}, \citenamefont {Ng}, \citenamefont {Nguyen},
  \citenamefont {Nguyen}, \citenamefont {Ni}, \citenamefont {Niu},
  \citenamefont {O'Brien}, \citenamefont {Oliver}, \citenamefont {Opremcak},
  \citenamefont {Ottosson}, \citenamefont {Petukhov}, \citenamefont {Pizzuto},
  \citenamefont {Platt}, \citenamefont {Potter}, \citenamefont {Pritchard},
  \citenamefont {Pryadko}, \citenamefont {Quintana}, \citenamefont
  {Ramachandran}, \citenamefont {Reagor}, \citenamefont {Redding},
  \citenamefont {Rhodes}, \citenamefont {Roberts}, \citenamefont {Rosenberg},
  \citenamefont {Rosenfeld}, \citenamefont {Roushan}, \citenamefont {Rubin},
  \citenamefont {Saei}, \citenamefont {Sank}, \citenamefont {Sankaragomathi},
  \citenamefont {Satzinger}, \citenamefont {Schurkus}, \citenamefont
  {Schuster}, \citenamefont {Senior}, \citenamefont {Shearn}, \citenamefont
  {Shorter}, \citenamefont {Shutty}, \citenamefont {Shvarts}, \citenamefont
  {Singh}, \citenamefont {Sivak}, \citenamefont {Skruzny}, \citenamefont
  {Small}, \citenamefont {Smelyanskiy}, \citenamefont {Smith}, \citenamefont
  {Somma}, \citenamefont {Springer}, \citenamefont {Sterling}, \citenamefont
  {Strain}, \citenamefont {Suchard}, \citenamefont {Szasz}, \citenamefont
  {Sztein}, \citenamefont {Thor}, \citenamefont {Torres}, \citenamefont
  {Torunbalci}, \citenamefont {Vaishnav}, \citenamefont {Vargas}, \citenamefont
  {Vdovichev}, \citenamefont {Vidal}, \citenamefont {Villalonga}, \citenamefont
  {Heidweiller}, \citenamefont {Waltman}, \citenamefont {Wang}, \citenamefont
  {Ware}, \citenamefont {Weber}, \citenamefont {Weidel}, \citenamefont {White},
  \citenamefont {Wong}, \citenamefont {Woo}, \citenamefont {Xing},
  \citenamefont {Yao}, \citenamefont {Yeh}, \citenamefont {Ying}, \citenamefont
  {Yoo}, \citenamefont {Yosri}, \citenamefont {Young}, \citenamefont {Zalcman},
  \citenamefont {Zhang}, \citenamefont {Zhu},\ and\ \citenamefont
  {Zobrist}}]{GoogleWillow}%
  \BibitemOpen
  \bibfield  {author} {\bibinfo {author} {\bibfnamefont {R.}~\bibnamefont
  {Acharya}}, \bibinfo {author} {\bibfnamefont {D.~A.}\ \bibnamefont {Abanin}},
  \bibinfo {author} {\bibfnamefont {L.}~\bibnamefont {Aghababaie-Beni}},
  \bibinfo {author} {\bibfnamefont {I.}~\bibnamefont {Aleiner}}, \bibinfo
  {author} {\bibfnamefont {T.~I.}\ \bibnamefont {Andersen}}, \bibinfo {author}
  {\bibfnamefont {M.}~\bibnamefont {Ansmann}}, \bibinfo {author} {\bibfnamefont
  {F.}~\bibnamefont {Arute}}, \bibinfo {author} {\bibfnamefont
  {K.}~\bibnamefont {Arya}}, \bibinfo {author} {\bibfnamefont {A.}~\bibnamefont
  {Asfaw}}, \bibinfo {author} {\bibfnamefont {N.}~\bibnamefont
  {Astrakhantsev}}, \bibinfo {author} {\bibfnamefont {J.}~\bibnamefont
  {Atalaya}}, \bibinfo {author} {\bibfnamefont {R.}~\bibnamefont {Babbush}},
  \bibinfo {author} {\bibfnamefont {D.}~\bibnamefont {Bacon}}, \bibinfo
  {author} {\bibfnamefont {B.}~\bibnamefont {Ballard}}, \bibinfo {author}
  {\bibfnamefont {J.~C.}\ \bibnamefont {Bardin}}, \bibinfo {author}
  {\bibfnamefont {J.}~\bibnamefont {Bausch}}, \bibinfo {author} {\bibfnamefont
  {A.}~\bibnamefont {Bengtsson}}, \bibinfo {author} {\bibfnamefont
  {A.}~\bibnamefont {Bilmes}}, \bibinfo {author} {\bibfnamefont
  {S.}~\bibnamefont {Blackwell}}, \bibinfo {author} {\bibfnamefont
  {S.}~\bibnamefont {Boixo}}, \bibinfo {author} {\bibfnamefont
  {G.}~\bibnamefont {Bortoli}}, \bibinfo {author} {\bibfnamefont
  {A.}~\bibnamefont {Bourassa}}, \bibinfo {author} {\bibfnamefont
  {J.}~\bibnamefont {Bovaird}}, \bibinfo {author} {\bibfnamefont
  {L.}~\bibnamefont {Brill}}, \bibinfo {author} {\bibfnamefont
  {M.}~\bibnamefont {Broughton}}, \bibinfo {author} {\bibfnamefont {D.~A.}\
  \bibnamefont {Browne}}, \bibinfo {author} {\bibfnamefont {B.}~\bibnamefont
  {Buchea}}, \bibinfo {author} {\bibfnamefont {B.~B.}\ \bibnamefont {Buckley}},
  \bibinfo {author} {\bibfnamefont {D.~A.}\ \bibnamefont {Buell}}, \bibinfo
  {author} {\bibfnamefont {T.}~\bibnamefont {Burger}}, \bibinfo {author}
  {\bibfnamefont {B.}~\bibnamefont {Burkett}}, \bibinfo {author} {\bibfnamefont
  {N.}~\bibnamefont {Bushnell}}, \bibinfo {author} {\bibfnamefont
  {A.}~\bibnamefont {Cabrera}}, \bibinfo {author} {\bibfnamefont
  {J.}~\bibnamefont {Campero}}, \bibinfo {author} {\bibfnamefont {H.-S.}\
  \bibnamefont {Chang}}, \bibinfo {author} {\bibfnamefont {Y.}~\bibnamefont
  {Chen}}, \bibinfo {author} {\bibfnamefont {Z.}~\bibnamefont {Chen}}, \bibinfo
  {author} {\bibfnamefont {B.}~\bibnamefont {Chiaro}}, \bibinfo {author}
  {\bibfnamefont {D.}~\bibnamefont {Chik}}, \bibinfo {author} {\bibfnamefont
  {C.}~\bibnamefont {Chou}}, \bibinfo {author} {\bibfnamefont {J.}~\bibnamefont
  {Claes}}, \bibinfo {author} {\bibfnamefont {A.~Y.}\ \bibnamefont {Cleland}},
  \bibinfo {author} {\bibfnamefont {J.}~\bibnamefont {Cogan}}, \bibinfo
  {author} {\bibfnamefont {R.}~\bibnamefont {Collins}}, \bibinfo {author}
  {\bibfnamefont {P.}~\bibnamefont {Conner}}, \bibinfo {author} {\bibfnamefont
  {W.}~\bibnamefont {Courtney}}, \bibinfo {author} {\bibfnamefont {A.~L.}\
  \bibnamefont {Crook}}, \bibinfo {author} {\bibfnamefont {B.}~\bibnamefont
  {Curtin}}, \bibinfo {author} {\bibfnamefont {S.}~\bibnamefont {Das}},
  \bibinfo {author} {\bibfnamefont {A.}~\bibnamefont {Davies}}, \bibinfo
  {author} {\bibfnamefont {L.}~\bibnamefont {De~Lorenzo}}, \bibinfo {author}
  {\bibfnamefont {D.~M.}\ \bibnamefont {Debroy}}, \bibinfo {author}
  {\bibfnamefont {S.}~\bibnamefont {Demura}}, \bibinfo {author} {\bibfnamefont
  {M.}~\bibnamefont {Devoret}}, \bibinfo {author} {\bibfnamefont
  {A.}~\bibnamefont {Di~Paolo}}, \bibinfo {author} {\bibfnamefont
  {P.}~\bibnamefont {Donohoe}}, \bibinfo {author} {\bibfnamefont
  {I.}~\bibnamefont {Drozdov}}, \bibinfo {author} {\bibfnamefont
  {A.}~\bibnamefont {Dunsworth}}, \bibinfo {author} {\bibfnamefont
  {C.}~\bibnamefont {Earle}}, \bibinfo {author} {\bibfnamefont
  {T.}~\bibnamefont {Edlich}}, \bibinfo {author} {\bibfnamefont
  {A.}~\bibnamefont {Eickbusch}}, \bibinfo {author} {\bibfnamefont {A.~M.}\
  \bibnamefont {Elbag}}, \bibinfo {author} {\bibfnamefont {M.}~\bibnamefont
  {Elzouka}}, \bibinfo {author} {\bibfnamefont {C.}~\bibnamefont {Erickson}},
  \bibinfo {author} {\bibfnamefont {L.}~\bibnamefont {Faoro}}, \bibinfo
  {author} {\bibfnamefont {E.}~\bibnamefont {Farhi}}, \bibinfo {author}
  {\bibfnamefont {V.~S.}\ \bibnamefont {Ferreira}}, \bibinfo {author}
  {\bibfnamefont {L.~F.}\ \bibnamefont {Burgos}}, \bibinfo {author}
  {\bibfnamefont {E.}~\bibnamefont {Forati}}, \bibinfo {author} {\bibfnamefont
  {A.~G.}\ \bibnamefont {Fowler}}, \bibinfo {author} {\bibfnamefont
  {B.}~\bibnamefont {Foxen}}, \bibinfo {author} {\bibfnamefont
  {S.}~\bibnamefont {Ganjam}}, \bibinfo {author} {\bibfnamefont
  {G.}~\bibnamefont {Garcia}}, \bibinfo {author} {\bibfnamefont
  {R.}~\bibnamefont {Gasca}}, \bibinfo {author} {\bibfnamefont
  {E.}~\bibnamefont {Genois}}, \bibinfo {author} {\bibfnamefont
  {W.}~\bibnamefont {Giang}}, \bibinfo {author} {\bibfnamefont
  {C.}~\bibnamefont {Gidney}}, \bibinfo {author} {\bibfnamefont
  {D.}~\bibnamefont {Gilboa}}, \bibinfo {author} {\bibfnamefont
  {R.}~\bibnamefont {Gosula}}, \bibinfo {author} {\bibfnamefont {A.~G.}\
  \bibnamefont {Dau}}, \bibinfo {author} {\bibfnamefont {D.}~\bibnamefont
  {Graumann}}, \bibinfo {author} {\bibfnamefont {A.}~\bibnamefont {Greene}},
  \bibinfo {author} {\bibfnamefont {J.~A.}\ \bibnamefont {Gross}}, \bibinfo
  {author} {\bibfnamefont {S.}~\bibnamefont {Habegger}}, \bibinfo {author}
  {\bibfnamefont {J.}~\bibnamefont {Hall}}, \bibinfo {author} {\bibfnamefont
  {M.~C.}\ \bibnamefont {Hamilton}}, \bibinfo {author} {\bibfnamefont
  {M.}~\bibnamefont {Hansen}}, \bibinfo {author} {\bibfnamefont {M.~P.}\
  \bibnamefont {Harrigan}}, \bibinfo {author} {\bibfnamefont {S.~D.}\
  \bibnamefont {Harrington}}, \bibinfo {author} {\bibfnamefont {F.~J.~H.}\
  \bibnamefont {Heras}}, \bibinfo {author} {\bibfnamefont {S.}~\bibnamefont
  {Heslin}}, \bibinfo {author} {\bibfnamefont {P.}~\bibnamefont {Heu}},
  \bibinfo {author} {\bibfnamefont {O.}~\bibnamefont {Higgott}}, \bibinfo
  {author} {\bibfnamefont {G.}~\bibnamefont {Hill}}, \bibinfo {author}
  {\bibfnamefont {J.}~\bibnamefont {Hilton}}, \bibinfo {author} {\bibfnamefont
  {G.}~\bibnamefont {Holland}}, \bibinfo {author} {\bibfnamefont
  {S.}~\bibnamefont {Hong}}, \bibinfo {author} {\bibfnamefont {H.-Y.}\
  \bibnamefont {Huang}}, \bibinfo {author} {\bibfnamefont {A.}~\bibnamefont
  {Huff}}, \bibinfo {author} {\bibfnamefont {W.~J.}\ \bibnamefont {Huggins}},
  \bibinfo {author} {\bibfnamefont {L.~B.}\ \bibnamefont {Ioffe}}, \bibinfo
  {author} {\bibfnamefont {S.~V.}\ \bibnamefont {Isakov}}, \bibinfo {author}
  {\bibfnamefont {J.}~\bibnamefont {Iveland}}, \bibinfo {author} {\bibfnamefont
  {E.}~\bibnamefont {Jeffrey}}, \bibinfo {author} {\bibfnamefont
  {Z.}~\bibnamefont {Jiang}}, \bibinfo {author} {\bibfnamefont
  {C.}~\bibnamefont {Jones}}, \bibinfo {author} {\bibfnamefont
  {S.}~\bibnamefont {Jordan}}, \bibinfo {author} {\bibfnamefont
  {C.}~\bibnamefont {Joshi}}, \bibinfo {author} {\bibfnamefont
  {P.}~\bibnamefont {Juhas}}, \bibinfo {author} {\bibfnamefont
  {D.}~\bibnamefont {Kafri}}, \bibinfo {author} {\bibfnamefont
  {H.}~\bibnamefont {Kang}}, \bibinfo {author} {\bibfnamefont {A.~H.}\
  \bibnamefont {Karamlou}}, \bibinfo {author} {\bibfnamefont {K.}~\bibnamefont
  {Kechedzhi}}, \bibinfo {author} {\bibfnamefont {J.}~\bibnamefont {Kelly}},
  \bibinfo {author} {\bibfnamefont {T.}~\bibnamefont {Khaire}}, \bibinfo
  {author} {\bibfnamefont {T.}~\bibnamefont {Khattar}}, \bibinfo {author}
  {\bibfnamefont {M.}~\bibnamefont {Khezri}}, \bibinfo {author} {\bibfnamefont
  {S.}~\bibnamefont {Kim}}, \bibinfo {author} {\bibfnamefont {P.~V.}\
  \bibnamefont {Klimov}}, \bibinfo {author} {\bibfnamefont {A.~R.}\
  \bibnamefont {Klots}}, \bibinfo {author} {\bibfnamefont {B.}~\bibnamefont
  {Kobrin}}, \bibinfo {author} {\bibfnamefont {P.}~\bibnamefont {Kohli}},
  \bibinfo {author} {\bibfnamefont {A.~N.}\ \bibnamefont {Korotkov}}, \bibinfo
  {author} {\bibfnamefont {F.}~\bibnamefont {Kostritsa}}, \bibinfo {author}
  {\bibfnamefont {R.}~\bibnamefont {Kothari}}, \bibinfo {author} {\bibfnamefont
  {B.}~\bibnamefont {Kozlovskii}}, \bibinfo {author} {\bibfnamefont {J.~M.}\
  \bibnamefont {Kreikebaum}}, \bibinfo {author} {\bibfnamefont {V.~D.}\
  \bibnamefont {Kurilovich}}, \bibinfo {author} {\bibfnamefont
  {N.}~\bibnamefont {Lacroix}}, \bibinfo {author} {\bibfnamefont
  {D.}~\bibnamefont {Landhuis}}, \bibinfo {author} {\bibfnamefont
  {T.}~\bibnamefont {Lange-Dei}}, \bibinfo {author} {\bibfnamefont {B.~W.}\
  \bibnamefont {Langley}}, \bibinfo {author} {\bibfnamefont {P.}~\bibnamefont
  {Laptev}}, \bibinfo {author} {\bibfnamefont {K.-M.}\ \bibnamefont {Lau}},
  \bibinfo {author} {\bibfnamefont {L.}~\bibnamefont {Le~Guevel}}, \bibinfo
  {author} {\bibfnamefont {J.}~\bibnamefont {Ledford}}, \bibinfo {author}
  {\bibfnamefont {J.}~\bibnamefont {Lee}}, \bibinfo {author} {\bibfnamefont
  {K.}~\bibnamefont {Lee}}, \bibinfo {author} {\bibfnamefont {Y.~D.}\
  \bibnamefont {Lensky}}, \bibinfo {author} {\bibfnamefont {S.}~\bibnamefont
  {Leon}}, \bibinfo {author} {\bibfnamefont {B.~J.}\ \bibnamefont {Lester}},
  \bibinfo {author} {\bibfnamefont {W.~Y.}\ \bibnamefont {Li}}, \bibinfo
  {author} {\bibfnamefont {Y.}~\bibnamefont {Li}}, \bibinfo {author}
  {\bibfnamefont {A.~T.}\ \bibnamefont {Lill}}, \bibinfo {author}
  {\bibfnamefont {W.}~\bibnamefont {Liu}}, \bibinfo {author} {\bibfnamefont
  {W.~P.}\ \bibnamefont {Livingston}}, \bibinfo {author} {\bibfnamefont
  {A.}~\bibnamefont {Locharla}}, \bibinfo {author} {\bibfnamefont
  {E.}~\bibnamefont {Lucero}}, \bibinfo {author} {\bibfnamefont
  {D.}~\bibnamefont {Lundahl}}, \bibinfo {author} {\bibfnamefont
  {A.}~\bibnamefont {Lunt}}, \bibinfo {author} {\bibfnamefont {S.}~\bibnamefont
  {Madhuk}}, \bibinfo {author} {\bibfnamefont {F.~D.}\ \bibnamefont {Malone}},
  \bibinfo {author} {\bibfnamefont {A.}~\bibnamefont {Maloney}}, \bibinfo
  {author} {\bibfnamefont {S.}~\bibnamefont {Mandr\`{a}}}, \bibinfo {author}
  {\bibfnamefont {J.}~\bibnamefont {Manyika}}, \bibinfo {author} {\bibfnamefont
  {L.~S.}\ \bibnamefont {Martin}}, \bibinfo {author} {\bibfnamefont
  {O.}~\bibnamefont {Martin}}, \bibinfo {author} {\bibfnamefont
  {S.}~\bibnamefont {Martin}}, \bibinfo {author} {\bibfnamefont
  {C.}~\bibnamefont {Maxfield}}, \bibinfo {author} {\bibfnamefont {J.~R.}\
  \bibnamefont {McClean}}, \bibinfo {author} {\bibfnamefont {M.}~\bibnamefont
  {McEwen}}, \bibinfo {author} {\bibfnamefont {S.}~\bibnamefont {Meeks}},
  \bibinfo {author} {\bibfnamefont {A.}~\bibnamefont {Megrant}}, \bibinfo
  {author} {\bibfnamefont {X.}~\bibnamefont {Mi}}, \bibinfo {author}
  {\bibfnamefont {K.~C.}\ \bibnamefont {Miao}}, \bibinfo {author}
  {\bibfnamefont {A.}~\bibnamefont {Mieszala}}, \bibinfo {author}
  {\bibfnamefont {R.}~\bibnamefont {Molavi}}, \bibinfo {author} {\bibfnamefont
  {S.}~\bibnamefont {Molina}}, \bibinfo {author} {\bibfnamefont
  {S.}~\bibnamefont {Montazeri}}, \bibinfo {author} {\bibfnamefont
  {A.}~\bibnamefont {Morvan}}, \bibinfo {author} {\bibfnamefont
  {R.}~\bibnamefont {Movassagh}}, \bibinfo {author} {\bibfnamefont
  {W.}~\bibnamefont {Mruczkiewicz}}, \bibinfo {author} {\bibfnamefont
  {O.}~\bibnamefont {Naaman}}, \bibinfo {author} {\bibfnamefont
  {M.}~\bibnamefont {Neeley}}, \bibinfo {author} {\bibfnamefont
  {C.}~\bibnamefont {Neill}}, \bibinfo {author} {\bibfnamefont
  {A.}~\bibnamefont {Nersisyan}}, \bibinfo {author} {\bibfnamefont
  {H.}~\bibnamefont {Neven}}, \bibinfo {author} {\bibfnamefont
  {M.}~\bibnamefont {Newman}}, \bibinfo {author} {\bibfnamefont {J.~H.}\
  \bibnamefont {Ng}}, \bibinfo {author} {\bibfnamefont {A.}~\bibnamefont
  {Nguyen}}, \bibinfo {author} {\bibfnamefont {M.}~\bibnamefont {Nguyen}},
  \bibinfo {author} {\bibfnamefont {C.-H.}\ \bibnamefont {Ni}}, \bibinfo
  {author} {\bibfnamefont {M.~Y.}\ \bibnamefont {Niu}}, \bibinfo {author}
  {\bibfnamefont {T.~E.}\ \bibnamefont {O'Brien}}, \bibinfo {author}
  {\bibfnamefont {W.~D.}\ \bibnamefont {Oliver}}, \bibinfo {author}
  {\bibfnamefont {A.}~\bibnamefont {Opremcak}}, \bibinfo {author}
  {\bibfnamefont {K.}~\bibnamefont {Ottosson}}, \bibinfo {author}
  {\bibfnamefont {A.}~\bibnamefont {Petukhov}}, \bibinfo {author}
  {\bibfnamefont {A.}~\bibnamefont {Pizzuto}}, \bibinfo {author} {\bibfnamefont
  {J.}~\bibnamefont {Platt}}, \bibinfo {author} {\bibfnamefont
  {R.}~\bibnamefont {Potter}}, \bibinfo {author} {\bibfnamefont
  {O.}~\bibnamefont {Pritchard}}, \bibinfo {author} {\bibfnamefont {L.~P.}\
  \bibnamefont {Pryadko}}, \bibinfo {author} {\bibfnamefont {C.}~\bibnamefont
  {Quintana}}, \bibinfo {author} {\bibfnamefont {G.}~\bibnamefont
  {Ramachandran}}, \bibinfo {author} {\bibfnamefont {M.~J.}\ \bibnamefont
  {Reagor}}, \bibinfo {author} {\bibfnamefont {J.}~\bibnamefont {Redding}},
  \bibinfo {author} {\bibfnamefont {D.~M.}\ \bibnamefont {Rhodes}}, \bibinfo
  {author} {\bibfnamefont {G.}~\bibnamefont {Roberts}}, \bibinfo {author}
  {\bibfnamefont {E.}~\bibnamefont {Rosenberg}}, \bibinfo {author}
  {\bibfnamefont {E.}~\bibnamefont {Rosenfeld}}, \bibinfo {author}
  {\bibfnamefont {P.}~\bibnamefont {Roushan}}, \bibinfo {author} {\bibfnamefont
  {N.~C.}\ \bibnamefont {Rubin}}, \bibinfo {author} {\bibfnamefont
  {N.}~\bibnamefont {Saei}}, \bibinfo {author} {\bibfnamefont {D.}~\bibnamefont
  {Sank}}, \bibinfo {author} {\bibfnamefont {K.}~\bibnamefont
  {Sankaragomathi}}, \bibinfo {author} {\bibfnamefont {K.~J.}\ \bibnamefont
  {Satzinger}}, \bibinfo {author} {\bibfnamefont {H.~F.}\ \bibnamefont
  {Schurkus}}, \bibinfo {author} {\bibfnamefont {C.}~\bibnamefont {Schuster}},
  \bibinfo {author} {\bibfnamefont {A.~W.}\ \bibnamefont {Senior}}, \bibinfo
  {author} {\bibfnamefont {M.~J.}\ \bibnamefont {Shearn}}, \bibinfo {author}
  {\bibfnamefont {A.}~\bibnamefont {Shorter}}, \bibinfo {author} {\bibfnamefont
  {N.}~\bibnamefont {Shutty}}, \bibinfo {author} {\bibfnamefont
  {V.}~\bibnamefont {Shvarts}}, \bibinfo {author} {\bibfnamefont
  {S.}~\bibnamefont {Singh}}, \bibinfo {author} {\bibfnamefont
  {V.}~\bibnamefont {Sivak}}, \bibinfo {author} {\bibfnamefont
  {J.}~\bibnamefont {Skruzny}}, \bibinfo {author} {\bibfnamefont
  {S.}~\bibnamefont {Small}}, \bibinfo {author} {\bibfnamefont
  {V.}~\bibnamefont {Smelyanskiy}}, \bibinfo {author} {\bibfnamefont {W.~C.}\
  \bibnamefont {Smith}}, \bibinfo {author} {\bibfnamefont {R.~D.}\ \bibnamefont
  {Somma}}, \bibinfo {author} {\bibfnamefont {S.}~\bibnamefont {Springer}},
  \bibinfo {author} {\bibfnamefont {G.}~\bibnamefont {Sterling}}, \bibinfo
  {author} {\bibfnamefont {D.}~\bibnamefont {Strain}}, \bibinfo {author}
  {\bibfnamefont {J.}~\bibnamefont {Suchard}}, \bibinfo {author} {\bibfnamefont
  {A.}~\bibnamefont {Szasz}}, \bibinfo {author} {\bibfnamefont
  {A.}~\bibnamefont {Sztein}}, \bibinfo {author} {\bibfnamefont
  {D.}~\bibnamefont {Thor}}, \bibinfo {author} {\bibfnamefont {A.}~\bibnamefont
  {Torres}}, \bibinfo {author} {\bibfnamefont {M.~M.}\ \bibnamefont
  {Torunbalci}}, \bibinfo {author} {\bibfnamefont {A.}~\bibnamefont
  {Vaishnav}}, \bibinfo {author} {\bibfnamefont {J.}~\bibnamefont {Vargas}},
  \bibinfo {author} {\bibfnamefont {S.}~\bibnamefont {Vdovichev}}, \bibinfo
  {author} {\bibfnamefont {G.}~\bibnamefont {Vidal}}, \bibinfo {author}
  {\bibfnamefont {B.}~\bibnamefont {Villalonga}}, \bibinfo {author}
  {\bibfnamefont {C.~V.}\ \bibnamefont {Heidweiller}}, \bibinfo {author}
  {\bibfnamefont {S.}~\bibnamefont {Waltman}}, \bibinfo {author} {\bibfnamefont
  {S.~X.}\ \bibnamefont {Wang}}, \bibinfo {author} {\bibfnamefont
  {B.}~\bibnamefont {Ware}}, \bibinfo {author} {\bibfnamefont {K.}~\bibnamefont
  {Weber}}, \bibinfo {author} {\bibfnamefont {T.}~\bibnamefont {Weidel}},
  \bibinfo {author} {\bibfnamefont {T.}~\bibnamefont {White}}, \bibinfo
  {author} {\bibfnamefont {K.}~\bibnamefont {Wong}}, \bibinfo {author}
  {\bibfnamefont {B.~W.~K.}\ \bibnamefont {Woo}}, \bibinfo {author}
  {\bibfnamefont {C.}~\bibnamefont {Xing}}, \bibinfo {author} {\bibfnamefont
  {Z.~J.}\ \bibnamefont {Yao}}, \bibinfo {author} {\bibfnamefont
  {P.}~\bibnamefont {Yeh}}, \bibinfo {author} {\bibfnamefont {B.}~\bibnamefont
  {Ying}}, \bibinfo {author} {\bibfnamefont {J.}~\bibnamefont {Yoo}}, \bibinfo
  {author} {\bibfnamefont {N.}~\bibnamefont {Yosri}}, \bibinfo {author}
  {\bibfnamefont {G.}~\bibnamefont {Young}}, \bibinfo {author} {\bibfnamefont
  {A.}~\bibnamefont {Zalcman}}, \bibinfo {author} {\bibfnamefont
  {Y.}~\bibnamefont {Zhang}}, \bibinfo {author} {\bibfnamefont
  {N.}~\bibnamefont {Zhu}},\ and\ \bibinfo {author} {\bibfnamefont
  {N.}~\bibnamefont {Zobrist}},\ }\href
  {https://doi.org/10.1038/s41586-024-08449-y} {\bibfield  {journal} {\bibinfo
  {journal} {Nature}\ }\textbf {\bibinfo {volume} {638}},\ \bibinfo {pages}
  {920–926} (\bibinfo {year} {2024})}\BibitemShut {NoStop}%
\bibitem [{\citenamefont {Kim}\ \emph {et~al.}(2023)\citenamefont {Kim},
  \citenamefont {Eddins}, \citenamefont {Anand}, \citenamefont {Wei},
  \citenamefont {van~den Berg}, \citenamefont {Rosenblatt}, \citenamefont
  {Nayfeh}, \citenamefont {Wu}, \citenamefont {Zaletel}, \citenamefont
  {Temme},\ and\ \citenamefont {Kandala}}]{QuantumSupIBM}%
  \BibitemOpen
  \bibfield  {author} {\bibinfo {author} {\bibfnamefont {Y.}~\bibnamefont
  {Kim}}, \bibinfo {author} {\bibfnamefont {A.}~\bibnamefont {Eddins}},
  \bibinfo {author} {\bibfnamefont {S.}~\bibnamefont {Anand}}, \bibinfo
  {author} {\bibfnamefont {K.~X.}\ \bibnamefont {Wei}}, \bibinfo {author}
  {\bibfnamefont {E.}~\bibnamefont {van~den Berg}}, \bibinfo {author}
  {\bibfnamefont {S.}~\bibnamefont {Rosenblatt}}, \bibinfo {author}
  {\bibfnamefont {H.}~\bibnamefont {Nayfeh}}, \bibinfo {author} {\bibfnamefont
  {Y.}~\bibnamefont {Wu}}, \bibinfo {author} {\bibfnamefont {M.}~\bibnamefont
  {Zaletel}}, \bibinfo {author} {\bibfnamefont {K.}~\bibnamefont {Temme}},\
  and\ \bibinfo {author} {\bibfnamefont {A.}~\bibnamefont {Kandala}},\ }\href
  {https://doi.org/10.1038/s41586-023-06096-3} {\bibfield  {journal} {\bibinfo
  {journal} {Nature}\ }\textbf {\bibinfo {volume} {618}},\ \bibinfo {pages}
  {500} (\bibinfo {year} {2023})}\BibitemShut {NoStop}%
\bibitem [{\citenamefont {Pokharel}\ \emph {et~al.}(2026)\citenamefont
  {Pokharel}, \citenamefont {Pan}, \citenamefont {Aziz}, \citenamefont {Govia},
  \citenamefont {Ganeshan}, \citenamefont {Iadecola}, \citenamefont {Wilson},
  \citenamefont {Jones}, \citenamefont {Deshpande}, \citenamefont {Pixley},\
  and\ \citenamefont {Takita}}]{IBMAdaptive}%
  \BibitemOpen
  \bibfield  {author} {\bibinfo {author} {\bibfnamefont {B.}~\bibnamefont
  {Pokharel}}, \bibinfo {author} {\bibfnamefont {H.}~\bibnamefont {Pan}},
  \bibinfo {author} {\bibfnamefont {K.}~\bibnamefont {Aziz}}, \bibinfo {author}
  {\bibfnamefont {L.~C.~G.}\ \bibnamefont {Govia}}, \bibinfo {author}
  {\bibfnamefont {S.}~\bibnamefont {Ganeshan}}, \bibinfo {author}
  {\bibfnamefont {T.}~\bibnamefont {Iadecola}}, \bibinfo {author}
  {\bibfnamefont {J.~H.}\ \bibnamefont {Wilson}}, \bibinfo {author}
  {\bibfnamefont {B.~A.}\ \bibnamefont {Jones}}, \bibinfo {author}
  {\bibfnamefont {A.}~\bibnamefont {Deshpande}}, \bibinfo {author}
  {\bibfnamefont {J.~H.}\ \bibnamefont {Pixley}},\ and\ \bibinfo {author}
  {\bibfnamefont {M.}~\bibnamefont {Takita}},\ }\bibfield  {journal} {\bibinfo
  {journal} {Nature Physics}\ }\href
  {https://doi.org/10.1038/s41567-026-03470-6} {10.1038/s41567-026-03470-6}
  (\bibinfo {year} {2026})\BibitemShut {NoStop}%
\bibitem [{\citenamefont {Hasnaoui}\ \emph {et~al.}(2026)\citenamefont
  {Hasnaoui}, \citenamefont {Limongi}, \citenamefont {Giordani}, \citenamefont
  {Polacchi}, \citenamefont {Quaranta}, \citenamefont {Bernard}, \citenamefont
  {Sciarrino},\ and\ \citenamefont {Lobino}}]{OpticalQuantumComputingReview}%
  \BibitemOpen
  \bibfield  {author} {\bibinfo {author} {\bibfnamefont {H.}~\bibnamefont
  {Hasnaoui}}, \bibinfo {author} {\bibfnamefont {L.}~\bibnamefont {Limongi}},
  \bibinfo {author} {\bibfnamefont {T.}~\bibnamefont {Giordani}}, \bibinfo
  {author} {\bibfnamefont {B.}~\bibnamefont {Polacchi}}, \bibinfo {author}
  {\bibfnamefont {A.}~\bibnamefont {Quaranta}}, \bibinfo {author}
  {\bibfnamefont {M.}~\bibnamefont {Bernard}}, \bibinfo {author} {\bibfnamefont
  {F.}~\bibnamefont {Sciarrino}},\ and\ \bibinfo {author} {\bibfnamefont
  {M.}~\bibnamefont {Lobino}},\ }\href {https://doi.org/10.1063/5.0308063}
  {\bibfield  {journal} {\bibinfo  {journal} {Applied Physics Reviews}\
  }\textbf {\bibinfo {volume} {13}},\ \bibinfo {pages} {031306} (\bibinfo
  {year} {2026})}\BibitemShut {NoStop}%
\bibitem [{\citenamefont {Zhong}\ \emph {et~al.}(2020)\citenamefont {Zhong},
  \citenamefont {Wang}, \citenamefont {Deng}, \citenamefont {Chen},
  \citenamefont {Peng}, \citenamefont {Luo}, \citenamefont {Qin}, \citenamefont
  {Wu}, \citenamefont {Ding}, \citenamefont {Hu}, \citenamefont {Hu},
  \citenamefont {Yang}, \citenamefont {Zhang}, \citenamefont {Li},
  \citenamefont {Li}, \citenamefont {Jiang}, \citenamefont {Gan}, \citenamefont
  {Yang}, \citenamefont {You}, \citenamefont {Wang}, \citenamefont {Li},
  \citenamefont {Liu}, \citenamefont {Lu},\ and\ \citenamefont
  {Pan}}]{QuantumSupChina}%
  \BibitemOpen
  \bibfield  {author} {\bibinfo {author} {\bibfnamefont {H.-S.}\ \bibnamefont
  {Zhong}}, \bibinfo {author} {\bibfnamefont {H.}~\bibnamefont {Wang}},
  \bibinfo {author} {\bibfnamefont {Y.-H.}\ \bibnamefont {Deng}}, \bibinfo
  {author} {\bibfnamefont {M.-C.}\ \bibnamefont {Chen}}, \bibinfo {author}
  {\bibfnamefont {L.-C.}\ \bibnamefont {Peng}}, \bibinfo {author}
  {\bibfnamefont {Y.-H.}\ \bibnamefont {Luo}}, \bibinfo {author} {\bibfnamefont
  {J.}~\bibnamefont {Qin}}, \bibinfo {author} {\bibfnamefont {D.}~\bibnamefont
  {Wu}}, \bibinfo {author} {\bibfnamefont {X.}~\bibnamefont {Ding}}, \bibinfo
  {author} {\bibfnamefont {Y.}~\bibnamefont {Hu}}, \bibinfo {author}
  {\bibfnamefont {P.}~\bibnamefont {Hu}}, \bibinfo {author} {\bibfnamefont
  {X.-Y.}\ \bibnamefont {Yang}}, \bibinfo {author} {\bibfnamefont {W.-J.}\
  \bibnamefont {Zhang}}, \bibinfo {author} {\bibfnamefont {H.}~\bibnamefont
  {Li}}, \bibinfo {author} {\bibfnamefont {Y.}~\bibnamefont {Li}}, \bibinfo
  {author} {\bibfnamefont {X.}~\bibnamefont {Jiang}}, \bibinfo {author}
  {\bibfnamefont {L.}~\bibnamefont {Gan}}, \bibinfo {author} {\bibfnamefont
  {G.}~\bibnamefont {Yang}}, \bibinfo {author} {\bibfnamefont {L.}~\bibnamefont
  {You}}, \bibinfo {author} {\bibfnamefont {Z.}~\bibnamefont {Wang}}, \bibinfo
  {author} {\bibfnamefont {L.}~\bibnamefont {Li}}, \bibinfo {author}
  {\bibfnamefont {N.-L.}\ \bibnamefont {Liu}}, \bibinfo {author} {\bibfnamefont
  {C.-Y.}\ \bibnamefont {Lu}},\ and\ \bibinfo {author} {\bibfnamefont {J.-W.}\
  \bibnamefont {Pan}},\ }\href {https://doi.org/10.1126/science.abe8770}
  {\bibfield  {journal} {\bibinfo  {journal} {Science}\ }\textbf {\bibinfo
  {volume} {370}},\ \bibinfo {pages} {1460} (\bibinfo {year}
  {2020})}\BibitemShut {NoStop}%
\bibitem [{\citenamefont {Deng}\ \emph
  {et~al.}(2023{\natexlab{a}})\citenamefont {Deng}, \citenamefont {Gu},
  \citenamefont {Liu}, \citenamefont {Gong}, \citenamefont {Su}, \citenamefont
  {Zhang}, \citenamefont {Tang}, \citenamefont {Jia}, \citenamefont {Xu},
  \citenamefont {Chen}, \citenamefont {Zhong}, \citenamefont {Qin},
  \citenamefont {Wang}, \citenamefont {Peng}, \citenamefont {Yan},
  \citenamefont {Hu}, \citenamefont {Huang}, \citenamefont {Li}, \citenamefont
  {Li}, \citenamefont {Chen}, \citenamefont {Jiang}, \citenamefont {Gan},
  \citenamefont {Yang}, \citenamefont {You}, \citenamefont {Li}, \citenamefont
  {Liu}, \citenamefont {Renema}, \citenamefont {Lu},\ and\ \citenamefont
  {Pan}}]{Jiuzhang3}%
  \BibitemOpen
  \bibfield  {author} {\bibinfo {author} {\bibfnamefont {Y.-H.}\ \bibnamefont
  {Deng}}, \bibinfo {author} {\bibfnamefont {Y.-C.}\ \bibnamefont {Gu}},
  \bibinfo {author} {\bibfnamefont {H.-L.}\ \bibnamefont {Liu}}, \bibinfo
  {author} {\bibfnamefont {S.-Q.}\ \bibnamefont {Gong}}, \bibinfo {author}
  {\bibfnamefont {H.}~\bibnamefont {Su}}, \bibinfo {author} {\bibfnamefont
  {Z.-J.}\ \bibnamefont {Zhang}}, \bibinfo {author} {\bibfnamefont {H.-Y.}\
  \bibnamefont {Tang}}, \bibinfo {author} {\bibfnamefont {M.-H.}\ \bibnamefont
  {Jia}}, \bibinfo {author} {\bibfnamefont {J.-M.}\ \bibnamefont {Xu}},
  \bibinfo {author} {\bibfnamefont {M.-C.}\ \bibnamefont {Chen}}, \bibinfo
  {author} {\bibfnamefont {H.-S.}\ \bibnamefont {Zhong}}, \bibinfo {author}
  {\bibfnamefont {J.}~\bibnamefont {Qin}}, \bibinfo {author} {\bibfnamefont
  {H.}~\bibnamefont {Wang}}, \bibinfo {author} {\bibfnamefont {L.-C.}\
  \bibnamefont {Peng}}, \bibinfo {author} {\bibfnamefont {J.}~\bibnamefont
  {Yan}}, \bibinfo {author} {\bibfnamefont {Y.}~\bibnamefont {Hu}}, \bibinfo
  {author} {\bibfnamefont {J.}~\bibnamefont {Huang}}, \bibinfo {author}
  {\bibfnamefont {H.}~\bibnamefont {Li}}, \bibinfo {author} {\bibfnamefont
  {Y.}~\bibnamefont {Li}}, \bibinfo {author} {\bibfnamefont {Y.}~\bibnamefont
  {Chen}}, \bibinfo {author} {\bibfnamefont {X.}~\bibnamefont {Jiang}},
  \bibinfo {author} {\bibfnamefont {L.}~\bibnamefont {Gan}}, \bibinfo {author}
  {\bibfnamefont {G.}~\bibnamefont {Yang}}, \bibinfo {author} {\bibfnamefont
  {L.}~\bibnamefont {You}}, \bibinfo {author} {\bibfnamefont {L.}~\bibnamefont
  {Li}}, \bibinfo {author} {\bibfnamefont {N.-L.}\ \bibnamefont {Liu}},
  \bibinfo {author} {\bibfnamefont {J.~J.}\ \bibnamefont {Renema}}, \bibinfo
  {author} {\bibfnamefont {C.-Y.}\ \bibnamefont {Lu}},\ and\ \bibinfo {author}
  {\bibfnamefont {J.-W.}\ \bibnamefont {Pan}},\ }\href
  {https://doi.org/10.1103/PhysRevLett.131.150601} {\bibfield  {journal}
  {\bibinfo  {journal} {Phys. Rev. Lett.}\ }\textbf {\bibinfo {volume} {131}},\
  \bibinfo {pages} {150601} (\bibinfo {year} {2023}{\natexlab{a}})}\BibitemShut
  {NoStop}%
\bibitem [{\citenamefont {Liu}\ \emph {et~al.}(2025)\citenamefont {Liu},
  \citenamefont {Su}, \citenamefont {Gong}, \citenamefont {Gu}, \citenamefont
  {Tang}, \citenamefont {Jia}, \citenamefont {Wei}, \citenamefont {Song},
  \citenamefont {Wang}, \citenamefont {Zheng}, \citenamefont {Chen},
  \citenamefont {Li}, \citenamefont {Ren}, \citenamefont {Zhu}, \citenamefont
  {Wang}, \citenamefont {Chen}, \citenamefont {Liu}, \citenamefont {Song},
  \citenamefont {Yang}, \citenamefont {Chen}, \citenamefont {An}, \citenamefont
  {Zhang}, \citenamefont {Gan}, \citenamefont {Yang}, \citenamefont {Xu},
  \citenamefont {He}, \citenamefont {Wang}, \citenamefont {Zhong},
  \citenamefont {Chen}, \citenamefont {Jiang}, \citenamefont {Li},
  \citenamefont {Liu}, \citenamefont {Deng}, \citenamefont {Su}, \citenamefont
  {Zhang}, \citenamefont {Lu},\ and\ \citenamefont {Pan}}]{Jiuzhang4}%
  \BibitemOpen
  \bibfield  {author} {\bibinfo {author} {\bibfnamefont {H.-L.}\ \bibnamefont
  {Liu}}, \bibinfo {author} {\bibfnamefont {H.}~\bibnamefont {Su}}, \bibinfo
  {author} {\bibfnamefont {S.-Q.}\ \bibnamefont {Gong}}, \bibinfo {author}
  {\bibfnamefont {Y.-C.}\ \bibnamefont {Gu}}, \bibinfo {author} {\bibfnamefont
  {H.-Y.}\ \bibnamefont {Tang}}, \bibinfo {author} {\bibfnamefont {M.-H.}\
  \bibnamefont {Jia}}, \bibinfo {author} {\bibfnamefont {Q.}~\bibnamefont
  {Wei}}, \bibinfo {author} {\bibfnamefont {Y.}~\bibnamefont {Song}}, \bibinfo
  {author} {\bibfnamefont {D.}~\bibnamefont {Wang}}, \bibinfo {author}
  {\bibfnamefont {M.}~\bibnamefont {Zheng}}, \bibinfo {author} {\bibfnamefont
  {F.}~\bibnamefont {Chen}}, \bibinfo {author} {\bibfnamefont {L.}~\bibnamefont
  {Li}}, \bibinfo {author} {\bibfnamefont {S.}~\bibnamefont {Ren}}, \bibinfo
  {author} {\bibfnamefont {X.}~\bibnamefont {Zhu}}, \bibinfo {author}
  {\bibfnamefont {M.}~\bibnamefont {Wang}}, \bibinfo {author} {\bibfnamefont
  {Y.}~\bibnamefont {Chen}}, \bibinfo {author} {\bibfnamefont {Y.}~\bibnamefont
  {Liu}}, \bibinfo {author} {\bibfnamefont {L.}~\bibnamefont {Song}}, \bibinfo
  {author} {\bibfnamefont {P.}~\bibnamefont {Yang}}, \bibinfo {author}
  {\bibfnamefont {J.}~\bibnamefont {Chen}}, \bibinfo {author} {\bibfnamefont
  {H.}~\bibnamefont {An}}, \bibinfo {author} {\bibfnamefont {L.}~\bibnamefont
  {Zhang}}, \bibinfo {author} {\bibfnamefont {L.}~\bibnamefont {Gan}}, \bibinfo
  {author} {\bibfnamefont {G.}~\bibnamefont {Yang}}, \bibinfo {author}
  {\bibfnamefont {J.-M.}\ \bibnamefont {Xu}}, \bibinfo {author} {\bibfnamefont
  {Y.-M.}\ \bibnamefont {He}}, \bibinfo {author} {\bibfnamefont
  {H.}~\bibnamefont {Wang}}, \bibinfo {author} {\bibfnamefont {H.-S.}\
  \bibnamefont {Zhong}}, \bibinfo {author} {\bibfnamefont {M.-C.}\ \bibnamefont
  {Chen}}, \bibinfo {author} {\bibfnamefont {X.}~\bibnamefont {Jiang}},
  \bibinfo {author} {\bibfnamefont {L.}~\bibnamefont {Li}}, \bibinfo {author}
  {\bibfnamefont {N.-L.}\ \bibnamefont {Liu}}, \bibinfo {author} {\bibfnamefont
  {Y.-H.}\ \bibnamefont {Deng}}, \bibinfo {author} {\bibfnamefont {X.-L.}\
  \bibnamefont {Su}}, \bibinfo {author} {\bibfnamefont {Q.}~\bibnamefont
  {Zhang}}, \bibinfo {author} {\bibfnamefont {C.-Y.}\ \bibnamefont {Lu}},\ and\
  \bibinfo {author} {\bibfnamefont {J.-W.}\ \bibnamefont {Pan}},\ }\href@noop
  {} {\bibinfo {title} {Robust quantum computational advantage with
  programmable 3050-photon gaussian boson sampling}} (\bibinfo {year} {2025}),\
  \Eprint {https://arxiv.org/abs/2508.09092} {arXiv:2508.09092 [quant-ph]}
  \BibitemShut {NoStop}%
\bibitem [{\citenamefont {Madsen}\ \emph {et~al.}(2022)\citenamefont {Madsen},
  \citenamefont {Laudenbach}, \citenamefont {Falamarzi.Askarani}, \citenamefont
  {Rortais}, \citenamefont {Vincent}, \citenamefont {Bulmer}, \citenamefont
  {Miatto}, \citenamefont {Neuhaus}, \citenamefont {Helt}, \citenamefont
  {Collins}, \citenamefont {Lita}, \citenamefont {Gerrits}, \citenamefont
  {Nam}, \citenamefont {Vaidya}, \citenamefont {Menotti}, \citenamefont
  {Dhand}, \citenamefont {Vernon}, \citenamefont {Quesada},\ and\ \citenamefont
  {Lavoie}}]{XanaduQAdv}%
  \BibitemOpen
  \bibfield  {author} {\bibinfo {author} {\bibfnamefont {L.~S.}\ \bibnamefont
  {Madsen}}, \bibinfo {author} {\bibfnamefont {F.}~\bibnamefont {Laudenbach}},
  \bibinfo {author} {\bibfnamefont {M.}~\bibnamefont {Falamarzi.Askarani}},
  \bibinfo {author} {\bibfnamefont {F.}~\bibnamefont {Rortais}}, \bibinfo
  {author} {\bibfnamefont {T.}~\bibnamefont {Vincent}}, \bibinfo {author}
  {\bibfnamefont {J.~F.~F.}\ \bibnamefont {Bulmer}}, \bibinfo {author}
  {\bibfnamefont {F.~M.}\ \bibnamefont {Miatto}}, \bibinfo {author}
  {\bibfnamefont {L.}~\bibnamefont {Neuhaus}}, \bibinfo {author} {\bibfnamefont
  {L.~G.}\ \bibnamefont {Helt}}, \bibinfo {author} {\bibfnamefont {M.~J.}\
  \bibnamefont {Collins}}, \bibinfo {author} {\bibfnamefont {A.~E.}\
  \bibnamefont {Lita}}, \bibinfo {author} {\bibfnamefont {T.}~\bibnamefont
  {Gerrits}}, \bibinfo {author} {\bibfnamefont {S.~W.}\ \bibnamefont {Nam}},
  \bibinfo {author} {\bibfnamefont {V.~D.}\ \bibnamefont {Vaidya}}, \bibinfo
  {author} {\bibfnamefont {M.}~\bibnamefont {Menotti}}, \bibinfo {author}
  {\bibfnamefont {I.}~\bibnamefont {Dhand}}, \bibinfo {author} {\bibfnamefont
  {Z.}~\bibnamefont {Vernon}}, \bibinfo {author} {\bibfnamefont
  {N.}~\bibnamefont {Quesada}},\ and\ \bibinfo {author} {\bibfnamefont
  {J.}~\bibnamefont {Lavoie}},\ }\href
  {https://doi.org/10.1038/s41586-022-04725-x} {\bibfield  {journal} {\bibinfo
  {journal} {Nature}\ }\textbf {\bibinfo {volume} {606}},\ \bibinfo {pages}
  {75} (\bibinfo {year} {2022})}\BibitemShut {NoStop}%
\bibitem [{\citenamefont {Aaronson}\ and\ \citenamefont
  {Arkhipov}(2011)}]{BosonSampling1}%
  \BibitemOpen
  \bibfield  {author} {\bibinfo {author} {\bibfnamefont {S.}~\bibnamefont
  {Aaronson}}\ and\ \bibinfo {author} {\bibfnamefont {A.}~\bibnamefont
  {Arkhipov}},\ }in\ \href {https://doi.org/10.1145/1993636.1993682} {\emph
  {\bibinfo {booktitle} {Proceedings of the 43rd annual {ACM} symposium on
  Theory of computing - {STOC} {\textquotesingle}11}}}\ (\bibinfo  {publisher}
  {{ACM} Press},\ \bibinfo {year} {2011})\BibitemShut {NoStop}%
\bibitem [{\citenamefont {Brod}\ \emph {et~al.}(2019)\citenamefont {Brod},
  \citenamefont {Galv{\~{a}}o}, \citenamefont {Crespi}, \citenamefont
  {Osellame}, \citenamefont {Spagnolo},\ and\ \citenamefont
  {Sciarrino}}]{BosonSampling2}%
  \BibitemOpen
  \bibfield  {author} {\bibinfo {author} {\bibfnamefont {D.~J.}\ \bibnamefont
  {Brod}}, \bibinfo {author} {\bibfnamefont {E.~F.}\ \bibnamefont
  {Galv{\~{a}}o}}, \bibinfo {author} {\bibfnamefont {A.}~\bibnamefont
  {Crespi}}, \bibinfo {author} {\bibfnamefont {R.}~\bibnamefont {Osellame}},
  \bibinfo {author} {\bibfnamefont {N.}~\bibnamefont {Spagnolo}},\ and\
  \bibinfo {author} {\bibfnamefont {F.}~\bibnamefont {Sciarrino}},\ }\href
  {https://doi.org/10.1117/1.ap.1.3.034001} {\bibfield  {journal} {\bibinfo
  {journal} {Adv. Photonics}\ }\textbf {\bibinfo {volume} {1}},\ \bibinfo
  {pages} {034001} (\bibinfo {year} {2019})}\BibitemShut {NoStop}%
\bibitem [{\citenamefont {Lund}\ \emph {et~al.}(2014)\citenamefont {Lund},
  \citenamefont {Laing}, \citenamefont {Rahimi-Keshari}, \citenamefont
  {Rudolph}, \citenamefont {O'Brien},\ and\ \citenamefont
  {Ralph}}]{GBSHardness1}%
  \BibitemOpen
  \bibfield  {author} {\bibinfo {author} {\bibfnamefont {A.}~\bibnamefont
  {Lund}}, \bibinfo {author} {\bibfnamefont {A.}~\bibnamefont {Laing}},
  \bibinfo {author} {\bibfnamefont {S.}~\bibnamefont {Rahimi-Keshari}},
  \bibinfo {author} {\bibfnamefont {T.}~\bibnamefont {Rudolph}}, \bibinfo
  {author} {\bibfnamefont {J.}~\bibnamefont {O'Brien}},\ and\ \bibinfo {author}
  {\bibfnamefont {T.}~\bibnamefont {Ralph}},\ }\href
  {https://doi.org/10.1103/physrevlett.113.100502} {\bibfield  {journal}
  {\bibinfo  {journal} {Phys. Rev. Lett.}\ }\textbf {\bibinfo {volume} {113}},\
  \bibinfo {pages} {100502} (\bibinfo {year} {2014})}\BibitemShut {NoStop}%
\bibitem [{\citenamefont {Hamilton}\ \emph {et~al.}(2017)\citenamefont
  {Hamilton}, \citenamefont {Kruse}, \citenamefont {Sansoni}, \citenamefont
  {Barkhofen}, \citenamefont {Silberhorn},\ and\ \citenamefont
  {Jex}}]{GBSHardness2}%
  \BibitemOpen
  \bibfield  {author} {\bibinfo {author} {\bibfnamefont {C.~S.}\ \bibnamefont
  {Hamilton}}, \bibinfo {author} {\bibfnamefont {R.}~\bibnamefont {Kruse}},
  \bibinfo {author} {\bibfnamefont {L.}~\bibnamefont {Sansoni}}, \bibinfo
  {author} {\bibfnamefont {S.}~\bibnamefont {Barkhofen}}, \bibinfo {author}
  {\bibfnamefont {C.}~\bibnamefont {Silberhorn}},\ and\ \bibinfo {author}
  {\bibfnamefont {I.}~\bibnamefont {Jex}},\ }\href
  {https://doi.org/10.1103/physrevlett.119.170501} {\bibfield  {journal}
  {\bibinfo  {journal} {Phys. Rev. Lett.}\ }\textbf {\bibinfo {volume} {119}},\
  \bibinfo {pages} {170501} (\bibinfo {year} {2017})}\BibitemShut {NoStop}%
\bibitem [{\citenamefont {Deshpande}\ \emph {et~al.}(2022)\citenamefont
  {Deshpande}, \citenamefont {Mehta}, \citenamefont {Vincent}, \citenamefont
  {Quesada}, \citenamefont {Hinsche}, \citenamefont {Ioannou}, \citenamefont
  {Madsen}, \citenamefont {Lavoie}, \citenamefont {Qi}, \citenamefont {Eisert},
  \citenamefont {Hangleiter}, \citenamefont {Fefferman},\ and\ \citenamefont
  {Dhand}}]{GBSHardness3}%
  \BibitemOpen
  \bibfield  {author} {\bibinfo {author} {\bibfnamefont {A.}~\bibnamefont
  {Deshpande}}, \bibinfo {author} {\bibfnamefont {A.}~\bibnamefont {Mehta}},
  \bibinfo {author} {\bibfnamefont {T.}~\bibnamefont {Vincent}}, \bibinfo
  {author} {\bibfnamefont {N.}~\bibnamefont {Quesada}}, \bibinfo {author}
  {\bibfnamefont {M.}~\bibnamefont {Hinsche}}, \bibinfo {author} {\bibfnamefont
  {M.}~\bibnamefont {Ioannou}}, \bibinfo {author} {\bibfnamefont
  {L.}~\bibnamefont {Madsen}}, \bibinfo {author} {\bibfnamefont
  {J.}~\bibnamefont {Lavoie}}, \bibinfo {author} {\bibfnamefont
  {H.}~\bibnamefont {Qi}}, \bibinfo {author} {\bibfnamefont {J.}~\bibnamefont
  {Eisert}}, \bibinfo {author} {\bibfnamefont {D.}~\bibnamefont {Hangleiter}},
  \bibinfo {author} {\bibfnamefont {B.}~\bibnamefont {Fefferman}},\ and\
  \bibinfo {author} {\bibfnamefont {I.}~\bibnamefont {Dhand}},\ }\href
  {https://doi.org/10.1126/sciadv.abi7894} {\bibfield  {journal} {\bibinfo
  {journal} {Sci. Adv.}\ }\textbf {\bibinfo {volume} {8}},\ \bibinfo {pages}
  {eabi7894} (\bibinfo {year} {2022})}\BibitemShut {NoStop}%
\bibitem [{\citenamefont {Yu}\ \emph {et~al.}(2023)\citenamefont {Yu},
  \citenamefont {Zhong}, \citenamefont {Fang}, \citenamefont {Patel},
  \citenamefont {Li}, \citenamefont {Liu}, \citenamefont {Li}, \citenamefont
  {Xu}, \citenamefont {Sagona-Stophel}, \citenamefont {Mer}, \citenamefont
  {Thomas}, \citenamefont {Meng}, \citenamefont {Li}, \citenamefont {Yang},
  \citenamefont {Wang}, \citenamefont {Guo}, \citenamefont {Zhang},
  \citenamefont {Tranmer}, \citenamefont {Dong}, \citenamefont {Wang},
  \citenamefont {Tang}, \citenamefont {Li}, \citenamefont {Walmsley},\ and\
  \citenamefont {Guo}}]{yu2023universal}%
  \BibitemOpen
  \bibfield  {author} {\bibinfo {author} {\bibfnamefont {S.}~\bibnamefont
  {Yu}}, \bibinfo {author} {\bibfnamefont {Z.-P.}\ \bibnamefont {Zhong}},
  \bibinfo {author} {\bibfnamefont {Y.}~\bibnamefont {Fang}}, \bibinfo {author}
  {\bibfnamefont {R.~B.}\ \bibnamefont {Patel}}, \bibinfo {author}
  {\bibfnamefont {Q.-P.}\ \bibnamefont {Li}}, \bibinfo {author} {\bibfnamefont
  {W.}~\bibnamefont {Liu}}, \bibinfo {author} {\bibfnamefont {Z.}~\bibnamefont
  {Li}}, \bibinfo {author} {\bibfnamefont {L.}~\bibnamefont {Xu}}, \bibinfo
  {author} {\bibfnamefont {S.}~\bibnamefont {Sagona-Stophel}}, \bibinfo
  {author} {\bibfnamefont {E.}~\bibnamefont {Mer}}, \bibinfo {author}
  {\bibfnamefont {S.~E.}\ \bibnamefont {Thomas}}, \bibinfo {author}
  {\bibfnamefont {Y.}~\bibnamefont {Meng}}, \bibinfo {author} {\bibfnamefont
  {Z.-P.}\ \bibnamefont {Li}}, \bibinfo {author} {\bibfnamefont {Y.-Z.}\
  \bibnamefont {Yang}}, \bibinfo {author} {\bibfnamefont {Z.-A.}\ \bibnamefont
  {Wang}}, \bibinfo {author} {\bibfnamefont {N.-J.}\ \bibnamefont {Guo}},
  \bibinfo {author} {\bibfnamefont {W.-H.}\ \bibnamefont {Zhang}}, \bibinfo
  {author} {\bibfnamefont {G.~K.}\ \bibnamefont {Tranmer}}, \bibinfo {author}
  {\bibfnamefont {Y.}~\bibnamefont {Dong}}, \bibinfo {author} {\bibfnamefont
  {Y.-T.}\ \bibnamefont {Wang}}, \bibinfo {author} {\bibfnamefont {J.-S.}\
  \bibnamefont {Tang}}, \bibinfo {author} {\bibfnamefont {C.-F.}\ \bibnamefont
  {Li}}, \bibinfo {author} {\bibfnamefont {I.~A.}\ \bibnamefont {Walmsley}},\
  and\ \bibinfo {author} {\bibfnamefont {G.-C.}\ \bibnamefont {Guo}},\ }\href
  {https://doi.org/10.1038/s43588-023-00526-y} {\bibfield  {journal} {\bibinfo
  {journal} {Nat. Comput. Sci.}\ }\textbf {\bibinfo {volume} {3}},\ \bibinfo
  {pages} {839} (\bibinfo {year} {2023})}\BibitemShut {NoStop}%
\bibitem [{\citenamefont {Deng}\ \emph
  {et~al.}(2023{\natexlab{b}})\citenamefont {Deng}, \citenamefont {Gong},
  \citenamefont {Gu}, \citenamefont {Zhang}, \citenamefont {Liu}, \citenamefont
  {Su}, \citenamefont {Tang}, \citenamefont {Xu}, \citenamefont {Jia},
  \citenamefont {Chen}, \citenamefont {Zhong}, \citenamefont {Wang},
  \citenamefont {Yan}, \citenamefont {Hu}, \citenamefont {Huang}, \citenamefont
  {Zhang}, \citenamefont {Li}, \citenamefont {Jiang}, \citenamefont {You},
  \citenamefont {Wang}, \citenamefont {Li}, \citenamefont {Liu}, \citenamefont
  {Lu},\ and\ \citenamefont {Pan}}]{Graphs_GBS_china}%
  \BibitemOpen
  \bibfield  {author} {\bibinfo {author} {\bibfnamefont {Y.-H.}\ \bibnamefont
  {Deng}}, \bibinfo {author} {\bibfnamefont {S.-Q.}\ \bibnamefont {Gong}},
  \bibinfo {author} {\bibfnamefont {Y.-C.}\ \bibnamefont {Gu}}, \bibinfo
  {author} {\bibfnamefont {Z.-J.}\ \bibnamefont {Zhang}}, \bibinfo {author}
  {\bibfnamefont {H.-L.}\ \bibnamefont {Liu}}, \bibinfo {author} {\bibfnamefont
  {H.}~\bibnamefont {Su}}, \bibinfo {author} {\bibfnamefont {H.-Y.}\
  \bibnamefont {Tang}}, \bibinfo {author} {\bibfnamefont {J.-M.}\ \bibnamefont
  {Xu}}, \bibinfo {author} {\bibfnamefont {M.-H.}\ \bibnamefont {Jia}},
  \bibinfo {author} {\bibfnamefont {M.-C.}\ \bibnamefont {Chen}}, \bibinfo
  {author} {\bibfnamefont {H.-S.}\ \bibnamefont {Zhong}}, \bibinfo {author}
  {\bibfnamefont {H.}~\bibnamefont {Wang}}, \bibinfo {author} {\bibfnamefont
  {J.}~\bibnamefont {Yan}}, \bibinfo {author} {\bibfnamefont {Y.}~\bibnamefont
  {Hu}}, \bibinfo {author} {\bibfnamefont {J.}~\bibnamefont {Huang}}, \bibinfo
  {author} {\bibfnamefont {W.-J.}\ \bibnamefont {Zhang}}, \bibinfo {author}
  {\bibfnamefont {H.}~\bibnamefont {Li}}, \bibinfo {author} {\bibfnamefont
  {X.}~\bibnamefont {Jiang}}, \bibinfo {author} {\bibfnamefont
  {L.}~\bibnamefont {You}}, \bibinfo {author} {\bibfnamefont {Z.}~\bibnamefont
  {Wang}}, \bibinfo {author} {\bibfnamefont {L.}~\bibnamefont {Li}}, \bibinfo
  {author} {\bibfnamefont {N.-L.}\ \bibnamefont {Liu}}, \bibinfo {author}
  {\bibfnamefont {C.-Y.}\ \bibnamefont {Lu}},\ and\ \bibinfo {author}
  {\bibfnamefont {J.-W.}\ \bibnamefont {Pan}},\ }\href
  {https://doi.org/10.1103/PhysRevLett.130.190601} {\bibfield  {journal}
  {\bibinfo  {journal} {Phys. Rev. Lett.}\ }\textbf {\bibinfo {volume} {130}},\
  \bibinfo {pages} {190601} (\bibinfo {year} {2023}{\natexlab{b}})}\BibitemShut
  {NoStop}%
\bibitem [{\citenamefont {Rohde}(2015)}]{LoopLinear}%
  \BibitemOpen
  \bibfield  {author} {\bibinfo {author} {\bibfnamefont {P.~P.}\ \bibnamefont
  {Rohde}},\ }\href {https://doi.org/10.1103/PhysRevA.91.012306} {\bibfield
  {journal} {\bibinfo  {journal} {Phys. Rev. A}\ }\textbf {\bibinfo {volume}
  {91}},\ \bibinfo {pages} {012306} (\bibinfo {year} {2015})}\BibitemShut
  {NoStop}%
\bibitem [{\citenamefont {Bartlett}\ \emph {et~al.}(2021)\citenamefont
  {Bartlett}, \citenamefont {Dutt},\ and\ \citenamefont {Fan}}]{BartlettLoop}%
  \BibitemOpen
  \bibfield  {author} {\bibinfo {author} {\bibfnamefont {B.}~\bibnamefont
  {Bartlett}}, \bibinfo {author} {\bibfnamefont {A.}~\bibnamefont {Dutt}},\
  and\ \bibinfo {author} {\bibfnamefont {S.}~\bibnamefont {Fan}},\ }\href
  {https://doi.org/10.1364/OPTICA.424258} {\bibfield  {journal} {\bibinfo
  {journal} {Optica}\ }\textbf {\bibinfo {volume} {8}},\ \bibinfo {pages}
  {1515} (\bibinfo {year} {2021})}\BibitemShut {NoStop}%
\bibitem [{\citenamefont {Humphreys}\ \emph {et~al.}(2013)\citenamefont
  {Humphreys}, \citenamefont {Metcalf}, \citenamefont {Spring}, \citenamefont
  {Moore}, \citenamefont {Jin}, \citenamefont {Barbieri}, \citenamefont
  {Kolthammer},\ and\ \citenamefont {Walmsley}}]{TimeBinCPHASE}%
  \BibitemOpen
  \bibfield  {author} {\bibinfo {author} {\bibfnamefont {P.~C.}\ \bibnamefont
  {Humphreys}}, \bibinfo {author} {\bibfnamefont {B.~J.}\ \bibnamefont
  {Metcalf}}, \bibinfo {author} {\bibfnamefont {J.~B.}\ \bibnamefont {Spring}},
  \bibinfo {author} {\bibfnamefont {M.}~\bibnamefont {Moore}}, \bibinfo
  {author} {\bibfnamefont {X.-M.}\ \bibnamefont {Jin}}, \bibinfo {author}
  {\bibfnamefont {M.}~\bibnamefont {Barbieri}}, \bibinfo {author}
  {\bibfnamefont {W.~S.}\ \bibnamefont {Kolthammer}},\ and\ \bibinfo {author}
  {\bibfnamefont {I.~A.}\ \bibnamefont {Walmsley}},\ }\href
  {https://doi.org/10.1103/PhysRevLett.111.150501} {\bibfield  {journal}
  {\bibinfo  {journal} {Phys. Rev. Lett.}\ }\textbf {\bibinfo {volume} {111}},\
  \bibinfo {pages} {150501} (\bibinfo {year} {2013})}\BibitemShut {NoStop}%
\bibitem [{\citenamefont {Carosini}\ \emph {et~al.}(2024)\citenamefont
  {Carosini}, \citenamefont {Oddi}, \citenamefont {Giorgino}, \citenamefont
  {Hansen}, \citenamefont {Seron}, \citenamefont {Piacentini}, \citenamefont
  {Guggemos}, \citenamefont {Agresti}, \citenamefont {Loredo},\ and\
  \citenamefont {Walther}}]{TimeBinCarosini}%
  \BibitemOpen
  \bibfield  {author} {\bibinfo {author} {\bibfnamefont {L.}~\bibnamefont
  {Carosini}}, \bibinfo {author} {\bibfnamefont {V.}~\bibnamefont {Oddi}},
  \bibinfo {author} {\bibfnamefont {F.}~\bibnamefont {Giorgino}}, \bibinfo
  {author} {\bibfnamefont {L.~M.}\ \bibnamefont {Hansen}}, \bibinfo {author}
  {\bibfnamefont {B.}~\bibnamefont {Seron}}, \bibinfo {author} {\bibfnamefont
  {S.}~\bibnamefont {Piacentini}}, \bibinfo {author} {\bibfnamefont
  {T.}~\bibnamefont {Guggemos}}, \bibinfo {author} {\bibfnamefont
  {I.}~\bibnamefont {Agresti}}, \bibinfo {author} {\bibfnamefont {J.~C.}\
  \bibnamefont {Loredo}},\ and\ \bibinfo {author} {\bibfnamefont
  {P.}~\bibnamefont {Walther}},\ }\href
  {https://doi.org/10.1126/sciadv.adj0993} {\bibfield  {journal} {\bibinfo
  {journal} {Science Advances}\ }\textbf {\bibinfo {volume} {10}},\ \bibinfo
  {pages} {eadj0993} (\bibinfo {year} {2024})}\BibitemShut {NoStop}%
\bibitem [{\citenamefont {Bartlett}\ and\ \citenamefont
  {Fan}(2020)}]{BartlettUniversal}%
  \BibitemOpen
  \bibfield  {author} {\bibinfo {author} {\bibfnamefont {B.}~\bibnamefont
  {Bartlett}}\ and\ \bibinfo {author} {\bibfnamefont {S.}~\bibnamefont {Fan}},\
  }\href {https://doi.org/10.1103/PhysRevA.101.042319} {\bibfield  {journal}
  {\bibinfo  {journal} {Phys. Rev. A}\ }\textbf {\bibinfo {volume} {101}},\
  \bibinfo {pages} {042319} (\bibinfo {year} {2020})}\BibitemShut {NoStop}%
\bibitem [{\citenamefont {Kok}\ \emph {et~al.}(2007)\citenamefont {Kok},
  \citenamefont {Munro}, \citenamefont {Nemoto}, \citenamefont {Ralph},
  \citenamefont {Dowling},\ and\ \citenamefont {Milburn}}]{DualRail1}%
  \BibitemOpen
  \bibfield  {author} {\bibinfo {author} {\bibfnamefont {P.}~\bibnamefont
  {Kok}}, \bibinfo {author} {\bibfnamefont {W.~J.}\ \bibnamefont {Munro}},
  \bibinfo {author} {\bibfnamefont {K.}~\bibnamefont {Nemoto}}, \bibinfo
  {author} {\bibfnamefont {T.~C.}\ \bibnamefont {Ralph}}, \bibinfo {author}
  {\bibfnamefont {J.~P.}\ \bibnamefont {Dowling}},\ and\ \bibinfo {author}
  {\bibfnamefont {G.~J.}\ \bibnamefont {Milburn}},\ }\href
  {https://doi.org/10.1103/RevModPhys.79.135} {\bibfield  {journal} {\bibinfo
  {journal} {Rev. Mod. Phys.}\ }\textbf {\bibinfo {volume} {79}},\ \bibinfo
  {pages} {135} (\bibinfo {year} {2007})}\BibitemShut {NoStop}%
\bibitem [{\citenamefont {Flamini}\ \emph {et~al.}(2018)\citenamefont
  {Flamini}, \citenamefont {Spagnolo},\ and\ \citenamefont
  {Sciarrino}}]{flamini18}%
  \BibitemOpen
  \bibfield  {author} {\bibinfo {author} {\bibfnamefont {F.}~\bibnamefont
  {Flamini}}, \bibinfo {author} {\bibfnamefont {N.}~\bibnamefont {Spagnolo}},\
  and\ \bibinfo {author} {\bibfnamefont {F.}~\bibnamefont {Sciarrino}},\ }\href
  {https://doi.org/10.1088/1361-6633/aad5b2} {\bibfield  {journal} {\bibinfo
  {journal} {Rep. Progr. Phys.}\ }\textbf {\bibinfo {volume} {82}},\ \bibinfo
  {pages} {016001} (\bibinfo {year} {2018})}\BibitemShut {NoStop}%
\bibitem [{\citenamefont {Allen}\ \emph {et~al.}(1992)\citenamefont {Allen},
  \citenamefont {Beijersbergen}, \citenamefont {Spreeuw},\ and\ \citenamefont
  {Woerdman}}]{allen92}%
  \BibitemOpen
  \bibfield  {author} {\bibinfo {author} {\bibfnamefont {L.}~\bibnamefont
  {Allen}}, \bibinfo {author} {\bibfnamefont {M.~W.}\ \bibnamefont
  {Beijersbergen}}, \bibinfo {author} {\bibfnamefont {R.~J.~C.}\ \bibnamefont
  {Spreeuw}},\ and\ \bibinfo {author} {\bibfnamefont {J.~P.}\ \bibnamefont
  {Woerdman}},\ }\href {https://doi.org/10.1103/PhysRevA.45.8185} {\bibfield
  {journal} {\bibinfo  {journal} {Phys. Rev. A}\ }\textbf {\bibinfo {volume}
  {45}},\ \bibinfo {pages} {8185} (\bibinfo {year} {1992})}\BibitemShut
  {NoStop}%
\bibitem [{\citenamefont {Lukens}\ and\ \citenamefont
  {Lougovski}(2017)}]{lukens17}%
  \BibitemOpen
  \bibfield  {author} {\bibinfo {author} {\bibfnamefont {J.~M.}\ \bibnamefont
  {Lukens}}\ and\ \bibinfo {author} {\bibfnamefont {P.}~\bibnamefont
  {Lougovski}},\ }\href {https://doi.org/10.1364/OPTICA.4.000008} {\bibfield
  {journal} {\bibinfo  {journal} {Optica}\ }\textbf {\bibinfo {volume} {4}},\
  \bibinfo {pages} {8} (\bibinfo {year} {2017})}\BibitemShut {NoStop}%
\bibitem [{\citenamefont {Fukui}\ and\ \citenamefont
  {Takeda}(2022)}]{QuantuCompCV}%
  \BibitemOpen
  \bibfield  {author} {\bibinfo {author} {\bibfnamefont {K.}~\bibnamefont
  {Fukui}}\ and\ \bibinfo {author} {\bibfnamefont {S.}~\bibnamefont {Takeda}},\
  }\href {https://doi.org/10.1088/1361-6455/ac489c} {\bibfield  {journal}
  {\bibinfo  {journal} {J. Phys. B: At. Mol. Opt. Phys.}\ }\textbf {\bibinfo
  {volume} {55}},\ \bibinfo {pages} {012001} (\bibinfo {year}
  {2022})}\BibitemShut {NoStop}%
\bibitem [{\citenamefont {Gottesman}\ \emph {et~al.}(2001)\citenamefont
  {Gottesman}, \citenamefont {Kitaev},\ and\ \citenamefont
  {Preskill}}]{GKP_Orig}%
  \BibitemOpen
  \bibfield  {author} {\bibinfo {author} {\bibfnamefont {D.}~\bibnamefont
  {Gottesman}}, \bibinfo {author} {\bibfnamefont {A.}~\bibnamefont {Kitaev}},\
  and\ \bibinfo {author} {\bibfnamefont {J.}~\bibnamefont {Preskill}},\ }\href
  {https://doi.org/10.1103/PhysRevA.64.012310} {\bibfield  {journal} {\bibinfo
  {journal} {Phys. Rev. A}\ }\textbf {\bibinfo {volume} {64}},\ \bibinfo
  {pages} {012310} (\bibinfo {year} {2001})}\BibitemShut {NoStop}%
\bibitem [{\citenamefont {Zhang}\ \emph {et~al.}(2021)\citenamefont {Zhang},
  \citenamefont {Zhao}, \citenamefont {Wu},\ and\ \citenamefont
  {Guo}}]{GKP_Error1}%
  \BibitemOpen
  \bibfield  {author} {\bibinfo {author} {\bibfnamefont {J.}~\bibnamefont
  {Zhang}}, \bibinfo {author} {\bibfnamefont {J.}~\bibnamefont {Zhao}},
  \bibinfo {author} {\bibfnamefont {Y.-C.}\ \bibnamefont {Wu}},\ and\ \bibinfo
  {author} {\bibfnamefont {G.-P.}\ \bibnamefont {Guo}},\ }\href
  {https://doi.org/10.1103/PhysRevA.104.062434} {\bibfield  {journal} {\bibinfo
   {journal} {Phys. Rev. A}\ }\textbf {\bibinfo {volume} {104}},\ \bibinfo
  {pages} {062434} (\bibinfo {year} {2021})}\BibitemShut {NoStop}%
\bibitem [{\citenamefont {Grimsmo}\ and\ \citenamefont
  {Puri}(2021)}]{GKP_Error2}%
  \BibitemOpen
  \bibfield  {author} {\bibinfo {author} {\bibfnamefont {A.~L.}\ \bibnamefont
  {Grimsmo}}\ and\ \bibinfo {author} {\bibfnamefont {S.}~\bibnamefont {Puri}},\
  }\href {https://doi.org/10.1103/PRXQuantum.2.020101} {\bibfield  {journal}
  {\bibinfo  {journal} {PRX Quantum}\ }\textbf {\bibinfo {volume} {2}},\
  \bibinfo {pages} {020101} (\bibinfo {year} {2021})}\BibitemShut {NoStop}%
\bibitem [{\citenamefont {Pfister}(2019)}]{CVPfister}%
  \BibitemOpen
  \bibfield  {author} {\bibinfo {author} {\bibfnamefont {O.}~\bibnamefont
  {Pfister}},\ }\href {https://doi.org/10.1088/1361-6455/ab526f} {\bibfield
  {journal} {\bibinfo  {journal} {J. Phys. B: At. Mol. Opt. Phys.}\ }\textbf
  {\bibinfo {volume} {53}},\ \bibinfo {pages} {012001} (\bibinfo {year}
  {2019})}\BibitemShut {NoStop}%
\bibitem [{\citenamefont {Zhan}\ \emph {et~al.}(2024)\citenamefont {Zhan},
  \citenamefont {Zhang}, \citenamefont {Erbanni}, \citenamefont {Burger},
  \citenamefont {Wan}, \citenamefont {Jiang}, \citenamefont {Chae},
  \citenamefont {Liu}, \citenamefont {Poletti},\ and\ \citenamefont
  {Kwek}}]{LoopLinearMultiTime}%
  \BibitemOpen
  \bibfield  {author} {\bibinfo {author} {\bibfnamefont {Y.}~\bibnamefont
  {Zhan}}, \bibinfo {author} {\bibfnamefont {H.}~\bibnamefont {Zhang}},
  \bibinfo {author} {\bibfnamefont {R.}~\bibnamefont {Erbanni}}, \bibinfo
  {author} {\bibfnamefont {A.}~\bibnamefont {Burger}}, \bibinfo {author}
  {\bibfnamefont {L.}~\bibnamefont {Wan}}, \bibinfo {author} {\bibfnamefont
  {X.}~\bibnamefont {Jiang}}, \bibinfo {author} {\bibfnamefont
  {S.}~\bibnamefont {Chae}}, \bibinfo {author} {\bibfnamefont {A.}~\bibnamefont
  {Liu}}, \bibinfo {author} {\bibfnamefont {D.}~\bibnamefont {Poletti}},\ and\
  \bibinfo {author} {\bibfnamefont {L.~C.}\ \bibnamefont {Kwek}},\ }\href@noop
  {} {\bibinfo {title} {Loop quantum photonic chip for coherent multi-time-step
  evolution}} (\bibinfo {year} {2024}),\ \Eprint
  {https://arxiv.org/abs/2411.11307} {arXiv:2411.11307 [quant-ph]} \BibitemShut
  {NoStop}%
\bibitem [{\citenamefont {Liu}\ and\ \citenamefont
  {Wei}(2023)}]{LinearOpticalUniversalQuantumGate}%
  \BibitemOpen
  \bibfield  {author} {\bibinfo {author} {\bibfnamefont {W.-Q.}\ \bibnamefont
  {Liu}}\ and\ \bibinfo {author} {\bibfnamefont {H.-R.}\ \bibnamefont {Wei}},\
  }\href {https://doi.org/https://doi.org/10.1002/qute.202300009} {\bibfield
  {journal} {\bibinfo  {journal} {Advanced Quantum Technologies}\ }\textbf
  {\bibinfo {volume} {6}},\ \bibinfo {pages} {2300009} (\bibinfo {year}
  {2023})}\BibitemShut {NoStop}%
\bibitem [{\citenamefont {Knill}\ \emph {et~al.}(2001)\citenamefont {Knill},
  \citenamefont {Laflamme},\ and\ \citenamefont {Milburn}}]{KLMScheme}%
  \BibitemOpen
  \bibfield  {author} {\bibinfo {author} {\bibfnamefont {E.}~\bibnamefont
  {Knill}}, \bibinfo {author} {\bibfnamefont {R.}~\bibnamefont {Laflamme}},\
  and\ \bibinfo {author} {\bibfnamefont {G.~J.}\ \bibnamefont {Milburn}},\
  }\href {https://doi.org/10.1038/35051009} {\bibfield  {journal} {\bibinfo
  {journal} {Nature}\ }\textbf {\bibinfo {volume} {409}},\ \bibinfo {pages}
  {46–52} (\bibinfo {year} {2001})}\BibitemShut {NoStop}%
\bibitem [{\citenamefont {Raussendorf}\ and\ \citenamefont
  {Briegel}(2001)}]{MeasurementFirst}%
  \BibitemOpen
  \bibfield  {author} {\bibinfo {author} {\bibfnamefont {R.}~\bibnamefont
  {Raussendorf}}\ and\ \bibinfo {author} {\bibfnamefont {H.~J.}\ \bibnamefont
  {Briegel}},\ }\href {https://doi.org/10.1103/PhysRevLett.86.5188} {\bibfield
  {journal} {\bibinfo  {journal} {Phys. Rev. Lett.}\ }\textbf {\bibinfo
  {volume} {86}},\ \bibinfo {pages} {5188} (\bibinfo {year}
  {2001})}\BibitemShut {NoStop}%
\bibitem [{\citenamefont {Raussendorf}\ \emph {et~al.}(2003)\citenamefont
  {Raussendorf}, \citenamefont {Browne},\ and\ \citenamefont
  {Briegel}}]{MeasurementFirst2}%
  \BibitemOpen
  \bibfield  {author} {\bibinfo {author} {\bibfnamefont {R.}~\bibnamefont
  {Raussendorf}}, \bibinfo {author} {\bibfnamefont {D.~E.}\ \bibnamefont
  {Browne}},\ and\ \bibinfo {author} {\bibfnamefont {H.~J.}\ \bibnamefont
  {Briegel}},\ }\href {https://doi.org/10.1103/PhysRevA.68.022312} {\bibfield
  {journal} {\bibinfo  {journal} {Phys. Rev. A}\ }\textbf {\bibinfo {volume}
  {68}},\ \bibinfo {pages} {022312} (\bibinfo {year} {2003})}\BibitemShut
  {NoStop}%
\bibitem [{\citenamefont {Browne}\ and\ \citenamefont
  {Rudolph}(2005)}]{Measurement1}%
  \BibitemOpen
  \bibfield  {author} {\bibinfo {author} {\bibfnamefont {D.~E.}\ \bibnamefont
  {Browne}}\ and\ \bibinfo {author} {\bibfnamefont {T.}~\bibnamefont
  {Rudolph}},\ }\href {https://doi.org/10.1103/PhysRevLett.95.010501}
  {\bibfield  {journal} {\bibinfo  {journal} {Phys. Rev. Lett.}\ }\textbf
  {\bibinfo {volume} {95}},\ \bibinfo {pages} {010501} (\bibinfo {year}
  {2005})}\BibitemShut {NoStop}%
\bibitem [{\citenamefont {Wei}(2021)}]{MeasurementReview}%
  \BibitemOpen
  \bibfield  {author} {\bibinfo {author} {\bibfnamefont {T.-C.}\ \bibnamefont
  {Wei}},\ }\href {https://doi.org/10.1093/acrefore/9780190871994.013.31}
  {\bibinfo {title} {Measurement-based quantum computation}},\ \bibinfo
  {howpublished} {Oxford Research Encyclopedia of Physics} (\bibinfo {year}
  {2021})\BibitemShut {NoStop}%
\bibitem [{\citenamefont {O'Brien}(2007)}]{MeasurementPhotonic}%
  \BibitemOpen
  \bibfield  {author} {\bibinfo {author} {\bibfnamefont {J.~L.}\ \bibnamefont
  {O'Brien}},\ }\href {https://doi.org/10.1126/science.1142892} {\bibfield
  {journal} {\bibinfo  {journal} {Science}\ }\textbf {\bibinfo {volume}
  {318}},\ \bibinfo {pages} {1567} (\bibinfo {year} {2007})}\BibitemShut
  {NoStop}%
\bibitem [{\citenamefont {Bartolucci}\ \emph {et~al.}(2023)\citenamefont
  {Bartolucci}, \citenamefont {Birchall}, \citenamefont {Bomb{\'{\i}}n},
  \citenamefont {Cable}, \citenamefont {Dawson}, \citenamefont
  {Gimeno-Segovia}, \citenamefont {Johnston}, \citenamefont {Kieling},
  \citenamefont {Nickerson}, \citenamefont {Pant}, \citenamefont {Pastawski},
  \citenamefont {Rudolph},\ and\ \citenamefont {Sparrow}}]{PsiquantumFusion}%
  \BibitemOpen
  \bibfield  {author} {\bibinfo {author} {\bibfnamefont {S.}~\bibnamefont
  {Bartolucci}}, \bibinfo {author} {\bibfnamefont {P.}~\bibnamefont
  {Birchall}}, \bibinfo {author} {\bibfnamefont {H.}~\bibnamefont
  {Bomb{\'{\i}}n}}, \bibinfo {author} {\bibfnamefont {H.}~\bibnamefont
  {Cable}}, \bibinfo {author} {\bibfnamefont {C.}~\bibnamefont {Dawson}},
  \bibinfo {author} {\bibfnamefont {M.}~\bibnamefont {Gimeno-Segovia}},
  \bibinfo {author} {\bibfnamefont {E.}~\bibnamefont {Johnston}}, \bibinfo
  {author} {\bibfnamefont {K.}~\bibnamefont {Kieling}}, \bibinfo {author}
  {\bibfnamefont {N.}~\bibnamefont {Nickerson}}, \bibinfo {author}
  {\bibfnamefont {M.}~\bibnamefont {Pant}}, \bibinfo {author} {\bibfnamefont
  {F.}~\bibnamefont {Pastawski}}, \bibinfo {author} {\bibfnamefont
  {T.}~\bibnamefont {Rudolph}},\ and\ \bibinfo {author} {\bibfnamefont
  {C.}~\bibnamefont {Sparrow}},\ }\href
  {https://doi.org/10.1038/s41467-023-36493-1} {\bibfield  {journal} {\bibinfo
  {journal} {Nat. Commun.}\ }\textbf {\bibinfo {volume} {14}},\ \bibinfo
  {pages} {912} (\bibinfo {year} {2023})}\BibitemShut {NoStop}%
\bibitem [{\citenamefont {Sakaguchi}\ \emph {et~al.}(2023)\citenamefont
  {Sakaguchi}, \citenamefont {Konno}, \citenamefont {Hanamura}, \citenamefont
  {Asavanant}, \citenamefont {Takase}, \citenamefont {Ogawa}, \citenamefont
  {Marek}, \citenamefont {Filip}, \citenamefont {ichi Yoshikawa}, \citenamefont
  {Huntington}, \citenamefont {Yonezawa},\ and\ \citenamefont
  {Furusawa}}]{MeasurementSakaguchi}%
  \BibitemOpen
  \bibfield  {author} {\bibinfo {author} {\bibfnamefont {A.}~\bibnamefont
  {Sakaguchi}}, \bibinfo {author} {\bibfnamefont {S.}~\bibnamefont {Konno}},
  \bibinfo {author} {\bibfnamefont {F.}~\bibnamefont {Hanamura}}, \bibinfo
  {author} {\bibfnamefont {W.}~\bibnamefont {Asavanant}}, \bibinfo {author}
  {\bibfnamefont {K.}~\bibnamefont {Takase}}, \bibinfo {author} {\bibfnamefont
  {H.}~\bibnamefont {Ogawa}}, \bibinfo {author} {\bibfnamefont
  {P.}~\bibnamefont {Marek}}, \bibinfo {author} {\bibfnamefont
  {R.}~\bibnamefont {Filip}}, \bibinfo {author} {\bibfnamefont
  {J.}~\bibnamefont {ichi Yoshikawa}}, \bibinfo {author} {\bibfnamefont
  {E.}~\bibnamefont {Huntington}}, \bibinfo {author} {\bibfnamefont
  {H.}~\bibnamefont {Yonezawa}},\ and\ \bibinfo {author} {\bibfnamefont
  {A.}~\bibnamefont {Furusawa}},\ }\href
  {https://doi.org/10.1038/s41467-023-39195-w} {\bibfield  {journal} {\bibinfo
  {journal} {Nat. Commun.}\ }\textbf {\bibinfo {volume} {14}},\ \bibinfo
  {pages} {3817} (\bibinfo {year} {2023})}\BibitemShut {NoStop}%
\bibitem [{\citenamefont {Pankovich}\ \emph {et~al.}(2024)\citenamefont
  {Pankovich}, \citenamefont {Kan}, \citenamefont {Wan}, \citenamefont
  {Ostmann}, \citenamefont {Neville}, \citenamefont {Omkar}, \citenamefont
  {Sohbi},\ and\ \citenamefont {Br\'adler}}]{ORCAGHZ}%
  \BibitemOpen
  \bibfield  {author} {\bibinfo {author} {\bibfnamefont {B.}~\bibnamefont
  {Pankovich}}, \bibinfo {author} {\bibfnamefont {A.}~\bibnamefont {Kan}},
  \bibinfo {author} {\bibfnamefont {K.~H.}\ \bibnamefont {Wan}}, \bibinfo
  {author} {\bibfnamefont {M.}~\bibnamefont {Ostmann}}, \bibinfo {author}
  {\bibfnamefont {A.}~\bibnamefont {Neville}}, \bibinfo {author} {\bibfnamefont
  {S.}~\bibnamefont {Omkar}}, \bibinfo {author} {\bibfnamefont
  {A.}~\bibnamefont {Sohbi}},\ and\ \bibinfo {author} {\bibfnamefont
  {K.}~\bibnamefont {Br\'adler}},\ }\href
  {https://doi.org/10.1103/PhysRevLett.133.050604} {\bibfield  {journal}
  {\bibinfo  {journal} {Phys. Rev. Lett.}\ }\textbf {\bibinfo {volume} {133}},\
  \bibinfo {pages} {050604} (\bibinfo {year} {2024})}\BibitemShut {NoStop}%
\bibitem [{\citenamefont {Spagnolo}\ \emph {et~al.}(2023)\citenamefont
  {Spagnolo}, \citenamefont {Brod}, \citenamefont {Galvão},\ and\
  \citenamefont {Sciarrino}}]{NonlinearBSSpagnolo}%
  \BibitemOpen
  \bibfield  {author} {\bibinfo {author} {\bibfnamefont {N.}~\bibnamefont
  {Spagnolo}}, \bibinfo {author} {\bibfnamefont {D.~J.}\ \bibnamefont {Brod}},
  \bibinfo {author} {\bibfnamefont {E.~F.}\ \bibnamefont {Galvão}},\ and\
  \bibinfo {author} {\bibfnamefont {F.}~\bibnamefont {Sciarrino}},\ }\href
  {https://doi.org/10.1038/s41534-023-00676-x} {\bibfield  {journal} {\bibinfo
  {journal} {npj Quantum Information}\ }\textbf {\bibinfo {volume} {9}},\
  \bibinfo {pages} {3} (\bibinfo {year} {2023})}\BibitemShut {NoStop}%
\bibitem [{\citenamefont {Steinbrecher}\ \emph {et~al.}(2019)\citenamefont
  {Steinbrecher}, \citenamefont {Olson}, \citenamefont {Englund},\ and\
  \citenamefont {Carolan}}]{QONN_Orig}%
  \BibitemOpen
  \bibfield  {author} {\bibinfo {author} {\bibfnamefont {G.~R.}\ \bibnamefont
  {Steinbrecher}}, \bibinfo {author} {\bibfnamefont {J.~P.}\ \bibnamefont
  {Olson}}, \bibinfo {author} {\bibfnamefont {D.}~\bibnamefont {Englund}},\
  and\ \bibinfo {author} {\bibfnamefont {J.}~\bibnamefont {Carolan}},\ }\href
  {https://doi.org/10.1038/s41534-019-0174-7} {\bibfield  {journal} {\bibinfo
  {journal} {npj Quantum Inf.}\ }\textbf {\bibinfo {volume} {5}},\ \bibinfo
  {pages} {60} (\bibinfo {year} {2019})}\BibitemShut {NoStop}%
\bibitem [{\citenamefont {Stanev}\ \emph {et~al.}(2023)\citenamefont {Stanev},
  \citenamefont {Spagnolo},\ and\ \citenamefont {Sciarrino}}]{ClonerQONN}%
  \BibitemOpen
  \bibfield  {author} {\bibinfo {author} {\bibfnamefont {D.}~\bibnamefont
  {Stanev}}, \bibinfo {author} {\bibfnamefont {N.}~\bibnamefont {Spagnolo}},\
  and\ \bibinfo {author} {\bibfnamefont {F.}~\bibnamefont {Sciarrino}},\ }\href
  {https://doi.org/10.1103/PhysRevResearch.5.013139} {\bibfield  {journal}
  {\bibinfo  {journal} {Phys. Rev. Res.}\ }\textbf {\bibinfo {volume} {5}},\
  \bibinfo {pages} {013139} (\bibinfo {year} {2023})}\BibitemShut {NoStop}%
\bibitem [{\citenamefont {Ewaniuk}\ \emph {et~al.}(2023)\citenamefont
  {Ewaniuk}, \citenamefont {Carolan}, \citenamefont {Shastri},\ and\
  \citenamefont {Rotenberg}}]{QONN_Realistic}%
  \BibitemOpen
  \bibfield  {author} {\bibinfo {author} {\bibfnamefont {J.}~\bibnamefont
  {Ewaniuk}}, \bibinfo {author} {\bibfnamefont {J.}~\bibnamefont {Carolan}},
  \bibinfo {author} {\bibfnamefont {B.~J.}\ \bibnamefont {Shastri}},\ and\
  \bibinfo {author} {\bibfnamefont {N.}~\bibnamefont {Rotenberg}},\ }\href
  {https://doi.org/10.1002/qute.202200125} {\bibfield  {journal} {\bibinfo
  {journal} {Adv. Quantum Technol.}\ }\textbf {\bibinfo {volume} {6}},\
  \bibinfo {pages} {2200125} (\bibinfo {year} {2023})}\BibitemShut {NoStop}%
\bibitem [{\citenamefont {Scala}\ \emph {et~al.}(2024)\citenamefont {Scala},
  \citenamefont {Nigro},\ and\ \citenamefont {Gerace}}]{NonlinearCNOTPavia}%
  \BibitemOpen
  \bibfield  {author} {\bibinfo {author} {\bibfnamefont {F.}~\bibnamefont
  {Scala}}, \bibinfo {author} {\bibfnamefont {D.}~\bibnamefont {Nigro}},\ and\
  \bibinfo {author} {\bibfnamefont {D.}~\bibnamefont {Gerace}},\ }\href
  {https://doi.org/10.1038/s42005-024-01610-z} {\bibfield  {journal} {\bibinfo
  {journal} {Communications Physics}\ }\textbf {\bibinfo {volume} {7}},\
  \bibinfo {pages} {118} (\bibinfo {year} {2024})}\BibitemShut {NoStop}%
\bibitem [{\citenamefont {Nielsen}\ \emph {et~al.}(2025)\citenamefont
  {Nielsen}, \citenamefont {Wang}, \citenamefont {Deacon}, \citenamefont
  {Sund}, \citenamefont {Liu}, \citenamefont {Scholz}, \citenamefont {Wieck},
  \citenamefont {Ludwig}, \citenamefont {Midolo}, \citenamefont {Sørensen},
  \citenamefont {Paesani},\ and\ \citenamefont
  {Lodahl}}]{PhotonicNonlinearity}%
  \BibitemOpen
  \bibfield  {author} {\bibinfo {author} {\bibfnamefont {K.~H.}\ \bibnamefont
  {Nielsen}}, \bibinfo {author} {\bibfnamefont {Y.}~\bibnamefont {Wang}},
  \bibinfo {author} {\bibfnamefont {E.~C.~R.}\ \bibnamefont {Deacon}}, \bibinfo
  {author} {\bibfnamefont {P.~I.}\ \bibnamefont {Sund}}, \bibinfo {author}
  {\bibfnamefont {Z.}~\bibnamefont {Liu}}, \bibinfo {author} {\bibfnamefont
  {S.}~\bibnamefont {Scholz}}, \bibinfo {author} {\bibfnamefont {A.~D.}\
  \bibnamefont {Wieck}}, \bibinfo {author} {\bibfnamefont {A.}~\bibnamefont
  {Ludwig}}, \bibinfo {author} {\bibfnamefont {L.}~\bibnamefont {Midolo}},
  \bibinfo {author} {\bibfnamefont {A.~S.}\ \bibnamefont {Sørensen}}, \bibinfo
  {author} {\bibfnamefont {S.}~\bibnamefont {Paesani}},\ and\ \bibinfo {author}
  {\bibfnamefont {P.}~\bibnamefont {Lodahl}},\ }\href
  {https://doi.org/10.1038/s41467-025-66205-w} {\bibfield  {journal} {\bibinfo
  {journal} {Nature Communications}\ }\textbf {\bibinfo {volume} {16}},\
  \bibinfo {pages} {11397} (\bibinfo {year} {2025})}\BibitemShut {NoStop}%
\bibitem [{\citenamefont {Nielsen}\ \emph {et~al.}(2026)\citenamefont
  {Nielsen}, \citenamefont {Corminboeuf}, \citenamefont {Tissot}, \citenamefont
  {Pettersson}, \citenamefont {Scholz}, \citenamefont {Ludwig}, \citenamefont
  {Midolo}, \citenamefont {Sørensen}, \citenamefont {Lodahl}, \citenamefont
  {Wang},\ and\ \citenamefont {Paesani}}]{PhotonSortingNonlinearity}%
  \BibitemOpen
  \bibfield  {author} {\bibinfo {author} {\bibfnamefont {K.~H.}\ \bibnamefont
  {Nielsen}}, \bibinfo {author} {\bibfnamefont {E.}~\bibnamefont
  {Corminboeuf}}, \bibinfo {author} {\bibfnamefont {B.}~\bibnamefont {Tissot}},
  \bibinfo {author} {\bibfnamefont {L.~A.}\ \bibnamefont {Pettersson}},
  \bibinfo {author} {\bibfnamefont {S.}~\bibnamefont {Scholz}}, \bibinfo
  {author} {\bibfnamefont {A.}~\bibnamefont {Ludwig}}, \bibinfo {author}
  {\bibfnamefont {L.}~\bibnamefont {Midolo}}, \bibinfo {author} {\bibfnamefont
  {A.~S.}\ \bibnamefont {Sørensen}}, \bibinfo {author} {\bibfnamefont
  {P.}~\bibnamefont {Lodahl}}, \bibinfo {author} {\bibfnamefont
  {Y.}~\bibnamefont {Wang}},\ and\ \bibinfo {author} {\bibfnamefont
  {S.}~\bibnamefont {Paesani}},\ }\href@noop {} {\bibinfo {title} {Photon
  sorting with a quantum emitter}} (\bibinfo {year} {2026}),\ \Eprint
  {https://arxiv.org/abs/2604.21758} {arXiv:2604.21758 [quant-ph]} \BibitemShut
  {NoStop}%
\bibitem [{\citenamefont {Bourassa}\ \emph {et~al.}(2021)\citenamefont
  {Bourassa}, \citenamefont {Alexander}, \citenamefont {Vasmer}, \citenamefont
  {Patil}, \citenamefont {Tzitrin}, \citenamefont {Matsuura}, \citenamefont
  {Su}, \citenamefont {Baragiola}, \citenamefont {Guha}, \citenamefont
  {Dauphinais}, \citenamefont {Sabapathy}, \citenamefont {Menicucci},\ and\
  \citenamefont {Dhand}}]{XanaduGKP}%
  \BibitemOpen
  \bibfield  {author} {\bibinfo {author} {\bibfnamefont {J.~E.}\ \bibnamefont
  {Bourassa}}, \bibinfo {author} {\bibfnamefont {R.~N.}\ \bibnamefont
  {Alexander}}, \bibinfo {author} {\bibfnamefont {M.}~\bibnamefont {Vasmer}},
  \bibinfo {author} {\bibfnamefont {A.}~\bibnamefont {Patil}}, \bibinfo
  {author} {\bibfnamefont {I.}~\bibnamefont {Tzitrin}}, \bibinfo {author}
  {\bibfnamefont {T.}~\bibnamefont {Matsuura}}, \bibinfo {author}
  {\bibfnamefont {D.}~\bibnamefont {Su}}, \bibinfo {author} {\bibfnamefont
  {B.~Q.}\ \bibnamefont {Baragiola}}, \bibinfo {author} {\bibfnamefont
  {S.}~\bibnamefont {Guha}}, \bibinfo {author} {\bibfnamefont {G.}~\bibnamefont
  {Dauphinais}}, \bibinfo {author} {\bibfnamefont {K.~K.}\ \bibnamefont
  {Sabapathy}}, \bibinfo {author} {\bibfnamefont {N.~C.}\ \bibnamefont
  {Menicucci}},\ and\ \bibinfo {author} {\bibfnamefont {I.}~\bibnamefont
  {Dhand}},\ }\href {https://doi.org/10.22331/q-2021-02-04-392} {\bibfield
  {journal} {\bibinfo  {journal} {Quantum}\ }\textbf {\bibinfo {volume} {5}},\
  \bibinfo {pages} {392} (\bibinfo {year} {2021})}\BibitemShut {NoStop}%
\bibitem [{\citenamefont {de~Gliniasty}\ \emph {et~al.}(2024)\citenamefont
  {de~Gliniasty}, \citenamefont {Hilaire}, \citenamefont {Emeriau},
  \citenamefont {Wein}, \citenamefont {Salavrakos},\ and\ \citenamefont
  {Mansfield}}]{QuandelaSpin}%
  \BibitemOpen
  \bibfield  {author} {\bibinfo {author} {\bibfnamefont {G.}~\bibnamefont
  {de~Gliniasty}}, \bibinfo {author} {\bibfnamefont {P.}~\bibnamefont
  {Hilaire}}, \bibinfo {author} {\bibfnamefont {P.-E.}\ \bibnamefont
  {Emeriau}}, \bibinfo {author} {\bibfnamefont {S.~C.}\ \bibnamefont {Wein}},
  \bibinfo {author} {\bibfnamefont {A.}~\bibnamefont {Salavrakos}},\ and\
  \bibinfo {author} {\bibfnamefont {S.}~\bibnamefont {Mansfield}},\ }\href
  {https://doi.org/10.22331/q-2024-07-24-1423} {\bibfield  {journal} {\bibinfo
  {journal} {Quantum}\ }\textbf {\bibinfo {volume} {8}},\ \bibinfo {pages}
  {1423} (\bibinfo {year} {2024})}\BibitemShut {NoStop}%
\bibitem [{\citenamefont {Aghaee~Rad}\ \emph {et~al.}(2025)\citenamefont
  {Aghaee~Rad}, \citenamefont {Ainsworth}, \citenamefont {Alexander},
  \citenamefont {Altieri}, \citenamefont {Askarani}, \citenamefont {Baby},
  \citenamefont {Banchi}, \citenamefont {Baragiola}, \citenamefont {Bourassa},
  \citenamefont {Chadwick}, \citenamefont {Charania}, \citenamefont {Chen},
  \citenamefont {Collins}, \citenamefont {Contu}, \citenamefont {D’Arcy},
  \citenamefont {Dauphinais}, \citenamefont {De~Prins}, \citenamefont
  {Deschenes}, \citenamefont {Di~Luch}, \citenamefont {Duque}, \citenamefont
  {Edke}, \citenamefont {Fayer}, \citenamefont {Ferracin}, \citenamefont
  {Ferretti}, \citenamefont {Gefaell}, \citenamefont {Glancy}, \citenamefont
  {González-Arciniegas}, \citenamefont {Grainge}, \citenamefont {Han},
  \citenamefont {Hastrup}, \citenamefont {Helt}, \citenamefont {Hillmann},
  \citenamefont {Hundal}, \citenamefont {Izumi}, \citenamefont {Jaeken},
  \citenamefont {Jonas}, \citenamefont {Kocsis}, \citenamefont {Krasnokutska},
  \citenamefont {Larsen}, \citenamefont {Laskowski}, \citenamefont
  {Laudenbach}, \citenamefont {Lavoie}, \citenamefont {Li}, \citenamefont
  {Lomonte}, \citenamefont {Lopetegui}, \citenamefont {Luey}, \citenamefont
  {Lund}, \citenamefont {Ma}, \citenamefont {Madsen}, \citenamefont {Mahler},
  \citenamefont {Mantilla~Calderón}, \citenamefont {Menotti}, \citenamefont
  {Miatto}, \citenamefont {Morrison}, \citenamefont {Nadkarni}, \citenamefont
  {Nakamura}, \citenamefont {Neuhaus}, \citenamefont {Niu}, \citenamefont
  {Noro}, \citenamefont {Papirov}, \citenamefont {Pesah}, \citenamefont
  {Phillips}, \citenamefont {Plick}, \citenamefont {Rogalsky}, \citenamefont
  {Rortais}, \citenamefont {Sabines-Chesterking}, \citenamefont {Safavi-Bayat},
  \citenamefont {Sazhaev}, \citenamefont {Seymour}, \citenamefont
  {Rezaei~Shad}, \citenamefont {Silverman}, \citenamefont {Srinivasan},
  \citenamefont {Stephan}, \citenamefont {Tang}, \citenamefont {Tasker},
  \citenamefont {Teo}, \citenamefont {Then}, \citenamefont {Tremblay},
  \citenamefont {Tzitrin}, \citenamefont {Vaidya}, \citenamefont {Vasmer},
  \citenamefont {Vernon}, \citenamefont {Villalobos}, \citenamefont {Walshe},
  \citenamefont {Weil}, \citenamefont {Xin}, \citenamefont {Yan}, \citenamefont
  {Yao}, \citenamefont {Zamani~Abnili},\ and\ \citenamefont
  {Zhang}}]{XanaduAurora}%
  \BibitemOpen
  \bibfield  {author} {\bibinfo {author} {\bibfnamefont {H.}~\bibnamefont
  {Aghaee~Rad}}, \bibinfo {author} {\bibfnamefont {T.}~\bibnamefont
  {Ainsworth}}, \bibinfo {author} {\bibfnamefont {R.~N.}\ \bibnamefont
  {Alexander}}, \bibinfo {author} {\bibfnamefont {B.}~\bibnamefont {Altieri}},
  \bibinfo {author} {\bibfnamefont {M.~F.}\ \bibnamefont {Askarani}}, \bibinfo
  {author} {\bibfnamefont {R.}~\bibnamefont {Baby}}, \bibinfo {author}
  {\bibfnamefont {L.}~\bibnamefont {Banchi}}, \bibinfo {author} {\bibfnamefont
  {B.~Q.}\ \bibnamefont {Baragiola}}, \bibinfo {author} {\bibfnamefont {J.~E.}\
  \bibnamefont {Bourassa}}, \bibinfo {author} {\bibfnamefont {R.~S.}\
  \bibnamefont {Chadwick}}, \bibinfo {author} {\bibfnamefont {I.}~\bibnamefont
  {Charania}}, \bibinfo {author} {\bibfnamefont {H.}~\bibnamefont {Chen}},
  \bibinfo {author} {\bibfnamefont {M.~J.}\ \bibnamefont {Collins}}, \bibinfo
  {author} {\bibfnamefont {P.}~\bibnamefont {Contu}}, \bibinfo {author}
  {\bibfnamefont {N.}~\bibnamefont {D’Arcy}}, \bibinfo {author}
  {\bibfnamefont {G.}~\bibnamefont {Dauphinais}}, \bibinfo {author}
  {\bibfnamefont {R.}~\bibnamefont {De~Prins}}, \bibinfo {author}
  {\bibfnamefont {D.}~\bibnamefont {Deschenes}}, \bibinfo {author}
  {\bibfnamefont {I.}~\bibnamefont {Di~Luch}}, \bibinfo {author} {\bibfnamefont
  {S.}~\bibnamefont {Duque}}, \bibinfo {author} {\bibfnamefont
  {P.}~\bibnamefont {Edke}}, \bibinfo {author} {\bibfnamefont {S.~E.}\
  \bibnamefont {Fayer}}, \bibinfo {author} {\bibfnamefont {S.}~\bibnamefont
  {Ferracin}}, \bibinfo {author} {\bibfnamefont {H.}~\bibnamefont {Ferretti}},
  \bibinfo {author} {\bibfnamefont {J.}~\bibnamefont {Gefaell}}, \bibinfo
  {author} {\bibfnamefont {S.}~\bibnamefont {Glancy}}, \bibinfo {author}
  {\bibfnamefont {C.}~\bibnamefont {González-Arciniegas}}, \bibinfo {author}
  {\bibfnamefont {T.}~\bibnamefont {Grainge}}, \bibinfo {author} {\bibfnamefont
  {Z.}~\bibnamefont {Han}}, \bibinfo {author} {\bibfnamefont {J.}~\bibnamefont
  {Hastrup}}, \bibinfo {author} {\bibfnamefont {L.~G.}\ \bibnamefont {Helt}},
  \bibinfo {author} {\bibfnamefont {T.}~\bibnamefont {Hillmann}}, \bibinfo
  {author} {\bibfnamefont {J.}~\bibnamefont {Hundal}}, \bibinfo {author}
  {\bibfnamefont {S.}~\bibnamefont {Izumi}}, \bibinfo {author} {\bibfnamefont
  {T.}~\bibnamefont {Jaeken}}, \bibinfo {author} {\bibfnamefont
  {M.}~\bibnamefont {Jonas}}, \bibinfo {author} {\bibfnamefont
  {S.}~\bibnamefont {Kocsis}}, \bibinfo {author} {\bibfnamefont
  {I.}~\bibnamefont {Krasnokutska}}, \bibinfo {author} {\bibfnamefont {M.~V.}\
  \bibnamefont {Larsen}}, \bibinfo {author} {\bibfnamefont {P.}~\bibnamefont
  {Laskowski}}, \bibinfo {author} {\bibfnamefont {F.}~\bibnamefont
  {Laudenbach}}, \bibinfo {author} {\bibfnamefont {J.}~\bibnamefont {Lavoie}},
  \bibinfo {author} {\bibfnamefont {M.}~\bibnamefont {Li}}, \bibinfo {author}
  {\bibfnamefont {E.}~\bibnamefont {Lomonte}}, \bibinfo {author} {\bibfnamefont
  {C.~E.}\ \bibnamefont {Lopetegui}}, \bibinfo {author} {\bibfnamefont
  {B.}~\bibnamefont {Luey}}, \bibinfo {author} {\bibfnamefont {A.~P.}\
  \bibnamefont {Lund}}, \bibinfo {author} {\bibfnamefont {C.}~\bibnamefont
  {Ma}}, \bibinfo {author} {\bibfnamefont {L.~S.}\ \bibnamefont {Madsen}},
  \bibinfo {author} {\bibfnamefont {D.~H.}\ \bibnamefont {Mahler}}, \bibinfo
  {author} {\bibfnamefont {L.}~\bibnamefont {Mantilla~Calderón}}, \bibinfo
  {author} {\bibfnamefont {M.}~\bibnamefont {Menotti}}, \bibinfo {author}
  {\bibfnamefont {F.~M.}\ \bibnamefont {Miatto}}, \bibinfo {author}
  {\bibfnamefont {B.}~\bibnamefont {Morrison}}, \bibinfo {author}
  {\bibfnamefont {P.~J.}\ \bibnamefont {Nadkarni}}, \bibinfo {author}
  {\bibfnamefont {T.}~\bibnamefont {Nakamura}}, \bibinfo {author}
  {\bibfnamefont {L.}~\bibnamefont {Neuhaus}}, \bibinfo {author} {\bibfnamefont
  {Z.}~\bibnamefont {Niu}}, \bibinfo {author} {\bibfnamefont {R.}~\bibnamefont
  {Noro}}, \bibinfo {author} {\bibfnamefont {K.}~\bibnamefont {Papirov}},
  \bibinfo {author} {\bibfnamefont {A.}~\bibnamefont {Pesah}}, \bibinfo
  {author} {\bibfnamefont {D.~S.}\ \bibnamefont {Phillips}}, \bibinfo {author}
  {\bibfnamefont {W.~N.}\ \bibnamefont {Plick}}, \bibinfo {author}
  {\bibfnamefont {T.}~\bibnamefont {Rogalsky}}, \bibinfo {author}
  {\bibfnamefont {F.}~\bibnamefont {Rortais}}, \bibinfo {author} {\bibfnamefont
  {J.}~\bibnamefont {Sabines-Chesterking}}, \bibinfo {author} {\bibfnamefont
  {S.}~\bibnamefont {Safavi-Bayat}}, \bibinfo {author} {\bibfnamefont
  {E.}~\bibnamefont {Sazhaev}}, \bibinfo {author} {\bibfnamefont
  {M.}~\bibnamefont {Seymour}}, \bibinfo {author} {\bibfnamefont
  {K.}~\bibnamefont {Rezaei~Shad}}, \bibinfo {author} {\bibfnamefont
  {M.}~\bibnamefont {Silverman}}, \bibinfo {author} {\bibfnamefont {S.~A.}\
  \bibnamefont {Srinivasan}}, \bibinfo {author} {\bibfnamefont
  {M.}~\bibnamefont {Stephan}}, \bibinfo {author} {\bibfnamefont {Q.~Y.}\
  \bibnamefont {Tang}}, \bibinfo {author} {\bibfnamefont {J.~F.}\ \bibnamefont
  {Tasker}}, \bibinfo {author} {\bibfnamefont {Y.~S.}\ \bibnamefont {Teo}},
  \bibinfo {author} {\bibfnamefont {R.~B.}\ \bibnamefont {Then}}, \bibinfo
  {author} {\bibfnamefont {J.~E.}\ \bibnamefont {Tremblay}}, \bibinfo {author}
  {\bibfnamefont {I.}~\bibnamefont {Tzitrin}}, \bibinfo {author} {\bibfnamefont
  {V.~D.}\ \bibnamefont {Vaidya}}, \bibinfo {author} {\bibfnamefont
  {M.}~\bibnamefont {Vasmer}}, \bibinfo {author} {\bibfnamefont
  {Z.}~\bibnamefont {Vernon}}, \bibinfo {author} {\bibfnamefont {L.~F. S.
  S.~M.}\ \bibnamefont {Villalobos}}, \bibinfo {author} {\bibfnamefont {B.~W.}\
  \bibnamefont {Walshe}}, \bibinfo {author} {\bibfnamefont {R.}~\bibnamefont
  {Weil}}, \bibinfo {author} {\bibfnamefont {X.}~\bibnamefont {Xin}}, \bibinfo
  {author} {\bibfnamefont {X.}~\bibnamefont {Yan}}, \bibinfo {author}
  {\bibfnamefont {Y.}~\bibnamefont {Yao}}, \bibinfo {author} {\bibfnamefont
  {M.}~\bibnamefont {Zamani~Abnili}},\ and\ \bibinfo {author} {\bibfnamefont
  {Y.}~\bibnamefont {Zhang}},\ }\href
  {https://doi.org/10.1038/s41586-024-08406-9} {\bibfield  {journal} {\bibinfo
  {journal} {Nature}\ }\textbf {\bibinfo {volume} {638}},\ \bibinfo {pages}
  {912} (\bibinfo {year} {2025})}\BibitemShut {NoStop}%
\bibitem [{\citenamefont {Alexander}\ \emph {et~al.}(2024)\citenamefont
  {Alexander}, \citenamefont {Bahgat}, \citenamefont {Benyamini}, \citenamefont
  {Black}, \citenamefont {Bonneau}, \citenamefont {Burgos}, \citenamefont
  {Burridge}, \citenamefont {Campbell}, \citenamefont {Catalano}, \citenamefont
  {Ceballos}, \citenamefont {Chang}, \citenamefont {Chung}, \citenamefont
  {Danesh}, \citenamefont {Dauer}, \citenamefont {Davis}, \citenamefont
  {Dudley}, \citenamefont {Er-Xuan}, \citenamefont {Fargas}, \citenamefont
  {Farsi}, \citenamefont {Fenrich}, \citenamefont {Frazer}, \citenamefont
  {Fukami}, \citenamefont {Ganesan}, \citenamefont {Gibson}, \citenamefont
  {Gimeno-Segovia}, \citenamefont {Goeldi}, \citenamefont {Goley},
  \citenamefont {Haislmaier}, \citenamefont {Halimi}, \citenamefont {Hansen},
  \citenamefont {Hardy}, \citenamefont {Horng}, \citenamefont {House},
  \citenamefont {Hu}, \citenamefont {Jadidi}, \citenamefont {Johansson},
  \citenamefont {Jones}, \citenamefont {Kamineni}, \citenamefont {Kelez},
  \citenamefont {Koustuban}, \citenamefont {Kovall}, \citenamefont {Krogen},
  \citenamefont {Kumar}, \citenamefont {Liang}, \citenamefont {LiCausi},
  \citenamefont {Llewellyn}, \citenamefont {Lokovic}, \citenamefont {Lovelady},
  \citenamefont {Manfrinato}, \citenamefont {Melnichuk}, \citenamefont {Souza},
  \citenamefont {Mendoza}, \citenamefont {Moores}, \citenamefont {Mukherjee},
  \citenamefont {Munns}, \citenamefont {Musalem}, \citenamefont {Najafi},
  \citenamefont {O'Brien}, \citenamefont {Ortmann}, \citenamefont {Pai},
  \citenamefont {Park}, \citenamefont {Peng}, \citenamefont {Penthorn},
  \citenamefont {Peterson}, \citenamefont {Poush}, \citenamefont {Pryde},
  \citenamefont {Ramprasad}, \citenamefont {Ray}, \citenamefont {Rodriguez},
  \citenamefont {Roxworthy}, \citenamefont {Rudolph}, \citenamefont {Saunders},
  \citenamefont {Shadbolt}, \citenamefont {Shah}, \citenamefont {Shin},
  \citenamefont {Smith}, \citenamefont {Sohn}, \citenamefont {Sohn},
  \citenamefont {Son}, \citenamefont {Sparrow}, \citenamefont {Staffaroni},
  \citenamefont {Stavrakas}, \citenamefont {Sukumaran}, \citenamefont
  {Tamborini}, \citenamefont {Thompson}, \citenamefont {Tran}, \citenamefont
  {Triplet}, \citenamefont {Tung}, \citenamefont {Vert}, \citenamefont
  {Vidrighin}, \citenamefont {Vorobeichik}, \citenamefont {Weigel},
  \citenamefont {Wingert}, \citenamefont {Wooding},\ and\ \citenamefont
  {Zhou}}]{PsiQuantumFab}%
  \BibitemOpen
  \bibfield  {author} {\bibinfo {author} {\bibfnamefont {K.}~\bibnamefont
  {Alexander}}, \bibinfo {author} {\bibfnamefont {A.}~\bibnamefont {Bahgat}},
  \bibinfo {author} {\bibfnamefont {A.}~\bibnamefont {Benyamini}}, \bibinfo
  {author} {\bibfnamefont {D.}~\bibnamefont {Black}}, \bibinfo {author}
  {\bibfnamefont {D.}~\bibnamefont {Bonneau}}, \bibinfo {author} {\bibfnamefont
  {S.}~\bibnamefont {Burgos}}, \bibinfo {author} {\bibfnamefont
  {B.}~\bibnamefont {Burridge}}, \bibinfo {author} {\bibfnamefont
  {G.}~\bibnamefont {Campbell}}, \bibinfo {author} {\bibfnamefont
  {G.}~\bibnamefont {Catalano}}, \bibinfo {author} {\bibfnamefont
  {A.}~\bibnamefont {Ceballos}}, \bibinfo {author} {\bibfnamefont {C.-M.}\
  \bibnamefont {Chang}}, \bibinfo {author} {\bibfnamefont {C.}~\bibnamefont
  {Chung}}, \bibinfo {author} {\bibfnamefont {F.}~\bibnamefont {Danesh}},
  \bibinfo {author} {\bibfnamefont {T.}~\bibnamefont {Dauer}}, \bibinfo
  {author} {\bibfnamefont {M.}~\bibnamefont {Davis}}, \bibinfo {author}
  {\bibfnamefont {E.}~\bibnamefont {Dudley}}, \bibinfo {author} {\bibfnamefont
  {P.}~\bibnamefont {Er-Xuan}}, \bibinfo {author} {\bibfnamefont
  {J.}~\bibnamefont {Fargas}}, \bibinfo {author} {\bibfnamefont
  {A.}~\bibnamefont {Farsi}}, \bibinfo {author} {\bibfnamefont
  {C.}~\bibnamefont {Fenrich}}, \bibinfo {author} {\bibfnamefont
  {J.}~\bibnamefont {Frazer}}, \bibinfo {author} {\bibfnamefont
  {M.}~\bibnamefont {Fukami}}, \bibinfo {author} {\bibfnamefont
  {Y.}~\bibnamefont {Ganesan}}, \bibinfo {author} {\bibfnamefont
  {G.}~\bibnamefont {Gibson}}, \bibinfo {author} {\bibfnamefont
  {M.}~\bibnamefont {Gimeno-Segovia}}, \bibinfo {author} {\bibfnamefont
  {S.}~\bibnamefont {Goeldi}}, \bibinfo {author} {\bibfnamefont
  {P.}~\bibnamefont {Goley}}, \bibinfo {author} {\bibfnamefont
  {R.}~\bibnamefont {Haislmaier}}, \bibinfo {author} {\bibfnamefont
  {S.}~\bibnamefont {Halimi}}, \bibinfo {author} {\bibfnamefont
  {P.}~\bibnamefont {Hansen}}, \bibinfo {author} {\bibfnamefont
  {S.}~\bibnamefont {Hardy}}, \bibinfo {author} {\bibfnamefont
  {J.}~\bibnamefont {Horng}}, \bibinfo {author} {\bibfnamefont
  {M.}~\bibnamefont {House}}, \bibinfo {author} {\bibfnamefont
  {H.}~\bibnamefont {Hu}}, \bibinfo {author} {\bibfnamefont {M.}~\bibnamefont
  {Jadidi}}, \bibinfo {author} {\bibfnamefont {H.}~\bibnamefont {Johansson}},
  \bibinfo {author} {\bibfnamefont {T.}~\bibnamefont {Jones}}, \bibinfo
  {author} {\bibfnamefont {V.}~\bibnamefont {Kamineni}}, \bibinfo {author}
  {\bibfnamefont {N.}~\bibnamefont {Kelez}}, \bibinfo {author} {\bibfnamefont
  {R.}~\bibnamefont {Koustuban}}, \bibinfo {author} {\bibfnamefont
  {G.}~\bibnamefont {Kovall}}, \bibinfo {author} {\bibfnamefont
  {P.}~\bibnamefont {Krogen}}, \bibinfo {author} {\bibfnamefont
  {N.}~\bibnamefont {Kumar}}, \bibinfo {author} {\bibfnamefont
  {Y.}~\bibnamefont {Liang}}, \bibinfo {author} {\bibfnamefont
  {N.}~\bibnamefont {LiCausi}}, \bibinfo {author} {\bibfnamefont
  {D.}~\bibnamefont {Llewellyn}}, \bibinfo {author} {\bibfnamefont
  {K.}~\bibnamefont {Lokovic}}, \bibinfo {author} {\bibfnamefont
  {M.}~\bibnamefont {Lovelady}}, \bibinfo {author} {\bibfnamefont
  {V.}~\bibnamefont {Manfrinato}}, \bibinfo {author} {\bibfnamefont
  {A.}~\bibnamefont {Melnichuk}}, \bibinfo {author} {\bibfnamefont
  {M.}~\bibnamefont {Souza}}, \bibinfo {author} {\bibfnamefont
  {G.}~\bibnamefont {Mendoza}}, \bibinfo {author} {\bibfnamefont
  {B.}~\bibnamefont {Moores}}, \bibinfo {author} {\bibfnamefont
  {S.}~\bibnamefont {Mukherjee}}, \bibinfo {author} {\bibfnamefont
  {J.}~\bibnamefont {Munns}}, \bibinfo {author} {\bibfnamefont {F.-X.}\
  \bibnamefont {Musalem}}, \bibinfo {author} {\bibfnamefont {F.}~\bibnamefont
  {Najafi}}, \bibinfo {author} {\bibfnamefont {J.~L.}\ \bibnamefont {O'Brien}},
  \bibinfo {author} {\bibfnamefont {J.~E.}\ \bibnamefont {Ortmann}}, \bibinfo
  {author} {\bibfnamefont {S.}~\bibnamefont {Pai}}, \bibinfo {author}
  {\bibfnamefont {B.}~\bibnamefont {Park}}, \bibinfo {author} {\bibfnamefont
  {H.-T.}\ \bibnamefont {Peng}}, \bibinfo {author} {\bibfnamefont
  {N.}~\bibnamefont {Penthorn}}, \bibinfo {author} {\bibfnamefont
  {B.}~\bibnamefont {Peterson}}, \bibinfo {author} {\bibfnamefont
  {M.}~\bibnamefont {Poush}}, \bibinfo {author} {\bibfnamefont {G.~J.}\
  \bibnamefont {Pryde}}, \bibinfo {author} {\bibfnamefont {T.}~\bibnamefont
  {Ramprasad}}, \bibinfo {author} {\bibfnamefont {G.}~\bibnamefont {Ray}},
  \bibinfo {author} {\bibfnamefont {A.}~\bibnamefont {Rodriguez}}, \bibinfo
  {author} {\bibfnamefont {B.}~\bibnamefont {Roxworthy}}, \bibinfo {author}
  {\bibfnamefont {T.}~\bibnamefont {Rudolph}}, \bibinfo {author} {\bibfnamefont
  {D.~J.}\ \bibnamefont {Saunders}}, \bibinfo {author} {\bibfnamefont
  {P.}~\bibnamefont {Shadbolt}}, \bibinfo {author} {\bibfnamefont
  {D.}~\bibnamefont {Shah}}, \bibinfo {author} {\bibfnamefont {H.}~\bibnamefont
  {Shin}}, \bibinfo {author} {\bibfnamefont {J.}~\bibnamefont {Smith}},
  \bibinfo {author} {\bibfnamefont {B.}~\bibnamefont {Sohn}}, \bibinfo {author}
  {\bibfnamefont {Y.-I.}\ \bibnamefont {Sohn}}, \bibinfo {author}
  {\bibfnamefont {G.}~\bibnamefont {Son}}, \bibinfo {author} {\bibfnamefont
  {C.}~\bibnamefont {Sparrow}}, \bibinfo {author} {\bibfnamefont
  {M.}~\bibnamefont {Staffaroni}}, \bibinfo {author} {\bibfnamefont
  {C.}~\bibnamefont {Stavrakas}}, \bibinfo {author} {\bibfnamefont
  {V.}~\bibnamefont {Sukumaran}}, \bibinfo {author} {\bibfnamefont
  {D.}~\bibnamefont {Tamborini}}, \bibinfo {author} {\bibfnamefont {M.~G.}\
  \bibnamefont {Thompson}}, \bibinfo {author} {\bibfnamefont {K.}~\bibnamefont
  {Tran}}, \bibinfo {author} {\bibfnamefont {M.}~\bibnamefont {Triplet}},
  \bibinfo {author} {\bibfnamefont {M.}~\bibnamefont {Tung}}, \bibinfo {author}
  {\bibfnamefont {A.}~\bibnamefont {Vert}}, \bibinfo {author} {\bibfnamefont
  {M.~D.}\ \bibnamefont {Vidrighin}}, \bibinfo {author} {\bibfnamefont
  {I.}~\bibnamefont {Vorobeichik}}, \bibinfo {author} {\bibfnamefont
  {P.}~\bibnamefont {Weigel}}, \bibinfo {author} {\bibfnamefont
  {M.}~\bibnamefont {Wingert}}, \bibinfo {author} {\bibfnamefont
  {J.}~\bibnamefont {Wooding}},\ and\ \bibinfo {author} {\bibfnamefont
  {X.}~\bibnamefont {Zhou}},\ }\href@noop {} {\bibinfo {title} {A
  manufacturable platform for photonic quantum computing}} (\bibinfo {year}
  {2024}),\ \Eprint {https://arxiv.org/abs/2404.17570} {arXiv:2404.17570
  [quant-ph]} \BibitemShut {NoStop}%
\bibitem [{\citenamefont {Maring}\ \emph {et~al.}(2024)\citenamefont {Maring},
  \citenamefont {Fyrillas}, \citenamefont {Pont}, \citenamefont {Ivanov},
  \citenamefont {Stepanov}, \citenamefont {Margaria}, \citenamefont {Hease},
  \citenamefont {Pishchagin}, \citenamefont {Lemaître}, \citenamefont
  {Sagnes}, \citenamefont {Au}, \citenamefont {Boissier}, \citenamefont
  {Bertasi}, \citenamefont {Baert}, \citenamefont {Valdivia}, \citenamefont
  {Billard}, \citenamefont {Acar}, \citenamefont {Brieussel}, \citenamefont
  {Mezher}, \citenamefont {Wein}, \citenamefont {Salavrakos}, \citenamefont
  {Sinnott}, \citenamefont {Fioretto}, \citenamefont {Emeriau}, \citenamefont
  {Belabas}, \citenamefont {Mansfield}, \citenamefont {Senellart},
  \citenamefont {Senellart},\ and\ \citenamefont {Somaschi}}]{QuandelaAscella}%
  \BibitemOpen
  \bibfield  {author} {\bibinfo {author} {\bibfnamefont {N.}~\bibnamefont
  {Maring}}, \bibinfo {author} {\bibfnamefont {A.}~\bibnamefont {Fyrillas}},
  \bibinfo {author} {\bibfnamefont {M.}~\bibnamefont {Pont}}, \bibinfo {author}
  {\bibfnamefont {E.}~\bibnamefont {Ivanov}}, \bibinfo {author} {\bibfnamefont
  {P.}~\bibnamefont {Stepanov}}, \bibinfo {author} {\bibfnamefont
  {N.}~\bibnamefont {Margaria}}, \bibinfo {author} {\bibfnamefont
  {W.}~\bibnamefont {Hease}}, \bibinfo {author} {\bibfnamefont
  {A.}~\bibnamefont {Pishchagin}}, \bibinfo {author} {\bibfnamefont
  {A.}~\bibnamefont {Lemaître}}, \bibinfo {author} {\bibfnamefont
  {I.}~\bibnamefont {Sagnes}}, \bibinfo {author} {\bibfnamefont {T.~H.}\
  \bibnamefont {Au}}, \bibinfo {author} {\bibfnamefont {S.}~\bibnamefont
  {Boissier}}, \bibinfo {author} {\bibfnamefont {E.}~\bibnamefont {Bertasi}},
  \bibinfo {author} {\bibfnamefont {A.}~\bibnamefont {Baert}}, \bibinfo
  {author} {\bibfnamefont {M.}~\bibnamefont {Valdivia}}, \bibinfo {author}
  {\bibfnamefont {M.}~\bibnamefont {Billard}}, \bibinfo {author} {\bibfnamefont
  {O.}~\bibnamefont {Acar}}, \bibinfo {author} {\bibfnamefont {A.}~\bibnamefont
  {Brieussel}}, \bibinfo {author} {\bibfnamefont {R.}~\bibnamefont {Mezher}},
  \bibinfo {author} {\bibfnamefont {S.~C.}\ \bibnamefont {Wein}}, \bibinfo
  {author} {\bibfnamefont {A.}~\bibnamefont {Salavrakos}}, \bibinfo {author}
  {\bibfnamefont {P.}~\bibnamefont {Sinnott}}, \bibinfo {author} {\bibfnamefont
  {D.~A.}\ \bibnamefont {Fioretto}}, \bibinfo {author} {\bibfnamefont {P.-E.}\
  \bibnamefont {Emeriau}}, \bibinfo {author} {\bibfnamefont {N.}~\bibnamefont
  {Belabas}}, \bibinfo {author} {\bibfnamefont {S.}~\bibnamefont {Mansfield}},
  \bibinfo {author} {\bibfnamefont {P.}~\bibnamefont {Senellart}}, \bibinfo
  {author} {\bibfnamefont {J.}~\bibnamefont {Senellart}},\ and\ \bibinfo
  {author} {\bibfnamefont {N.}~\bibnamefont {Somaschi}},\ }\href
  {https://doi.org/10.1038/s41566-024-01403-4} {\bibfield  {journal} {\bibinfo
  {journal} {Nature Photonics}\ }\textbf {\bibinfo {volume} {18}},\ \bibinfo
  {pages} {603–609} (\bibinfo {year} {2024})}\BibitemShut {NoStop}%
\bibitem [{\citenamefont {Di~Micco}\ \emph {et~al.}(2026)\citenamefont
  {Di~Micco}, \citenamefont {Hoch}, \citenamefont {Ciorra}, \citenamefont
  {de~Salles}, \citenamefont {Spagnolo}, \citenamefont {Giordani},
  \citenamefont {Carvacho}, \citenamefont {Di~Giano}, \citenamefont {Gardina},
  \citenamefont {Crespi}, \citenamefont {Ceccarelli}, \citenamefont
  {Osellame},\ and\ \citenamefont {Sciarrino}}]{DiMicco2026BosonSampling128}%
  \BibitemOpen
  \bibfield  {author} {\bibinfo {author} {\bibfnamefont {S.}~\bibnamefont
  {Di~Micco}}, \bibinfo {author} {\bibfnamefont {F.}~\bibnamefont {Hoch}},
  \bibinfo {author} {\bibfnamefont {A.}~\bibnamefont {Ciorra}}, \bibinfo
  {author} {\bibfnamefont {D.~C.}\ \bibnamefont {de~Salles}}, \bibinfo {author}
  {\bibfnamefont {N.}~\bibnamefont {Spagnolo}}, \bibinfo {author}
  {\bibfnamefont {T.}~\bibnamefont {Giordani}}, \bibinfo {author}
  {\bibfnamefont {G.}~\bibnamefont {Carvacho}}, \bibinfo {author}
  {\bibfnamefont {N.}~\bibnamefont {Di~Giano}}, \bibinfo {author}
  {\bibfnamefont {M.}~\bibnamefont {Gardina}}, \bibinfo {author} {\bibfnamefont
  {A.}~\bibnamefont {Crespi}}, \bibinfo {author} {\bibfnamefont
  {F.}~\bibnamefont {Ceccarelli}}, \bibinfo {author} {\bibfnamefont
  {R.}~\bibnamefont {Osellame}},\ and\ \bibinfo {author} {\bibfnamefont
  {F.}~\bibnamefont {Sciarrino}},\ }\href@noop {} {\bibinfo {title} {Boson
  sampling with a reconfigurable 128 modes 3d integrated photonic circuit}}
  (\bibinfo {year} {2026}),\ \Eprint {https://arxiv.org/abs/2605.04162}
  {arXiv:2605.04162 [quant-ph]} \BibitemShut {NoStop}%
\bibitem [{\citenamefont {Pont}\ \emph {et~al.}(2022)\citenamefont {Pont},
  \citenamefont {Albiero}, \citenamefont {Thomas}, \citenamefont {Spagnolo},
  \citenamefont {Ceccarelli}, \citenamefont {Corrielli}, \citenamefont
  {Brieussel}, \citenamefont {Somaschi}, \citenamefont {Huet}, \citenamefont
  {Harouri}, \citenamefont {Lema\^{\i}tre}, \citenamefont {Sagnes},
  \citenamefont {Belabas}, \citenamefont {Sciarrino}, \citenamefont {Osellame},
  \citenamefont {Senellart},\ and\ \citenamefont {Crespi}}]{Pont22}%
  \BibitemOpen
  \bibfield  {author} {\bibinfo {author} {\bibfnamefont {M.}~\bibnamefont
  {Pont}}, \bibinfo {author} {\bibfnamefont {R.}~\bibnamefont {Albiero}},
  \bibinfo {author} {\bibfnamefont {S.~E.}\ \bibnamefont {Thomas}}, \bibinfo
  {author} {\bibfnamefont {N.}~\bibnamefont {Spagnolo}}, \bibinfo {author}
  {\bibfnamefont {F.}~\bibnamefont {Ceccarelli}}, \bibinfo {author}
  {\bibfnamefont {G.}~\bibnamefont {Corrielli}}, \bibinfo {author}
  {\bibfnamefont {A.}~\bibnamefont {Brieussel}}, \bibinfo {author}
  {\bibfnamefont {N.}~\bibnamefont {Somaschi}}, \bibinfo {author}
  {\bibfnamefont {H.}~\bibnamefont {Huet}}, \bibinfo {author} {\bibfnamefont
  {A.}~\bibnamefont {Harouri}}, \bibinfo {author} {\bibfnamefont
  {A.}~\bibnamefont {Lema\^{\i}tre}}, \bibinfo {author} {\bibfnamefont
  {I.}~\bibnamefont {Sagnes}}, \bibinfo {author} {\bibfnamefont
  {N.}~\bibnamefont {Belabas}}, \bibinfo {author} {\bibfnamefont
  {F.}~\bibnamefont {Sciarrino}}, \bibinfo {author} {\bibfnamefont
  {R.}~\bibnamefont {Osellame}}, \bibinfo {author} {\bibfnamefont
  {P.}~\bibnamefont {Senellart}},\ and\ \bibinfo {author} {\bibfnamefont
  {A.}~\bibnamefont {Crespi}},\ }\href
  {https://doi.org/10.1103/PhysRevX.12.031033} {\bibfield  {journal} {\bibinfo
  {journal} {Phys. Rev. X}\ }\textbf {\bibinfo {volume} {12}},\ \bibinfo
  {pages} {031033} (\bibinfo {year} {2022})}\BibitemShut {NoStop}%
\bibitem [{\citenamefont {Rodari}\ \emph {et~al.}(2024)\citenamefont {Rodari},
  \citenamefont {Novo}, \citenamefont {Albiero}, \citenamefont {Suprano},
  \citenamefont {Tavares}, \citenamefont {Caruccio}, \citenamefont {Hoch},
  \citenamefont {Giordani}, \citenamefont {Carvacho}, \citenamefont {Gardina},
  \citenamefont {{Di Giano}}, \citenamefont {{Di Giorgio}}, \citenamefont
  {Corrielli}, \citenamefont {Ceccarelli}, \citenamefont {Osellame},
  \citenamefont {Spagnolo}, \citenamefont {Galv\~{a}o},\ and\ \citenamefont
  {Sciarrino}}]{rodari2024}%
  \BibitemOpen
  \bibfield  {author} {\bibinfo {author} {\bibfnamefont {G.}~\bibnamefont
  {Rodari}}, \bibinfo {author} {\bibfnamefont {L.}~\bibnamefont {Novo}},
  \bibinfo {author} {\bibfnamefont {R.}~\bibnamefont {Albiero}}, \bibinfo
  {author} {\bibfnamefont {A.}~\bibnamefont {Suprano}}, \bibinfo {author}
  {\bibfnamefont {C.~T.}\ \bibnamefont {Tavares}}, \bibinfo {author}
  {\bibfnamefont {E.}~\bibnamefont {Caruccio}}, \bibinfo {author}
  {\bibfnamefont {F.}~\bibnamefont {Hoch}}, \bibinfo {author} {\bibfnamefont
  {T.}~\bibnamefont {Giordani}}, \bibinfo {author} {\bibfnamefont
  {G.}~\bibnamefont {Carvacho}}, \bibinfo {author} {\bibfnamefont
  {M.}~\bibnamefont {Gardina}}, \bibinfo {author} {\bibfnamefont
  {N.}~\bibnamefont {{Di Giano}}}, \bibinfo {author} {\bibfnamefont
  {S.}~\bibnamefont {{Di Giorgio}}}, \bibinfo {author} {\bibfnamefont
  {G.}~\bibnamefont {Corrielli}}, \bibinfo {author} {\bibfnamefont
  {F.}~\bibnamefont {Ceccarelli}}, \bibinfo {author} {\bibfnamefont
  {R.}~\bibnamefont {Osellame}}, \bibinfo {author} {\bibfnamefont
  {N.}~\bibnamefont {Spagnolo}}, \bibinfo {author} {\bibfnamefont {E.~F.}\
  \bibnamefont {Galv\~{a}o}},\ and\ \bibinfo {author} {\bibfnamefont
  {F.}~\bibnamefont {Sciarrino}},\ }\href@noop {} {\bibinfo {title}
  {Semi-device independent characterization of multiphoton
  indistinguishability}} (\bibinfo {year} {2024}),\ \Eprint
  {https://arxiv.org/abs/2404.18636} {arXiv:2404.18636 [quant-ph]} \BibitemShut
  {NoStop}%
\bibitem [{\citenamefont {Das}\ \emph {et~al.}(2022)\citenamefont {Das},
  \citenamefont {Vittal},\ and\ \citenamefont {Qureshi}}]{swap1}%
  \BibitemOpen
  \bibfield  {author} {\bibinfo {author} {\bibfnamefont {P.}~\bibnamefont
  {Das}}, \bibinfo {author} {\bibfnamefont {S.~K.}\ \bibnamefont {Vittal}},\
  and\ \bibinfo {author} {\bibfnamefont {M.}~\bibnamefont {Qureshi}},\
  }\href@noop {} {\bibinfo {title} {Foresight: Reducing swaps in nisq programs
  via adaptive multi-candidate evaluations}} (\bibinfo {year} {2022}),\ \Eprint
  {https://arxiv.org/abs/2204.13142} {arXiv:2204.13142 [quant-ph]} \BibitemShut
  {NoStop}%
\bibitem [{\citenamefont {Hashim}\ \emph {et~al.}(2022)\citenamefont {Hashim},
  \citenamefont {Rines}, \citenamefont {Omole}, \citenamefont {Naik},
  \citenamefont {Kreikebaum}, \citenamefont {Santiago}, \citenamefont {Chong},
  \citenamefont {Siddiqi},\ and\ \citenamefont {Gokhale}}]{swap2}%
  \BibitemOpen
  \bibfield  {author} {\bibinfo {author} {\bibfnamefont {A.}~\bibnamefont
  {Hashim}}, \bibinfo {author} {\bibfnamefont {R.}~\bibnamefont {Rines}},
  \bibinfo {author} {\bibfnamefont {V.}~\bibnamefont {Omole}}, \bibinfo
  {author} {\bibfnamefont {R.~K.}\ \bibnamefont {Naik}}, \bibinfo {author}
  {\bibfnamefont {J.~M.}\ \bibnamefont {Kreikebaum}}, \bibinfo {author}
  {\bibfnamefont {D.~I.}\ \bibnamefont {Santiago}}, \bibinfo {author}
  {\bibfnamefont {F.~T.}\ \bibnamefont {Chong}}, \bibinfo {author}
  {\bibfnamefont {I.}~\bibnamefont {Siddiqi}},\ and\ \bibinfo {author}
  {\bibfnamefont {P.}~\bibnamefont {Gokhale}},\ }\href
  {https://doi.org/10.1103/PhysRevResearch.4.033028} {\bibfield  {journal}
  {\bibinfo  {journal} {Phys. Rev. Res.}\ }\textbf {\bibinfo {volume} {4}},\
  \bibinfo {pages} {033028} (\bibinfo {year} {2022})}\BibitemShut {NoStop}%
\bibitem [{\citenamefont {Barenco}\ \emph {et~al.}(1995)\citenamefont
  {Barenco}, \citenamefont {Bennett}, \citenamefont {Cleve}, \citenamefont
  {DiVincenzo}, \citenamefont {Margolus}, \citenamefont {Shor}, \citenamefont
  {Sleator}, \citenamefont {Smolin},\ and\ \citenamefont
  {Weinfurter}}]{UniversalQuantumGates}%
  \BibitemOpen
  \bibfield  {author} {\bibinfo {author} {\bibfnamefont {A.}~\bibnamefont
  {Barenco}}, \bibinfo {author} {\bibfnamefont {C.~H.}\ \bibnamefont
  {Bennett}}, \bibinfo {author} {\bibfnamefont {R.}~\bibnamefont {Cleve}},
  \bibinfo {author} {\bibfnamefont {D.~P.}\ \bibnamefont {DiVincenzo}},
  \bibinfo {author} {\bibfnamefont {N.}~\bibnamefont {Margolus}}, \bibinfo
  {author} {\bibfnamefont {P.}~\bibnamefont {Shor}}, \bibinfo {author}
  {\bibfnamefont {T.}~\bibnamefont {Sleator}}, \bibinfo {author} {\bibfnamefont
  {J.~A.}\ \bibnamefont {Smolin}},\ and\ \bibinfo {author} {\bibfnamefont
  {H.}~\bibnamefont {Weinfurter}},\ }\href
  {https://doi.org/10.1103/PhysRevA.52.3457} {\bibfield  {journal} {\bibinfo
  {journal} {Phys. Rev. A}\ }\textbf {\bibinfo {volume} {52}},\ \bibinfo
  {pages} {3457} (\bibinfo {year} {1995})}\BibitemShut {NoStop}%
\bibitem [{\citenamefont {Clements}\ \emph {et~al.}(2016)\citenamefont
  {Clements}, \citenamefont {Humphreys}, \citenamefont {Metcalf}, \citenamefont
  {Kolthammer},\ and\ \citenamefont {Walmsley}}]{Clements2017Optimal}%
  \BibitemOpen
  \bibfield  {author} {\bibinfo {author} {\bibfnamefont {W.~R.}\ \bibnamefont
  {Clements}}, \bibinfo {author} {\bibfnamefont {P.~C.}\ \bibnamefont
  {Humphreys}}, \bibinfo {author} {\bibfnamefont {B.~J.}\ \bibnamefont
  {Metcalf}}, \bibinfo {author} {\bibfnamefont {W.~S.}\ \bibnamefont
  {Kolthammer}},\ and\ \bibinfo {author} {\bibfnamefont {I.~A.}\ \bibnamefont
  {Walmsley}},\ }\href {https://doi.org/10.1364/OPTICA.3.001460} {\bibfield
  {journal} {\bibinfo  {journal} {Optica}\ }\textbf {\bibinfo {volume} {3}},\
  \bibinfo {pages} {1460} (\bibinfo {year} {2016})}\BibitemShut {NoStop}%
\bibitem [{\citenamefont {Bu\v{z}ek}\ \emph {et~al.}(1997)\citenamefont
  {Bu\v{z}ek}, \citenamefont {Braunstein}, \citenamefont {Hillery},\ and\
  \citenamefont {Bru{\ss}}}]{QuantumClonerStructure}%
  \BibitemOpen
  \bibfield  {author} {\bibinfo {author} {\bibfnamefont {V.}~\bibnamefont
  {Bu\v{z}ek}}, \bibinfo {author} {\bibfnamefont {S.~L.}\ \bibnamefont
  {Braunstein}}, \bibinfo {author} {\bibfnamefont {M.}~\bibnamefont
  {Hillery}},\ and\ \bibinfo {author} {\bibfnamefont {D.}~\bibnamefont
  {Bru{\ss}}},\ }\href {https://api.semanticscholar.org/CorpusID:16909325}
  {\bibfield  {journal} {\bibinfo  {journal} {Phys. Rev. A}\ }\textbf {\bibinfo
  {volume} {56}},\ \bibinfo {pages} {3446} (\bibinfo {year}
  {1997})}\BibitemShut {NoStop}%
\bibitem [{\citenamefont {Lei}\ \emph {et~al.}(2023)\citenamefont {Lei},
  \citenamefont {Asadi}, \citenamefont {Zhong}, \citenamefont {Kuzmich},
  \citenamefont {Simon},\ and\ \citenamefont {Hosseini}}]{QuantumMemory1}%
  \BibitemOpen
  \bibfield  {author} {\bibinfo {author} {\bibfnamefont {Y.}~\bibnamefont
  {Lei}}, \bibinfo {author} {\bibfnamefont {F.~K.}\ \bibnamefont {Asadi}},
  \bibinfo {author} {\bibfnamefont {T.}~\bibnamefont {Zhong}}, \bibinfo
  {author} {\bibfnamefont {A.}~\bibnamefont {Kuzmich}}, \bibinfo {author}
  {\bibfnamefont {C.}~\bibnamefont {Simon}},\ and\ \bibinfo {author}
  {\bibfnamefont {M.}~\bibnamefont {Hosseini}},\ }\href
  {https://doi.org/10.1364/OPTICA.493732} {\bibfield  {journal} {\bibinfo
  {journal} {Optica}\ }\textbf {\bibinfo {volume} {10}},\ \bibinfo {pages}
  {1511} (\bibinfo {year} {2023})}\BibitemShut {NoStop}%
\bibitem [{\citenamefont {Calderbank}\ and\ \citenamefont
  {Shor}(1996)}]{CSSCode1}%
  \BibitemOpen
  \bibfield  {author} {\bibinfo {author} {\bibfnamefont {A.~R.}\ \bibnamefont
  {Calderbank}}\ and\ \bibinfo {author} {\bibfnamefont {P.~W.}\ \bibnamefont
  {Shor}},\ }\href {https://doi.org/10.1103/PhysRevA.54.1098} {\bibfield
  {journal} {\bibinfo  {journal} {Phys. Rev. A}\ }\textbf {\bibinfo {volume}
  {54}},\ \bibinfo {pages} {1098} (\bibinfo {year} {1996})}\BibitemShut
  {NoStop}%
\bibitem [{\citenamefont {Steane}(1996)}]{CSSCode2}%
  \BibitemOpen
  \bibfield  {author} {\bibinfo {author} {\bibfnamefont {A.}~\bibnamefont
  {Steane}},\ }\href {https://doi.org/10.1098/rspa.1996.0136} {\bibfield
  {journal} {\bibinfo  {journal} {Proc. R. Soc. A: Math. Phys. Eng. Sci.}\
  }\textbf {\bibinfo {volume} {452}},\ \bibinfo {pages} {2551–2577} (\bibinfo
  {year} {1996})}\BibitemShut {NoStop}%
\bibitem [{\citenamefont {Bacon}(2006)}]{BaconCode}%
  \BibitemOpen
  \bibfield  {author} {\bibinfo {author} {\bibfnamefont {D.}~\bibnamefont
  {Bacon}},\ }\href {https://doi.org/10.1103/physreva.73.012340} {\bibfield
  {journal} {\bibinfo  {journal} {Phys. Rev. A}\ }\textbf {\bibinfo {volume}
  {73}},\ \bibinfo {pages} {012340} (\bibinfo {year} {2006})}\BibitemShut
  {NoStop}%
\bibitem [{\citenamefont {Kitaev}(2003)}]{ToricCode}%
  \BibitemOpen
  \bibfield  {author} {\bibinfo {author} {\bibfnamefont {A.}~\bibnamefont
  {Kitaev}},\ }\href
  {https://doi.org/https://doi.org/10.1016/S0003-4916(02)00018-0} {\bibfield
  {journal} {\bibinfo  {journal} {Ann. Phys.}\ }\textbf {\bibinfo {volume}
  {303}},\ \bibinfo {pages} {2} (\bibinfo {year} {2003})}\BibitemShut {NoStop}%
\bibitem [{\citenamefont {Fowler}\ \emph {et~al.}(2012)\citenamefont {Fowler},
  \citenamefont {Mariantoni}, \citenamefont {Martinis},\ and\ \citenamefont
  {Cleland}}]{SurfaceCodeBase}%
  \BibitemOpen
  \bibfield  {author} {\bibinfo {author} {\bibfnamefont {A.~G.}\ \bibnamefont
  {Fowler}}, \bibinfo {author} {\bibfnamefont {M.}~\bibnamefont {Mariantoni}},
  \bibinfo {author} {\bibfnamefont {J.~M.}\ \bibnamefont {Martinis}},\ and\
  \bibinfo {author} {\bibfnamefont {A.~N.}\ \bibnamefont {Cleland}},\ }\href
  {https://doi.org/10.1103/PhysRevA.86.032324} {\bibfield  {journal} {\bibinfo
  {journal} {Phys. Rev. A}\ }\textbf {\bibinfo {volume} {86}},\ \bibinfo
  {pages} {032324} (\bibinfo {year} {2012})}\BibitemShut {NoStop}%
\bibitem [{\citenamefont {Stace}\ \emph {et~al.}(2009)\citenamefont {Stace},
  \citenamefont {Barrett},\ and\ \citenamefont
  {Doherty}}]{SurfaceCodeCorrection50P1}%
  \BibitemOpen
  \bibfield  {author} {\bibinfo {author} {\bibfnamefont {T.~M.}\ \bibnamefont
  {Stace}}, \bibinfo {author} {\bibfnamefont {S.~D.}\ \bibnamefont {Barrett}},\
  and\ \bibinfo {author} {\bibfnamefont {A.~C.}\ \bibnamefont {Doherty}},\
  }\href {https://doi.org/10.1103/physrevlett.102.200501} {\bibfield  {journal}
  {\bibinfo  {journal} {Phys. Rev. Lett.}\ }\textbf {\bibinfo {volume} {102}},\
  \bibinfo {pages} {200501} (\bibinfo {year} {2009})}\BibitemShut {NoStop}%
\bibitem [{\citenamefont {Stricker}\ \emph {et~al.}(2020)\citenamefont
  {Stricker}, \citenamefont {Vodola}, \citenamefont {Erhard}, \citenamefont
  {Postler}, \citenamefont {Meth}, \citenamefont {Ringbauer}, \citenamefont
  {Schindler}, \citenamefont {Monz}, \citenamefont {M\"{u}ller},\ and\
  \citenamefont {Blatt}}]{SurfaceCodeCorrection50P2}%
  \BibitemOpen
  \bibfield  {author} {\bibinfo {author} {\bibfnamefont {R.}~\bibnamefont
  {Stricker}}, \bibinfo {author} {\bibfnamefont {D.}~\bibnamefont {Vodola}},
  \bibinfo {author} {\bibfnamefont {A.}~\bibnamefont {Erhard}}, \bibinfo
  {author} {\bibfnamefont {L.}~\bibnamefont {Postler}}, \bibinfo {author}
  {\bibfnamefont {M.}~\bibnamefont {Meth}}, \bibinfo {author} {\bibfnamefont
  {M.}~\bibnamefont {Ringbauer}}, \bibinfo {author} {\bibfnamefont
  {P.}~\bibnamefont {Schindler}}, \bibinfo {author} {\bibfnamefont
  {T.}~\bibnamefont {Monz}}, \bibinfo {author} {\bibfnamefont {M.}~\bibnamefont
  {M\"{u}ller}},\ and\ \bibinfo {author} {\bibfnamefont {R.}~\bibnamefont
  {Blatt}},\ }\href {https://doi.org/10.1038/s41586-020-2667-0} {\bibfield
  {journal} {\bibinfo  {journal} {Nature}\ }\textbf {\bibinfo {volume} {585}},\
  \bibinfo {pages} {207–210} (\bibinfo {year} {2020})}\BibitemShut {NoStop}%
\bibitem [{\citenamefont {Vallero}\ \emph {et~al.}(2024)\citenamefont
  {Vallero}, \citenamefont {Casagranda}, \citenamefont {Vella},\ and\
  \citenamefont {Rech}}]{SurfaceCodeConnectivity}%
  \BibitemOpen
  \bibfield  {author} {\bibinfo {author} {\bibfnamefont {M.}~\bibnamefont
  {Vallero}}, \bibinfo {author} {\bibfnamefont {G.}~\bibnamefont {Casagranda}},
  \bibinfo {author} {\bibfnamefont {F.}~\bibnamefont {Vella}},\ and\ \bibinfo
  {author} {\bibfnamefont {P.}~\bibnamefont {Rech}},\ }\href@noop {} {\bibinfo
  {title} {On the efficacy of surface codes in compensating for radiation
  events in superconducting devices}} (\bibinfo {year} {2024}),\ \Eprint
  {https://arxiv.org/abs/2407.10841} {arXiv:2407.10841 [quant-ph]} \BibitemShut
  {NoStop}%
\bibitem [{\citenamefont {Chabaud}\ \emph {et~al.}(2021)\citenamefont
  {Chabaud}, \citenamefont {Markham},\ and\ \citenamefont
  {Sohbi}}]{QML_Adaptive}%
  \BibitemOpen
  \bibfield  {author} {\bibinfo {author} {\bibfnamefont {U.}~\bibnamefont
  {Chabaud}}, \bibinfo {author} {\bibfnamefont {D.}~\bibnamefont {Markham}},\
  and\ \bibinfo {author} {\bibfnamefont {A.}~\bibnamefont {Sohbi}},\ }\href
  {https://doi.org/10.22331/q-2021-07-05-496} {\bibfield  {journal} {\bibinfo
  {journal} {Quantum}\ }\textbf {\bibinfo {volume} {5}},\ \bibinfo {pages}
  {496} (\bibinfo {year} {2021})}\BibitemShut {NoStop}%
\bibitem [{\citenamefont {Santori}\ \emph {et~al.}(2002)\citenamefont
  {Santori}, \citenamefont {Fattal}, \citenamefont {Vučković}, \citenamefont
  {Solomon},\ and\ \citenamefont {Yamamoto}}]{PhotonIndistinguishability1}%
  \BibitemOpen
  \bibfield  {author} {\bibinfo {author} {\bibfnamefont {C.}~\bibnamefont
  {Santori}}, \bibinfo {author} {\bibfnamefont {D.}~\bibnamefont {Fattal}},
  \bibinfo {author} {\bibfnamefont {J.}~\bibnamefont {Vučković}}, \bibinfo
  {author} {\bibfnamefont {G.~S.}\ \bibnamefont {Solomon}},\ and\ \bibinfo
  {author} {\bibfnamefont {Y.}~\bibnamefont {Yamamoto}},\ }\href
  {https://doi.org/10.1038/nature01086} {\bibfield  {journal} {\bibinfo
  {journal} {Nature}\ }\textbf {\bibinfo {volume} {419}},\ \bibinfo {pages}
  {594–597} (\bibinfo {year} {2002})}\BibitemShut {NoStop}%
\bibitem [{\citenamefont {Komza}\ \emph {et~al.}(2024)\citenamefont {Komza},
  \citenamefont {Samutpraphoot}, \citenamefont {Odeh}, \citenamefont {Tang},
  \citenamefont {Mathew}, \citenamefont {Chang}, \citenamefont {Song},
  \citenamefont {Kim}, \citenamefont {Xiong}, \citenamefont {Hautier},\ and\
  \citenamefont {Sipahigil}}]{PhotonIndistinguishability2}%
  \BibitemOpen
  \bibfield  {author} {\bibinfo {author} {\bibfnamefont {L.}~\bibnamefont
  {Komza}}, \bibinfo {author} {\bibfnamefont {P.}~\bibnamefont
  {Samutpraphoot}}, \bibinfo {author} {\bibfnamefont {M.}~\bibnamefont {Odeh}},
  \bibinfo {author} {\bibfnamefont {Y.-L.}\ \bibnamefont {Tang}}, \bibinfo
  {author} {\bibfnamefont {M.}~\bibnamefont {Mathew}}, \bibinfo {author}
  {\bibfnamefont {J.}~\bibnamefont {Chang}}, \bibinfo {author} {\bibfnamefont
  {H.}~\bibnamefont {Song}}, \bibinfo {author} {\bibfnamefont {M.-K.}\
  \bibnamefont {Kim}}, \bibinfo {author} {\bibfnamefont {Y.}~\bibnamefont
  {Xiong}}, \bibinfo {author} {\bibfnamefont {G.}~\bibnamefont {Hautier}},\
  and\ \bibinfo {author} {\bibfnamefont {A.}~\bibnamefont {Sipahigil}},\ }\href
  {https://doi.org/10.1038/s41467-024-51265-1} {\bibfield  {journal} {\bibinfo
  {journal} {Nat. Commun.}\ }\textbf {\bibinfo {volume} {15}},\ \bibinfo
  {pages} {6920} (\bibinfo {year} {2024})}\BibitemShut {NoStop}%
\bibitem [{\citenamefont {Hoffet}\ \emph {et~al.}(2024)\citenamefont {Hoffet},
  \citenamefont {Lowinski}, \citenamefont {Heller}, \citenamefont
  {Padr\'on-Brito},\ and\ \citenamefont
  {de~Riedmatten}}]{PhotonIndistinguishabilityDifferentSources}%
  \BibitemOpen
  \bibfield  {author} {\bibinfo {author} {\bibfnamefont {F.}~\bibnamefont
  {Hoffet}}, \bibinfo {author} {\bibfnamefont {J.}~\bibnamefont {Lowinski}},
  \bibinfo {author} {\bibfnamefont {L.}~\bibnamefont {Heller}}, \bibinfo
  {author} {\bibfnamefont {A.}~\bibnamefont {Padr\'on-Brito}},\ and\ \bibinfo
  {author} {\bibfnamefont {H.}~\bibnamefont {de~Riedmatten}},\ }\href
  {https://doi.org/10.1103/PRXQuantum.5.030305} {\bibfield  {journal} {\bibinfo
   {journal} {PRX Quantum}\ }\textbf {\bibinfo {volume} {5}},\ \bibinfo {pages}
  {030305} (\bibinfo {year} {2024})}\BibitemShut {NoStop}%
\bibitem [{\citenamefont {Clemens}\ \emph {et~al.}(2004)\citenamefont
  {Clemens}, \citenamefont {Siddiqui},\ and\ \citenamefont
  {Gea-Banacloche}}]{CorrelatedNoise}%
  \BibitemOpen
  \bibfield  {author} {\bibinfo {author} {\bibfnamefont {J.~P.}\ \bibnamefont
  {Clemens}}, \bibinfo {author} {\bibfnamefont {S.}~\bibnamefont {Siddiqui}},\
  and\ \bibinfo {author} {\bibfnamefont {J.}~\bibnamefont {Gea-Banacloche}},\
  }\href {https://doi.org/10.1103/PhysRevA.69.062313} {\bibfield  {journal}
  {\bibinfo  {journal} {Phys. Rev. A}\ }\textbf {\bibinfo {volume} {69}},\
  \bibinfo {pages} {062313} (\bibinfo {year} {2004})}\BibitemShut {NoStop}%
\bibitem [{\citenamefont {McCaskey}\ \emph {et~al.}(2018)\citenamefont
  {McCaskey}, \citenamefont {Dumitrescu}, \citenamefont {Liakh}, \citenamefont
  {Chen}, \citenamefont {Feng},\ and\ \citenamefont
  {Humble}}]{HardwareIndipendent}%
  \BibitemOpen
  \bibfield  {author} {\bibinfo {author} {\bibfnamefont {A.~J.}\ \bibnamefont
  {McCaskey}}, \bibinfo {author} {\bibfnamefont {E.~F.}\ \bibnamefont
  {Dumitrescu}}, \bibinfo {author} {\bibfnamefont {D.}~\bibnamefont {Liakh}},
  \bibinfo {author} {\bibfnamefont {M.}~\bibnamefont {Chen}}, \bibinfo {author}
  {\bibfnamefont {W.}~\bibnamefont {Feng}},\ and\ \bibinfo {author}
  {\bibfnamefont {T.~S.}\ \bibnamefont {Humble}},\ }\href
  {https://doi.org/10.1016/j.softx.2018.07.007} {\bibfield  {journal} {\bibinfo
   {journal} {SoftwareX}\ }\textbf {\bibinfo {volume} {7}},\ \bibinfo {pages}
  {245} (\bibinfo {year} {2018})}\BibitemShut {NoStop}%
\end{thebibliography}

%

\end{document}